\documentclass[12pt]{article}
\usepackage{times,latexsym,amssymb,alg,ifthen}
\usepackage{phcalc}
\usepackage{graphicx,color,geometry}
\newcommand{\ignore}[1]{}

\definecolor{snow}{rgb}{0.99609375,0.9765625,0.9765625}
\definecolor{ghost}{rgb}{0.96875,0.96875,0.99609375}
\definecolor{GhostWhite}{rgb}{0.96875,0.96875,0.99609375}
\definecolor{WhiteSmoke}{rgb}{0.95703125,0.95703125,0.95703125}
\definecolor{gainsboro}{rgb}{0.859375,0.859375,0.859375}
\definecolor{floral}{rgb}{0.99609375,0.9765625,0.9375}
\definecolor{FloralWhite}{rgb}{0.99609375,0.9765625,0.9375}
\definecolor{old}{rgb}{0.98828125,0.95703125,0.8984375}
\definecolor{OldLace}{rgb}{0.98828125,0.95703125,0.8984375}
\definecolor{linen}{rgb}{0.9765625,0.9375,0.8984375}
\definecolor{antique}{rgb}{0.9765625,0.91796875,0.83984375}
\definecolor{AntiqueWhite}{rgb}{0.9765625,0.91796875,0.83984375}
\definecolor{papaya}{rgb}{0.99609375,0.93359375,0.83203125}
\definecolor{PapayaWhip}{rgb}{0.99609375,0.93359375,0.83203125}
\definecolor{blanched}{rgb}{0.99609375,0.91796875,0.80078125}
\definecolor{BlanchedAlmond}{rgb}{0.99609375,0.91796875,0.80078125}
\definecolor{bisque}{rgb}{0.99609375,0.890625,0.765625}
\definecolor{peach}{rgb}{0.99609375,0.8515625,0.72265625}
\definecolor{PeachPuff}{rgb}{0.99609375,0.8515625,0.72265625}
\definecolor{navajo}{rgb}{0.99609375,0.8671875,0.67578125}
\definecolor{NavajoWhite}{rgb}{0.99609375,0.8671875,0.67578125}
\definecolor{moccasin}{rgb}{0.99609375,0.890625,0.70703125}
\definecolor{cornsilk}{rgb}{0.99609375,0.96875,0.859375}
\definecolor{ivory}{rgb}{0.99609375,0.99609375,0.9375}
\definecolor{lemon}{rgb}{0.99609375,0.9765625,0.80078125}
\definecolor{LemonChiffon}{rgb}{0.99609375,0.9765625,0.80078125}
\definecolor{seashell}{rgb}{0.99609375,0.95703125,0.9296875}
\definecolor{honeydew}{rgb}{0.9375,0.99609375,0.9375}
\definecolor{mint}{rgb}{0.95703125,0.99609375,0.9765625}
\definecolor{MintCream}{rgb}{0.95703125,0.99609375,0.9765625}
\definecolor{azure}{rgb}{0.9375,0.99609375,0.99609375}
\definecolor{alice}{rgb}{0.9375,0.96875,0.99609375}
\definecolor{AliceBlue}{rgb}{0.9375,0.96875,0.99609375}
\definecolor{lavender}{rgb}{0.99609375,0.9375,0.95703125}
\definecolor{LavenderBlush}{rgb}{0.99609375,0.9375,0.95703125}
\definecolor{misty}{rgb}{0.99609375,0.890625,0.87890625}
\definecolor{MistyRose}{rgb}{0.99609375,0.890625,0.87890625}
\definecolor{DarkSlateGray}{rgb}{0.18359375,0.30859375,0.30859375}
\definecolor{dim}{rgb}{0.41015625,0.41015625,0.41015625}
\definecolor{DimGray}{rgb}{0.41015625,0.41015625,0.41015625}
\definecolor{dim}{rgb}{0.41015625,0.41015625,0.41015625}
\definecolor{DimGrey}{rgb}{0.41015625,0.41015625,0.41015625}
\definecolor{SlateGray}{rgb}{0.4375,0.5,0.5625}
\definecolor{SlateGrey}{rgb}{0.4375,0.5,0.5625}
\definecolor{LightSlateGray}{rgb}{0.46484375,0.53125,0.59765625}
\definecolor{LightSlateGrey}{rgb}{0.46484375,0.53125,0.59765625}
\definecolor{gray}{rgb}{0.7421875,0.7421875,0.7421875}
\definecolor{grey}{rgb}{0.7421875,0.7421875,0.7421875}
\definecolor{LightGrey}{rgb}{0.82421875,0.82421875,0.82421875}
\definecolor{LightGray}{rgb}{0.82421875,0.82421875,0.82421875}
\definecolor{midnight}{rgb}{0.09765625,0.09765625,0.4375}
\definecolor{MidnightBlue}{rgb}{0.09765625,0.09765625,0.4375}
\definecolor{NavyBlue}{rgb}{0,0,0.5}
\definecolor{cornflower}{rgb}{0.390625,0.58203125,0.92578125}
\definecolor{CornflowerBlue}{rgb}{0.390625,0.58203125,0.92578125}
\definecolor{DarkSlateBlue}{rgb}{0.28125,0.23828125,0.54296875}
\definecolor{SlateBlue}{rgb}{0.4140625,0.3515625,0.80078125}
\definecolor{MediumSlateBlue}{rgb}{0.48046875,0.40625,0.9296875}
\definecolor{light}{rgb}{0.515625,0.4375,0.99609375}
\definecolor{LightSlateBlue}{rgb}{0.515625,0.4375,0.99609375}
\definecolor{MediumBlue}{rgb}{0,0,0.80078125}
\definecolor{royal}{rgb}{0.25390625,0.41015625,0.87890625}
\definecolor{RoyalBlue}{rgb}{0.25390625,0.41015625,0.87890625}
\definecolor{dodger}{rgb}{0.1171875,0.5625,0.99609375}
\definecolor{DodgerBlue}{rgb}{0.1171875,0.5625,0.99609375}
\definecolor{deep}{rgb}{0,0.74609375,0.99609375}
\definecolor{DeepSkyBlue}{rgb}{0,0.74609375,0.99609375}
\definecolor{sky}{rgb}{0.52734375,0.8046875,0.91796875}
\definecolor{SkyBlue}{rgb}{0.52734375,0.8046875,0.91796875}
\definecolor{LightSkyBlue}{rgb}{0.52734375,0.8046875,0.9765625}
\definecolor{steel}{rgb}{0.2734375,0.5078125,0.703125}
\definecolor{SteelBlue}{rgb}{0.2734375,0.5078125,0.703125}
\definecolor{LightSteelBlue}{rgb}{0.6875,0.765625,0.8671875}
\definecolor{LightBlue}{rgb}{0.67578125,0.84375,0.8984375}
\definecolor{powder}{rgb}{0.6875,0.875,0.8984375}
\definecolor{PowderBlue}{rgb}{0.6875,0.875,0.8984375}
\definecolor{PaleTurquoise}{rgb}{0.68359375,0.9296875,0.9296875}
\definecolor{DarkTurquoise}{rgb}{0,0.8046875,0.81640625}
\definecolor{MediumTurquoise}{rgb}{0.28125,0.81640625,0.796875}
\definecolor{turquoise}{rgb}{0.25,0.875,0.8125}
\definecolor{LightCyan}{rgb}{0.875,0.99609375,0.99609375}
\definecolor{cadet}{rgb}{0.37109375,0.6171875,0.625}
\definecolor{CadetBlue}{rgb}{0.37109375,0.6171875,0.625}
\definecolor{MediumAquamarine}{rgb}{0.3984375,0.80078125,0.6640625}
\definecolor{aquamarine}{rgb}{0.49609375,0.99609375,0.828125}
\definecolor{DarkGreen}{rgb}{0,0.390625,0}
\definecolor{DarkOliveGreen}{rgb}{0.33203125,0.41796875,0.18359375}
\definecolor{DarkSeaGreen}{rgb}{0.55859375,0.734375,0.55859375}
\definecolor{sea}{rgb}{0.1796875,0.54296875,0.33984375}
\definecolor{SeaGreen}{rgb}{0.1796875,0.54296875,0.33984375}
\definecolor{MediumSeaGreen}{rgb}{0.234375,0.69921875,0.44140625}
\definecolor{LightSeaGreen}{rgb}{0.125,0.6953125,0.6640625}
\definecolor{PaleGreen}{rgb}{0.59375,0.98046875,0.59375}
\definecolor{spring}{rgb}{0,0.99609375,0.49609375}
\definecolor{SpringGreen}{rgb}{0,0.99609375,0.49609375}
\definecolor{lawn}{rgb}{0.484375,0.984375,0}
\definecolor{LawnGreen}{rgb}{0.484375,0.984375,0}
\definecolor{chartreuse}{rgb}{0.49609375,0.99609375,0}
\definecolor{MediumSpringGreen}{rgb}{0,0.9765625,0.6015625}
\definecolor{GreenYellow}{rgb}{0.67578125,0.99609375,0.18359375}
\definecolor{lime}{rgb}{0.1953125,0.80078125,0.1953125}
\definecolor{LimeGreen}{rgb}{0.1953125,0.80078125,0.1953125}
\definecolor{YellowGreen}{rgb}{0.6015625,0.80078125,0.1953125}
\definecolor{forest}{rgb}{0.1328125,0.54296875,0.1328125}
\definecolor{ForestGreen}{rgb}{0.1328125,0.54296875,0.1328125}
\definecolor{olive}{rgb}{0.41796875,0.5546875,0.13671875}
\definecolor{OliveDrab}{rgb}{0.41796875,0.5546875,0.13671875}
\definecolor{DarkKhaki}{rgb}{0.73828125,0.71484375,0.41796875}
\definecolor{khaki}{rgb}{0.9375,0.8984375,0.546875}
\definecolor{PaleGoldenrod}{rgb}{0.9296875,0.90625,0.6640625}
\definecolor{LightGoldenrodYellow}{rgb}{0.9765625,0.9765625,0.8203125}
\definecolor{LightYellow}{rgb}{0.99609375,0.99609375,0.875}
\definecolor{gold}{rgb}{0.99609375,0.83984375,0}
\definecolor{LightGoldenrod}{rgb}{0.9296875,0.86328125,0.5078125}
\definecolor{goldenrod}{rgb}{0.8515625,0.64453125,0.125}
\definecolor{DarkGoldenrod}{rgb}{0.71875,0.5234375,0.04296875}
\definecolor{rosy}{rgb}{0.734375,0.55859375,0.55859375}
\definecolor{RosyBrown}{rgb}{0.734375,0.55859375,0.55859375}
\definecolor{indian}{rgb}{0.80078125,0.359375,0.359375}
\definecolor{IndianRed}{rgb}{0.80078125,0.359375,0.359375}
\definecolor{saddle}{rgb}{0.54296875,0.26953125,0.07421875}
\definecolor{SaddleBrown}{rgb}{0.54296875,0.26953125,0.07421875}
\definecolor{sienna}{rgb}{0.625,0.3203125,0.17578125}
\definecolor{peru}{rgb}{0.80078125,0.51953125,0.24609375}
\definecolor{burlywood}{rgb}{0.8671875,0.71875,0.52734375}
\definecolor{beige}{rgb}{0.95703125,0.95703125,0.859375}
\definecolor{wheat}{rgb}{0.95703125,0.8671875,0.69921875}
\definecolor{sandy}{rgb}{0.953125,0.640625,0.375}
\definecolor{SandyBrown}{rgb}{0.953125,0.640625,0.375}
\definecolor{tan}{rgb}{0.8203125,0.703125,0.546875}
\definecolor{chocolate}{rgb}{0.8203125,0.41015625,0.1171875}
\definecolor{firebrick}{rgb}{0.6953125,0.1328125,0.1328125}
\definecolor{brown}{rgb}{0.64453125,0.1640625,0.1640625}
\definecolor{DarkSalmon}{rgb}{0.91015625,0.5859375,0.4765625}
\definecolor{salmon}{rgb}{0.9765625,0.5,0.4453125}
\definecolor{LightSalmon}{rgb}{0.99609375,0.625,0.4765625}
\definecolor{orange}{rgb}{0.99609375,0.64453125,0}
\definecolor{DarkOrange}{rgb}{0.99609375,0.546875,0}
\definecolor{coral}{rgb}{0.99609375,0.49609375,0.3125}
\definecolor{LightCoral}{rgb}{0.9375,0.5,0.5}
\definecolor{tomato}{rgb}{0.99609375,0.38671875,0.27734375}
\definecolor{OrangeRed}{rgb}{0.99609375,0.26953125,0}
\definecolor{HotPink}{rgb}{0.99609375,0.41015625,0.703125}
\definecolor{DeepPink}{rgb}{0.99609375,0.078125,0.57421875}
\definecolor{pink}{rgb}{0.99609375,0.75,0.79296875}
\definecolor{LightPink}{rgb}{0.99609375,0.7109375,0.75390625}
\definecolor{PaleVioletRed}{rgb}{0.85546875,0.4375,0.57421875}
\definecolor{maroon}{rgb}{0.6875,0.1875,0.375}
\definecolor{MediumVioletRed}{rgb}{0.77734375,0.08203125,0.51953125}
\definecolor{violet}{rgb}{0.8125,0.125,0.5625}
\definecolor{VioletRed}{rgb}{0.8125,0.125,0.5625}
\definecolor{plum}{rgb}{0.86328125,0.625,0.86328125}
\definecolor{orchid}{rgb}{0.8515625,0.4375,0.8359375}
\definecolor{MediumOrchid}{rgb}{0.7265625,0.33203125,0.82421875}
\definecolor{DarkOrchid}{rgb}{0.59765625,0.1953125,0.796875}
\definecolor{DarkViolet}{rgb}{0.578125,0,0.82421875}
\definecolor{blue}{rgb}{0.5390625,0.16796875,0.8828125}
\definecolor{BlueViolet}{rgb}{0.5390625,0.16796875,0.8828125}
\definecolor{purple}{rgb}{0.625,0.125,0.9375}
\definecolor{MediumPurple}{rgb}{0.57421875,0.4375,0.85546875}
\definecolor{thistle}{rgb}{0.84375,0.74609375,0.84375}
\definecolor{snow1}{rgb}{0.99609375,0.9765625,0.9765625}
\definecolor{snow2}{rgb}{0.9296875,0.91015625,0.91015625}
\definecolor{snow3}{rgb}{0.80078125,0.78515625,0.78515625}
\definecolor{snow4}{rgb}{0.54296875,0.53515625,0.53515625}
\definecolor{seashell1}{rgb}{0.99609375,0.95703125,0.9296875}
\definecolor{seashell2}{rgb}{0.9296875,0.89453125,0.8671875}
\definecolor{seashell3}{rgb}{0.80078125,0.76953125,0.74609375}
\definecolor{seashell4}{rgb}{0.54296875,0.5234375,0.5078125}
\definecolor{AntiqueWhite1}{rgb}{0.99609375,0.93359375,0.85546875}
\definecolor{AntiqueWhite2}{rgb}{0.9296875,0.87109375,0.796875}
\definecolor{AntiqueWhite3}{rgb}{0.80078125,0.75,0.6875}
\definecolor{AntiqueWhite4}{rgb}{0.54296875,0.51171875,0.46875}
\definecolor{bisque1}{rgb}{0.99609375,0.890625,0.765625}
\definecolor{bisque2}{rgb}{0.9296875,0.83203125,0.71484375}
\definecolor{bisque3}{rgb}{0.80078125,0.71484375,0.6171875}
\definecolor{bisque4}{rgb}{0.54296875,0.48828125,0.41796875}
\definecolor{PeachPuff1}{rgb}{0.99609375,0.8515625,0.72265625}
\definecolor{PeachPuff2}{rgb}{0.9296875,0.79296875,0.67578125}
\definecolor{PeachPuff3}{rgb}{0.80078125,0.68359375,0.58203125}
\definecolor{PeachPuff4}{rgb}{0.54296875,0.46484375,0.39453125}
\definecolor{NavajoWhite1}{rgb}{0.99609375,0.8671875,0.67578125}
\definecolor{NavajoWhite2}{rgb}{0.9296875,0.80859375,0.62890625}
\definecolor{NavajoWhite3}{rgb}{0.80078125,0.69921875,0.54296875}
\definecolor{NavajoWhite4}{rgb}{0.54296875,0.47265625,0.3671875}
\definecolor{LemonChiffon1}{rgb}{0.99609375,0.9765625,0.80078125}
\definecolor{LemonChiffon2}{rgb}{0.9296875,0.91015625,0.74609375}
\definecolor{LemonChiffon3}{rgb}{0.80078125,0.78515625,0.64453125}
\definecolor{LemonChiffon4}{rgb}{0.54296875,0.53515625,0.4375}
\definecolor{cornsilk1}{rgb}{0.99609375,0.96875,0.859375}
\definecolor{cornsilk2}{rgb}{0.9296875,0.90625,0.80078125}
\definecolor{cornsilk3}{rgb}{0.80078125,0.78125,0.69140625}
\definecolor{cornsilk4}{rgb}{0.54296875,0.53125,0.46875}
\definecolor{ivory1}{rgb}{0.99609375,0.99609375,0.9375}
\definecolor{ivory2}{rgb}{0.9296875,0.9296875,0.875}
\definecolor{ivory3}{rgb}{0.80078125,0.80078125,0.75390625}
\definecolor{ivory4}{rgb}{0.54296875,0.54296875,0.51171875}
\definecolor{honeydew1}{rgb}{0.9375,0.99609375,0.9375}
\definecolor{honeydew2}{rgb}{0.875,0.9296875,0.875}
\definecolor{honeydew3}{rgb}{0.75390625,0.80078125,0.75390625}
\definecolor{honeydew4}{rgb}{0.51171875,0.54296875,0.51171875}
\definecolor{LavenderBlush1}{rgb}{0.99609375,0.9375,0.95703125}
\definecolor{LavenderBlush2}{rgb}{0.9296875,0.875,0.89453125}
\definecolor{LavenderBlush3}{rgb}{0.80078125,0.75390625,0.76953125}
\definecolor{LavenderBlush4}{rgb}{0.54296875,0.51171875,0.5234375}
\definecolor{MistyRose1}{rgb}{0.99609375,0.890625,0.87890625}
\definecolor{MistyRose2}{rgb}{0.9296875,0.83203125,0.8203125}
\definecolor{MistyRose3}{rgb}{0.80078125,0.71484375,0.70703125}
\definecolor{MistyRose4}{rgb}{0.54296875,0.48828125,0.48046875}
\definecolor{azure1}{rgb}{0.9375,0.99609375,0.99609375}
\definecolor{azure2}{rgb}{0.875,0.9296875,0.9296875}
\definecolor{azure3}{rgb}{0.75390625,0.80078125,0.80078125}
\definecolor{azure4}{rgb}{0.51171875,0.54296875,0.54296875}
\definecolor{SlateBlue1}{rgb}{0.51171875,0.43359375,0.99609375}
\definecolor{SlateBlue2}{rgb}{0.4765625,0.40234375,0.9296875}
\definecolor{SlateBlue3}{rgb}{0.41015625,0.34765625,0.80078125}
\definecolor{SlateBlue4}{rgb}{0.27734375,0.234375,0.54296875}
\definecolor{RoyalBlue1}{rgb}{0.28125,0.4609375,0.99609375}
\definecolor{RoyalBlue2}{rgb}{0.26171875,0.4296875,0.9296875}
\definecolor{RoyalBlue3}{rgb}{0.2265625,0.37109375,0.80078125}
\definecolor{RoyalBlue4}{rgb}{0.15234375,0.25,0.54296875}
\definecolor{blue1}{rgb}{0,0,0.99609375}
\definecolor{blue2}{rgb}{0,0,0.9296875}
\definecolor{blue3}{rgb}{0,0,0.80078125}
\definecolor{blue4}{rgb}{0,0,0.54296875}
\definecolor{DodgerBlue1}{rgb}{0.1171875,0.5625,0.99609375}
\definecolor{DodgerBlue2}{rgb}{0.109375,0.5234375,0.9296875}
\definecolor{DodgerBlue3}{rgb}{0.09375,0.453125,0.80078125}
\definecolor{DodgerBlue4}{rgb}{0.0625,0.3046875,0.54296875}
\definecolor{SteelBlue1}{rgb}{0.38671875,0.71875,0.99609375}
\definecolor{SteelBlue2}{rgb}{0.359375,0.671875,0.9296875}
\definecolor{SteelBlue3}{rgb}{0.30859375,0.578125,0.80078125}
\definecolor{SteelBlue4}{rgb}{0.2109375,0.390625,0.54296875}
\definecolor{DeepSkyBlue1}{rgb}{0,0.74609375,0.99609375}
\definecolor{DeepSkyBlue2}{rgb}{0,0.6953125,0.9296875}
\definecolor{DeepSkyBlue3}{rgb}{0,0.6015625,0.80078125}
\definecolor{DeepSkyBlue4}{rgb}{0,0.40625,0.54296875}
\definecolor{SkyBlue1}{rgb}{0.52734375,0.8046875,0.99609375}
\definecolor{SkyBlue2}{rgb}{0.4921875,0.75,0.9296875}
\definecolor{SkyBlue3}{rgb}{0.421875,0.6484375,0.80078125}
\definecolor{SkyBlue4}{rgb}{0.2890625,0.4375,0.54296875}
\definecolor{LightSkyBlue1}{rgb}{0.6875,0.8828125,0.99609375}
\definecolor{LightSkyBlue2}{rgb}{0.640625,0.82421875,0.9296875}
\definecolor{LightSkyBlue3}{rgb}{0.55078125,0.7109375,0.80078125}
\definecolor{LightSkyBlue4}{rgb}{0.375,0.48046875,0.54296875}
\definecolor{SlateGray1}{rgb}{0.7734375,0.8828125,0.99609375}
\definecolor{SlateGray2}{rgb}{0.72265625,0.82421875,0.9296875}
\definecolor{SlateGray3}{rgb}{0.62109375,0.7109375,0.80078125}
\definecolor{SlateGray4}{rgb}{0.421875,0.48046875,0.54296875}
\definecolor{LightSteelBlue1}{rgb}{0.7890625,0.87890625,0.99609375}
\definecolor{LightSteelBlue2}{rgb}{0.734375,0.8203125,0.9296875}
\definecolor{LightSteelBlue3}{rgb}{0.6328125,0.70703125,0.80078125}
\definecolor{LightSteelBlue4}{rgb}{0.4296875,0.48046875,0.54296875}
\definecolor{LightBlue1}{rgb}{0.74609375,0.93359375,0.99609375}
\definecolor{LightBlue2}{rgb}{0.6953125,0.87109375,0.9296875}
\definecolor{LightBlue3}{rgb}{0.6015625,0.75,0.80078125}
\definecolor{LightBlue4}{rgb}{0.40625,0.51171875,0.54296875}
\definecolor{LightCyan1}{rgb}{0.875,0.99609375,0.99609375}
\definecolor{LightCyan2}{rgb}{0.81640625,0.9296875,0.9296875}
\definecolor{LightCyan3}{rgb}{0.703125,0.80078125,0.80078125}
\definecolor{LightCyan4}{rgb}{0.4765625,0.54296875,0.54296875}
\definecolor{PaleTurquoise1}{rgb}{0.73046875,0.99609375,0.99609375}
\definecolor{PaleTurquoise2}{rgb}{0.6796875,0.9296875,0.9296875}
\definecolor{PaleTurquoise3}{rgb}{0.5859375,0.80078125,0.80078125}
\definecolor{PaleTurquoise4}{rgb}{0.3984375,0.54296875,0.54296875}
\definecolor{CadetBlue1}{rgb}{0.59375,0.95703125,0.99609375}
\definecolor{CadetBlue2}{rgb}{0.5546875,0.89453125,0.9296875}
\definecolor{CadetBlue3}{rgb}{0.4765625,0.76953125,0.80078125}
\definecolor{CadetBlue4}{rgb}{0.32421875,0.5234375,0.54296875}
\definecolor{turquoise1}{rgb}{0,0.95703125,0.99609375}
\definecolor{turquoise2}{rgb}{0,0.89453125,0.9296875}
\definecolor{turquoise3}{rgb}{0,0.76953125,0.80078125}
\definecolor{turquoise4}{rgb}{0,0.5234375,0.54296875}
\definecolor{cyan1}{rgb}{0,0.99609375,0.99609375}
\definecolor{cyan2}{rgb}{0,0.9296875,0.9296875}
\definecolor{cyan3}{rgb}{0,0.80078125,0.80078125}
\definecolor{cyan4}{rgb}{0,0.54296875,0.54296875}
\definecolor{DarkSlateGray1}{rgb}{0.58984375,0.99609375,0.99609375}
\definecolor{DarkSlateGray2}{rgb}{0.55078125,0.9296875,0.9296875}
\definecolor{DarkSlateGray3}{rgb}{0.47265625,0.80078125,0.80078125}
\definecolor{DarkSlateGray4}{rgb}{0.3203125,0.54296875,0.54296875}
\definecolor{aquamarine1}{rgb}{0.49609375,0.99609375,0.828125}
\definecolor{aquamarine2}{rgb}{0.4609375,0.9296875,0.7734375}
\definecolor{aquamarine3}{rgb}{0.3984375,0.80078125,0.6640625}
\definecolor{aquamarine4}{rgb}{0.26953125,0.54296875,0.453125}
\definecolor{DarkSeaGreen1}{rgb}{0.75390625,0.99609375,0.75390625}
\definecolor{DarkSeaGreen2}{rgb}{0.703125,0.9296875,0.703125}
\definecolor{DarkSeaGreen3}{rgb}{0.60546875,0.80078125,0.60546875}
\definecolor{DarkSeaGreen4}{rgb}{0.41015625,0.54296875,0.41015625}
\definecolor{SeaGreen1}{rgb}{0.328125,0.99609375,0.62109375}
\definecolor{SeaGreen2}{rgb}{0.3046875,0.9296875,0.578125}
\definecolor{SeaGreen3}{rgb}{0.26171875,0.80078125,0.5}
\definecolor{SeaGreen4}{rgb}{0.1796875,0.54296875,0.33984375}
\definecolor{PaleGreen1}{rgb}{0.6015625,0.99609375,0.6015625}
\definecolor{PaleGreen2}{rgb}{0.5625,0.9296875,0.5625}
\definecolor{PaleGreen3}{rgb}{0.484375,0.80078125,0.484375}
\definecolor{PaleGreen4}{rgb}{0.328125,0.54296875,0.328125}
\definecolor{SpringGreen1}{rgb}{0,0.99609375,0.49609375}
\definecolor{SpringGreen2}{rgb}{0,0.9296875,0.4609375}
\definecolor{SpringGreen3}{rgb}{0,0.80078125,0.3984375}
\definecolor{SpringGreen4}{rgb}{0,0.54296875,0.26953125}
\definecolor{green1}{rgb}{0,0.99609375,0}
\definecolor{green2}{rgb}{0,0.9296875,0}
\definecolor{green3}{rgb}{0,0.80078125,0}
\definecolor{green4}{rgb}{0,0.54296875,0}
\definecolor{chartreuse1}{rgb}{0.49609375,0.99609375,0}
\definecolor{chartreuse2}{rgb}{0.4609375,0.9296875,0}
\definecolor{chartreuse3}{rgb}{0.3984375,0.80078125,0}
\definecolor{chartreuse4}{rgb}{0.26953125,0.54296875,0}
\definecolor{OliveDrab1}{rgb}{0.75,0.99609375,0.2421875}
\definecolor{OliveDrab2}{rgb}{0.69921875,0.9296875,0.2265625}
\definecolor{OliveDrab3}{rgb}{0.6015625,0.80078125,0.1953125}
\definecolor{OliveDrab4}{rgb}{0.41015625,0.54296875,0.1328125}
\definecolor{DarkOliveGreen1}{rgb}{0.7890625,0.99609375,0.4375}
\definecolor{DarkOliveGreen2}{rgb}{0.734375,0.9296875,0.40625}
\definecolor{DarkOliveGreen3}{rgb}{0.6328125,0.80078125,0.3515625}
\definecolor{DarkOliveGreen4}{rgb}{0.4296875,0.54296875,0.23828125}
\definecolor{khaki1}{rgb}{0.99609375,0.9609375,0.55859375}
\definecolor{khaki2}{rgb}{0.9296875,0.8984375,0.51953125}
\definecolor{khaki3}{rgb}{0.80078125,0.7734375,0.44921875}
\definecolor{khaki4}{rgb}{0.54296875,0.5234375,0.3046875}
\definecolor{LightGoldenrod1}{rgb}{0.99609375,0.921875,0.54296875}
\definecolor{LightGoldenrod2}{rgb}{0.9296875,0.859375,0.5078125}
\definecolor{LightGoldenrod3}{rgb}{0.80078125,0.7421875,0.4375}
\definecolor{LightGoldenrod4}{rgb}{0.54296875,0.50390625,0.296875}
\definecolor{LightYellow1}{rgb}{0.99609375,0.99609375,0.875}
\definecolor{LightYellow2}{rgb}{0.9296875,0.9296875,0.81640625}
\definecolor{LightYellow3}{rgb}{0.80078125,0.80078125,0.703125}
\definecolor{LightYellow4}{rgb}{0.54296875,0.54296875,0.4765625}
\definecolor{yellow1}{rgb}{0.99609375,0.99609375,0}
\definecolor{yellow2}{rgb}{0.9296875,0.9296875,0}
\definecolor{yellow3}{rgb}{0.80078125,0.80078125,0}
\definecolor{yellow4}{rgb}{0.54296875,0.54296875,0}
\definecolor{gold1}{rgb}{0.99609375,0.83984375,0}
\definecolor{gold2}{rgb}{0.9296875,0.78515625,0}
\definecolor{gold3}{rgb}{0.80078125,0.67578125,0}
\definecolor{gold4}{rgb}{0.54296875,0.45703125,0}
\definecolor{goldenrod1}{rgb}{0.99609375,0.75390625,0.14453125}
\definecolor{goldenrod2}{rgb}{0.9296875,0.703125,0.1328125}
\definecolor{goldenrod3}{rgb}{0.80078125,0.60546875,0.11328125}
\definecolor{goldenrod4}{rgb}{0.54296875,0.41015625,0.078125}
\definecolor{DarkGoldenrod1}{rgb}{0.99609375,0.72265625,0.05859375}
\definecolor{DarkGoldenrod2}{rgb}{0.9296875,0.67578125,0.0546875}
\definecolor{DarkGoldenrod3}{rgb}{0.80078125,0.58203125,0.046875}
\definecolor{DarkGoldenrod4}{rgb}{0.54296875,0.39453125,0.03125}
\definecolor{RosyBrown1}{rgb}{0.99609375,0.75390625,0.75390625}
\definecolor{RosyBrown2}{rgb}{0.9296875,0.703125,0.703125}
\definecolor{RosyBrown3}{rgb}{0.80078125,0.60546875,0.60546875}
\definecolor{RosyBrown4}{rgb}{0.54296875,0.41015625,0.41015625}
\definecolor{IndianRed1}{rgb}{0.99609375,0.4140625,0.4140625}
\definecolor{IndianRed2}{rgb}{0.9296875,0.38671875,0.38671875}
\definecolor{IndianRed3}{rgb}{0.80078125,0.33203125,0.33203125}
\definecolor{IndianRed4}{rgb}{0.54296875,0.2265625,0.2265625}
\definecolor{sienna1}{rgb}{0.99609375,0.5078125,0.27734375}
\definecolor{sienna2}{rgb}{0.9296875,0.47265625,0.2578125}
\definecolor{sienna3}{rgb}{0.80078125,0.40625,0.22265625}
\definecolor{sienna4}{rgb}{0.54296875,0.27734375,0.1484375}
\definecolor{burlywood1}{rgb}{0.99609375,0.82421875,0.60546875}
\definecolor{burlywood2}{rgb}{0.9296875,0.76953125,0.56640625}
\definecolor{burlywood3}{rgb}{0.80078125,0.6640625,0.48828125}
\definecolor{burlywood4}{rgb}{0.54296875,0.44921875,0.33203125}
\definecolor{wheat1}{rgb}{0.99609375,0.90234375,0.7265625}
\definecolor{wheat2}{rgb}{0.9296875,0.84375,0.6796875}
\definecolor{wheat3}{rgb}{0.80078125,0.7265625,0.5859375}
\definecolor{wheat4}{rgb}{0.54296875,0.4921875,0.3984375}
\definecolor{tan1}{rgb}{0.99609375,0.64453125,0.30859375}
\definecolor{tan2}{rgb}{0.9296875,0.6015625,0.28515625}
\definecolor{tan3}{rgb}{0.80078125,0.51953125,0.24609375}
\definecolor{tan4}{rgb}{0.54296875,0.3515625,0.16796875}
\definecolor{chocolate1}{rgb}{0.99609375,0.49609375,0.140625}
\definecolor{chocolate2}{rgb}{0.9296875,0.4609375,0.12890625}
\definecolor{chocolate3}{rgb}{0.80078125,0.3984375,0.11328125}
\definecolor{chocolate4}{rgb}{0.54296875,0.26953125,0.07421875}
\definecolor{firebrick1}{rgb}{0.99609375,0.1875,0.1875}
\definecolor{firebrick2}{rgb}{0.9296875,0.171875,0.171875}
\definecolor{firebrick3}{rgb}{0.80078125,0.1484375,0.1484375}
\definecolor{firebrick4}{rgb}{0.54296875,0.1015625,0.1015625}
\definecolor{brown1}{rgb}{0.99609375,0.25,0.25}
\definecolor{brown2}{rgb}{0.9296875,0.23046875,0.23046875}
\definecolor{brown3}{rgb}{0.80078125,0.19921875,0.19921875}
\definecolor{brown4}{rgb}{0.54296875,0.13671875,0.13671875}
\definecolor{salmon1}{rgb}{0.99609375,0.546875,0.41015625}
\definecolor{salmon2}{rgb}{0.9296875,0.5078125,0.3828125}
\definecolor{salmon3}{rgb}{0.80078125,0.4375,0.328125}
\definecolor{salmon4}{rgb}{0.54296875,0.296875,0.22265625}
\definecolor{LightSalmon1}{rgb}{0.99609375,0.625,0.4765625}
\definecolor{LightSalmon2}{rgb}{0.9296875,0.58203125,0.4453125}
\definecolor{LightSalmon3}{rgb}{0.80078125,0.50390625,0.3828125}
\definecolor{LightSalmon4}{rgb}{0.54296875,0.33984375,0.2578125}
\definecolor{orange1}{rgb}{0.99609375,0.64453125,0}
\definecolor{orange2}{rgb}{0.9296875,0.6015625,0}
\definecolor{orange3}{rgb}{0.80078125,0.51953125,0}
\definecolor{orange4}{rgb}{0.54296875,0.3515625,0}
\definecolor{DarkOrange1}{rgb}{0.99609375,0.49609375,0}
\definecolor{DarkOrange2}{rgb}{0.9296875,0.4609375,0}
\definecolor{DarkOrange3}{rgb}{0.80078125,0.3984375,0}
\definecolor{DarkOrange4}{rgb}{0.54296875,0.26953125,0}
\definecolor{coral1}{rgb}{0.99609375,0.4453125,0.3359375}
\definecolor{coral2}{rgb}{0.9296875,0.4140625,0.3125}
\definecolor{coral3}{rgb}{0.80078125,0.35546875,0.26953125}
\definecolor{coral4}{rgb}{0.54296875,0.2421875,0.18359375}
\definecolor{tomato1}{rgb}{0.99609375,0.38671875,0.27734375}
\definecolor{tomato2}{rgb}{0.9296875,0.359375,0.2578125}
\definecolor{tomato3}{rgb}{0.80078125,0.30859375,0.22265625}
\definecolor{tomato4}{rgb}{0.54296875,0.2109375,0.1484375}
\definecolor{OrangeRed1}{rgb}{0.99609375,0.26953125,0}
\definecolor{OrangeRed2}{rgb}{0.9296875,0.25,0}
\definecolor{OrangeRed3}{rgb}{0.80078125,0.21484375,0}
\definecolor{OrangeRed4}{rgb}{0.54296875,0.14453125,0}
\definecolor{red1}{rgb}{0.99609375,0,0}
\definecolor{red2}{rgb}{0.9296875,0,0}
\definecolor{red3}{rgb}{0.80078125,0,0}
\definecolor{red4}{rgb}{0.54296875,0,0}
\definecolor{DeepPink1}{rgb}{0.99609375,0.078125,0.57421875}
\definecolor{DeepPink2}{rgb}{0.9296875,0.0703125,0.53515625}
\definecolor{DeepPink3}{rgb}{0.80078125,0.0625,0.4609375}
\definecolor{DeepPink4}{rgb}{0.54296875,0.0390625,0.3125}
\definecolor{HotPink1}{rgb}{0.99609375,0.4296875,0.703125}
\definecolor{HotPink2}{rgb}{0.9296875,0.4140625,0.65234375}
\definecolor{HotPink3}{rgb}{0.80078125,0.375,0.5625}
\definecolor{HotPink4}{rgb}{0.54296875,0.2265625,0.3828125}
\definecolor{pink1}{rgb}{0.99609375,0.70703125,0.76953125}
\definecolor{pink2}{rgb}{0.9296875,0.66015625,0.71875}
\definecolor{pink3}{rgb}{0.80078125,0.56640625,0.6171875}
\definecolor{pink4}{rgb}{0.54296875,0.38671875,0.421875}
\definecolor{LightPink1}{rgb}{0.99609375,0.6796875,0.72265625}
\definecolor{LightPink2}{rgb}{0.9296875,0.6328125,0.67578125}
\definecolor{LightPink3}{rgb}{0.80078125,0.546875,0.58203125}
\definecolor{LightPink4}{rgb}{0.54296875,0.37109375,0.39453125}
\definecolor{PaleVioletRed1}{rgb}{0.99609375,0.5078125,0.66796875}
\definecolor{PaleVioletRed2}{rgb}{0.9296875,0.47265625,0.62109375}
\definecolor{PaleVioletRed3}{rgb}{0.80078125,0.40625,0.53515625}
\definecolor{PaleVioletRed4}{rgb}{0.54296875,0.27734375,0.36328125}
\definecolor{maroon1}{rgb}{0.99609375,0.203125,0.69921875}
\definecolor{maroon2}{rgb}{0.9296875,0.1875,0.65234375}
\definecolor{maroon3}{rgb}{0.80078125,0.16015625,0.5625}
\definecolor{maroon4}{rgb}{0.54296875,0.109375,0.3828125}
\definecolor{VioletRed1}{rgb}{0.99609375,0.2421875,0.5859375}
\definecolor{VioletRed2}{rgb}{0.9296875,0.2265625,0.546875}
\definecolor{VioletRed3}{rgb}{0.80078125,0.1953125,0.46875}
\definecolor{VioletRed4}{rgb}{0.54296875,0.1328125,0.3203125}
\definecolor{magenta1}{rgb}{0.99609375,0,0.99609375}
\definecolor{magenta2}{rgb}{0.9296875,0,0.9296875}
\definecolor{magenta3}{rgb}{0.80078125,0,0.80078125}
\definecolor{magenta4}{rgb}{0.54296875,0,0.54296875}
\definecolor{orchid1}{rgb}{0.99609375,0.51171875,0.9765625}
\definecolor{orchid2}{rgb}{0.9296875,0.4765625,0.91015625}
\definecolor{orchid3}{rgb}{0.80078125,0.41015625,0.78515625}
\definecolor{orchid4}{rgb}{0.54296875,0.27734375,0.53515625}
\definecolor{plum1}{rgb}{0.99609375,0.73046875,0.99609375}
\definecolor{plum2}{rgb}{0.9296875,0.6796875,0.9296875}
\definecolor{plum3}{rgb}{0.80078125,0.5859375,0.80078125}
\definecolor{plum4}{rgb}{0.54296875,0.3984375,0.54296875}
\definecolor{MediumOrchid1}{rgb}{0.875,0.3984375,0.99609375}
\definecolor{MediumOrchid2}{rgb}{0.81640625,0.37109375,0.9296875}
\definecolor{MediumOrchid3}{rgb}{0.703125,0.3203125,0.80078125}
\definecolor{MediumOrchid4}{rgb}{0.4765625,0.21484375,0.54296875}
\definecolor{DarkOrchid1}{rgb}{0.74609375,0.2421875,0.99609375}
\definecolor{DarkOrchid2}{rgb}{0.6953125,0.2265625,0.9296875}
\definecolor{DarkOrchid3}{rgb}{0.6015625,0.1953125,0.80078125}
\definecolor{DarkOrchid4}{rgb}{0.40625,0.1328125,0.54296875}
\definecolor{purple1}{rgb}{0.60546875,0.1875,0.99609375}
\definecolor{purple2}{rgb}{0.56640625,0.171875,0.9296875}
\definecolor{purple3}{rgb}{0.48828125,0.1484375,0.80078125}
\definecolor{purple4}{rgb}{0.33203125,0.1015625,0.54296875}
\definecolor{MediumPurple1}{rgb}{0.66796875,0.5078125,0.99609375}
\definecolor{MediumPurple2}{rgb}{0.62109375,0.47265625,0.9296875}
\definecolor{MediumPurple3}{rgb}{0.53515625,0.40625,0.80078125}
\definecolor{MediumPurple4}{rgb}{0.36328125,0.27734375,0.54296875}
\definecolor{thistle1}{rgb}{0.99609375,0.87890625,0.99609375}
\definecolor{thistle2}{rgb}{0.9296875,0.8203125,0.9296875}
\definecolor{thistle3}{rgb}{0.80078125,0.70703125,0.80078125}
\definecolor{thistle4}{rgb}{0.54296875,0.48046875,0.54296875}
\definecolor{gray0}{rgb}{0,0,0}
\definecolor{grey0}{rgb}{0,0,0}
\definecolor{gray1}{rgb}{0.01171875,0.01171875,0.01171875}
\definecolor{grey1}{rgb}{0.01171875,0.01171875,0.01171875}
\definecolor{gray2}{rgb}{0.01953125,0.01953125,0.01953125}
\definecolor{grey2}{rgb}{0.01953125,0.01953125,0.01953125}
\definecolor{gray3}{rgb}{0.03125,0.03125,0.03125}
\definecolor{grey3}{rgb}{0.03125,0.03125,0.03125}
\definecolor{gray4}{rgb}{0.0390625,0.0390625,0.0390625}
\definecolor{grey4}{rgb}{0.0390625,0.0390625,0.0390625}
\definecolor{gray5}{rgb}{0.05078125,0.05078125,0.05078125}
\definecolor{grey5}{rgb}{0.05078125,0.05078125,0.05078125}
\definecolor{gray6}{rgb}{0.05859375,0.05859375,0.05859375}
\definecolor{grey6}{rgb}{0.05859375,0.05859375,0.05859375}
\definecolor{gray7}{rgb}{0.0703125,0.0703125,0.0703125}
\definecolor{grey7}{rgb}{0.0703125,0.0703125,0.0703125}
\definecolor{gray8}{rgb}{0.078125,0.078125,0.078125}
\definecolor{grey8}{rgb}{0.078125,0.078125,0.078125}
\definecolor{gray9}{rgb}{0.08984375,0.08984375,0.08984375}
\definecolor{grey9}{rgb}{0.08984375,0.08984375,0.08984375}
\definecolor{gray10}{rgb}{0.1015625,0.1015625,0.1015625}
\definecolor{grey10}{rgb}{0.1015625,0.1015625,0.1015625}
\definecolor{gray11}{rgb}{0.109375,0.109375,0.109375}
\definecolor{grey11}{rgb}{0.109375,0.109375,0.109375}
\definecolor{gray12}{rgb}{0.12109375,0.12109375,0.12109375}
\definecolor{grey12}{rgb}{0.12109375,0.12109375,0.12109375}
\definecolor{gray13}{rgb}{0.12890625,0.12890625,0.12890625}
\definecolor{grey13}{rgb}{0.12890625,0.12890625,0.12890625}
\definecolor{gray14}{rgb}{0.140625,0.140625,0.140625}
\definecolor{grey14}{rgb}{0.140625,0.140625,0.140625}
\definecolor{gray15}{rgb}{0.1484375,0.1484375,0.1484375}
\definecolor{grey15}{rgb}{0.1484375,0.1484375,0.1484375}
\definecolor{gray16}{rgb}{0.16015625,0.16015625,0.16015625}
\definecolor{grey16}{rgb}{0.16015625,0.16015625,0.16015625}
\definecolor{gray17}{rgb}{0.16796875,0.16796875,0.16796875}
\definecolor{grey17}{rgb}{0.16796875,0.16796875,0.16796875}
\definecolor{gray18}{rgb}{0.1796875,0.1796875,0.1796875}
\definecolor{grey18}{rgb}{0.1796875,0.1796875,0.1796875}
\definecolor{gray19}{rgb}{0.1875,0.1875,0.1875}
\definecolor{grey19}{rgb}{0.1875,0.1875,0.1875}
\definecolor{gray20}{rgb}{0.19921875,0.19921875,0.19921875}
\definecolor{grey20}{rgb}{0.19921875,0.19921875,0.19921875}
\definecolor{gray21}{rgb}{0.2109375,0.2109375,0.2109375}
\definecolor{grey21}{rgb}{0.2109375,0.2109375,0.2109375}
\definecolor{gray22}{rgb}{0.21875,0.21875,0.21875}
\definecolor{grey22}{rgb}{0.21875,0.21875,0.21875}
\definecolor{gray23}{rgb}{0.23046875,0.23046875,0.23046875}
\definecolor{grey23}{rgb}{0.23046875,0.23046875,0.23046875}
\definecolor{gray24}{rgb}{0.23828125,0.23828125,0.23828125}
\definecolor{grey24}{rgb}{0.23828125,0.23828125,0.23828125}
\definecolor{gray25}{rgb}{0.25,0.25,0.25}
\definecolor{grey25}{rgb}{0.25,0.25,0.25}
\definecolor{gray26}{rgb}{0.2578125,0.2578125,0.2578125}
\definecolor{grey26}{rgb}{0.2578125,0.2578125,0.2578125}
\definecolor{gray27}{rgb}{0.26953125,0.26953125,0.26953125}
\definecolor{grey27}{rgb}{0.26953125,0.26953125,0.26953125}
\definecolor{gray28}{rgb}{0.27734375,0.27734375,0.27734375}
\definecolor{grey28}{rgb}{0.27734375,0.27734375,0.27734375}
\definecolor{gray29}{rgb}{0.2890625,0.2890625,0.2890625}
\definecolor{grey29}{rgb}{0.2890625,0.2890625,0.2890625}
\definecolor{gray30}{rgb}{0.30078125,0.30078125,0.30078125}
\definecolor{grey30}{rgb}{0.30078125,0.30078125,0.30078125}
\definecolor{gray31}{rgb}{0.30859375,0.30859375,0.30859375}
\definecolor{grey31}{rgb}{0.30859375,0.30859375,0.30859375}
\definecolor{gray32}{rgb}{0.3203125,0.3203125,0.3203125}
\definecolor{grey32}{rgb}{0.3203125,0.3203125,0.3203125}
\definecolor{gray33}{rgb}{0.328125,0.328125,0.328125}
\definecolor{grey33}{rgb}{0.328125,0.328125,0.328125}
\definecolor{gray34}{rgb}{0.33984375,0.33984375,0.33984375}
\definecolor{grey34}{rgb}{0.33984375,0.33984375,0.33984375}
\definecolor{gray35}{rgb}{0.34765625,0.34765625,0.34765625}
\definecolor{grey35}{rgb}{0.34765625,0.34765625,0.34765625}
\definecolor{gray36}{rgb}{0.359375,0.359375,0.359375}
\definecolor{grey36}{rgb}{0.359375,0.359375,0.359375}
\definecolor{gray37}{rgb}{0.3671875,0.3671875,0.3671875}
\definecolor{grey37}{rgb}{0.3671875,0.3671875,0.3671875}
\definecolor{gray38}{rgb}{0.37890625,0.37890625,0.37890625}
\definecolor{grey38}{rgb}{0.37890625,0.37890625,0.37890625}
\definecolor{gray39}{rgb}{0.38671875,0.38671875,0.38671875}
\definecolor{grey39}{rgb}{0.38671875,0.38671875,0.38671875}
\definecolor{gray40}{rgb}{0.3984375,0.3984375,0.3984375}
\definecolor{grey40}{rgb}{0.3984375,0.3984375,0.3984375}
\definecolor{gray41}{rgb}{0.41015625,0.41015625,0.41015625}
\definecolor{grey41}{rgb}{0.41015625,0.41015625,0.41015625}
\definecolor{gray42}{rgb}{0.41796875,0.41796875,0.41796875}
\definecolor{grey42}{rgb}{0.41796875,0.41796875,0.41796875}
\definecolor{gray43}{rgb}{0.4296875,0.4296875,0.4296875}
\definecolor{grey43}{rgb}{0.4296875,0.4296875,0.4296875}
\definecolor{gray44}{rgb}{0.4375,0.4375,0.4375}
\definecolor{grey44}{rgb}{0.4375,0.4375,0.4375}
\definecolor{gray45}{rgb}{0.44921875,0.44921875,0.44921875}
\definecolor{grey45}{rgb}{0.44921875,0.44921875,0.44921875}
\definecolor{gray46}{rgb}{0.45703125,0.45703125,0.45703125}
\definecolor{grey46}{rgb}{0.45703125,0.45703125,0.45703125}
\definecolor{gray47}{rgb}{0.46875,0.46875,0.46875}
\definecolor{grey47}{rgb}{0.46875,0.46875,0.46875}
\definecolor{gray48}{rgb}{0.4765625,0.4765625,0.4765625}
\definecolor{grey48}{rgb}{0.4765625,0.4765625,0.4765625}
\definecolor{gray49}{rgb}{0.48828125,0.48828125,0.48828125}
\definecolor{grey49}{rgb}{0.48828125,0.48828125,0.48828125}
\definecolor{gray50}{rgb}{0.49609375,0.49609375,0.49609375}
\definecolor{grey50}{rgb}{0.49609375,0.49609375,0.49609375}
\definecolor{gray51}{rgb}{0.5078125,0.5078125,0.5078125}
\definecolor{grey51}{rgb}{0.5078125,0.5078125,0.5078125}
\definecolor{gray52}{rgb}{0.51953125,0.51953125,0.51953125}
\definecolor{grey52}{rgb}{0.51953125,0.51953125,0.51953125}
\definecolor{gray53}{rgb}{0.52734375,0.52734375,0.52734375}
\definecolor{grey53}{rgb}{0.52734375,0.52734375,0.52734375}
\definecolor{gray54}{rgb}{0.5390625,0.5390625,0.5390625}
\definecolor{grey54}{rgb}{0.5390625,0.5390625,0.5390625}
\definecolor{gray55}{rgb}{0.546875,0.546875,0.546875}
\definecolor{grey55}{rgb}{0.546875,0.546875,0.546875}
\definecolor{gray56}{rgb}{0.55859375,0.55859375,0.55859375}
\definecolor{grey56}{rgb}{0.55859375,0.55859375,0.55859375}
\definecolor{gray57}{rgb}{0.56640625,0.56640625,0.56640625}
\definecolor{grey57}{rgb}{0.56640625,0.56640625,0.56640625}
\definecolor{gray58}{rgb}{0.578125,0.578125,0.578125}
\definecolor{grey58}{rgb}{0.578125,0.578125,0.578125}
\definecolor{gray59}{rgb}{0.5859375,0.5859375,0.5859375}
\definecolor{grey59}{rgb}{0.5859375,0.5859375,0.5859375}
\definecolor{gray60}{rgb}{0.59765625,0.59765625,0.59765625}
\definecolor{grey60}{rgb}{0.59765625,0.59765625,0.59765625}
\definecolor{gray61}{rgb}{0.609375,0.609375,0.609375}
\definecolor{grey61}{rgb}{0.609375,0.609375,0.609375}
\definecolor{gray62}{rgb}{0.6171875,0.6171875,0.6171875}
\definecolor{grey62}{rgb}{0.6171875,0.6171875,0.6171875}
\definecolor{gray63}{rgb}{0.62890625,0.62890625,0.62890625}
\definecolor{grey63}{rgb}{0.62890625,0.62890625,0.62890625}
\definecolor{gray64}{rgb}{0.63671875,0.63671875,0.63671875}
\definecolor{grey64}{rgb}{0.63671875,0.63671875,0.63671875}
\definecolor{gray65}{rgb}{0.6484375,0.6484375,0.6484375}
\definecolor{grey65}{rgb}{0.6484375,0.6484375,0.6484375}
\definecolor{gray66}{rgb}{0.65625,0.65625,0.65625}
\definecolor{grey66}{rgb}{0.65625,0.65625,0.65625}
\definecolor{gray67}{rgb}{0.66796875,0.66796875,0.66796875}
\definecolor{grey67}{rgb}{0.66796875,0.66796875,0.66796875}
\definecolor{gray68}{rgb}{0.67578125,0.67578125,0.67578125}
\definecolor{grey68}{rgb}{0.67578125,0.67578125,0.67578125}
\definecolor{gray69}{rgb}{0.6875,0.6875,0.6875}
\definecolor{grey69}{rgb}{0.6875,0.6875,0.6875}
\definecolor{gray70}{rgb}{0.69921875,0.69921875,0.69921875}
\definecolor{grey70}{rgb}{0.69921875,0.69921875,0.69921875}
\definecolor{gray71}{rgb}{0.70703125,0.70703125,0.70703125}
\definecolor{grey71}{rgb}{0.70703125,0.70703125,0.70703125}
\definecolor{gray72}{rgb}{0.71875,0.71875,0.71875}
\definecolor{grey72}{rgb}{0.71875,0.71875,0.71875}
\definecolor{gray73}{rgb}{0.7265625,0.7265625,0.7265625}
\definecolor{grey73}{rgb}{0.7265625,0.7265625,0.7265625}
\definecolor{gray74}{rgb}{0.73828125,0.73828125,0.73828125}
\definecolor{grey74}{rgb}{0.73828125,0.73828125,0.73828125}
\definecolor{gray75}{rgb}{0.74609375,0.74609375,0.74609375}
\definecolor{grey75}{rgb}{0.74609375,0.74609375,0.74609375}
\definecolor{gray76}{rgb}{0.7578125,0.7578125,0.7578125}
\definecolor{grey76}{rgb}{0.7578125,0.7578125,0.7578125}
\definecolor{gray77}{rgb}{0.765625,0.765625,0.765625}
\definecolor{grey77}{rgb}{0.765625,0.765625,0.765625}
\definecolor{gray78}{rgb}{0.77734375,0.77734375,0.77734375}
\definecolor{grey78}{rgb}{0.77734375,0.77734375,0.77734375}
\definecolor{gray79}{rgb}{0.78515625,0.78515625,0.78515625}
\definecolor{grey79}{rgb}{0.78515625,0.78515625,0.78515625}
\definecolor{gray80}{rgb}{0.796875,0.796875,0.796875}
\definecolor{grey80}{rgb}{0.796875,0.796875,0.796875}
\definecolor{gray81}{rgb}{0.80859375,0.80859375,0.80859375}
\definecolor{grey81}{rgb}{0.80859375,0.80859375,0.80859375}
\definecolor{gray82}{rgb}{0.81640625,0.81640625,0.81640625}
\definecolor{grey82}{rgb}{0.81640625,0.81640625,0.81640625}
\definecolor{gray83}{rgb}{0.828125,0.828125,0.828125}
\definecolor{grey83}{rgb}{0.828125,0.828125,0.828125}
\definecolor{gray84}{rgb}{0.8359375,0.8359375,0.8359375}
\definecolor{grey84}{rgb}{0.8359375,0.8359375,0.8359375}
\definecolor{gray85}{rgb}{0.84765625,0.84765625,0.84765625}
\definecolor{grey85}{rgb}{0.84765625,0.84765625,0.84765625}
\definecolor{gray86}{rgb}{0.85546875,0.85546875,0.85546875}
\definecolor{grey86}{rgb}{0.85546875,0.85546875,0.85546875}
\definecolor{gray87}{rgb}{0.8671875,0.8671875,0.8671875}
\definecolor{grey87}{rgb}{0.8671875,0.8671875,0.8671875}
\definecolor{gray88}{rgb}{0.875,0.875,0.875}
\definecolor{grey88}{rgb}{0.875,0.875,0.875}
\definecolor{gray89}{rgb}{0.88671875,0.88671875,0.88671875}
\definecolor{grey89}{rgb}{0.88671875,0.88671875,0.88671875}
\definecolor{gray90}{rgb}{0.89453125,0.89453125,0.89453125}
\definecolor{grey90}{rgb}{0.89453125,0.89453125,0.89453125}
\definecolor{gray91}{rgb}{0.90625,0.90625,0.90625}
\definecolor{grey91}{rgb}{0.90625,0.90625,0.90625}
\definecolor{gray92}{rgb}{0.91796875,0.91796875,0.91796875}
\definecolor{grey92}{rgb}{0.91796875,0.91796875,0.91796875}
\definecolor{gray93}{rgb}{0.92578125,0.92578125,0.92578125}
\definecolor{grey93}{rgb}{0.92578125,0.92578125,0.92578125}
\definecolor{gray94}{rgb}{0.9375,0.9375,0.9375}
\definecolor{grey94}{rgb}{0.9375,0.9375,0.9375}
\definecolor{gray95}{rgb}{0.9453125,0.9453125,0.9453125}
\definecolor{grey95}{rgb}{0.9453125,0.9453125,0.9453125}
\definecolor{gray96}{rgb}{0.95703125,0.95703125,0.95703125}
\definecolor{grey96}{rgb}{0.95703125,0.95703125,0.95703125}
\definecolor{gray97}{rgb}{0.96484375,0.96484375,0.96484375}
\definecolor{grey97}{rgb}{0.96484375,0.96484375,0.96484375}
\definecolor{gray98}{rgb}{0.9765625,0.9765625,0.9765625}
\definecolor{grey98}{rgb}{0.9765625,0.9765625,0.9765625}
\definecolor{gray99}{rgb}{0.984375,0.984375,0.984375}
\definecolor{grey99}{rgb}{0.984375,0.984375,0.984375}
\definecolor{gray100}{rgb}{0.99609375,0.99609375,0.99609375}
\definecolor{grey100}{rgb}{0.99609375,0.99609375,0.99609375}
\definecolor{DarkGrey}{rgb}{0.66015625,0.66015625,0.66015625}
\definecolor{DarkGray}{rgb}{0.66015625,0.66015625,0.66015625}
\definecolor{DarkBlue}{rgb}{0,0,0.54296875}
\definecolor{DarkCyan}{rgb}{0,0.54296875,0.54296875}
\definecolor{DarkMagenta}{rgb}{0.54296875,0,0.54296875}
\definecolor{DarkRed}{rgb}{0.54296875,0,0}
\definecolor{LightGreen}{rgb}{0.5625,0.9296875,0.5625}

\usepackage{html}

\newcommand{\makecnts}{\tableofcontents\newpage}
\newcommand{\makeaddress}{

\noindent\textbf{Addresses:}
\begin{tabular}[t]{lcl}
E. Knill:&
Los Alamos National Laboratory&\htmladdnormallink{\texttt{knill@lanl.gov}}{mailto:knill@lanl.gov}\\
R. Laflamme:&
University of Waterloo and Perimeter Institute
&\htmladdnormallink{\texttt{laflamme@iqc.ca}}{mailto:laflamme@iqc.ca}
\\
H. Barnum: &Los Alamos National Laboratory&\htmladdnormallink{\texttt{barnum@lanl.gov}}{mailto:barnum@lanl.gov}
\\
D. Dalvit: &''&\htmladdnormallink{\texttt{dalvit@lanl.gov}}{mailto:dalvit@lanl.gov}
\\
J. Dziarmaga: &''&\htmladdnormallink{\texttt{jpd@lanl.gov}}{mailto:jpd@lanl.gov}
\\
J. Gubernatis: &''&\htmladdnormallink{\texttt{jg@lanl.gov}}{mailto:jg@lanl.gov}
\\
L. Gurvits: &''&\htmladdnormallink{\texttt{gurvits@lanl.gov}}{mailto:gurvits@lanl.gov}
\\
G. Ortiz: &''&\htmladdnormallink{\texttt{g\string_ortiz@lanl.gov}}{mailto:g_ortiz@lanl.gov}
\\
L. Viola: &''&\htmladdnormallink{\texttt{lviola@lanl.gov}}{mailto:lviola@lanl.gov}
\\
W. H. Zurek: &''&\htmladdnormallink{\texttt{whz@lanl.gov}}{mailto:whz@lanl.gov}
\end{tabular}
}

\newcommand{\latexignore}[1]{#1}

\newcommand{\seqIndex}[2]{#1_{#2}}
\newcommand{\assignfrom}{\leftarrow}

\newcommand{\isq}[1]{\ulcorner #1\urcorner}

\newcommand{\pcomment}[1]{\algbegin\mbox{\textbf{C: }}\textsl{#1}\\\algend}

\newcommand{\ealgend}{\algend\mbox{\textbf{end}}\\}
\newenvironment{pseudocode}[1]{\par
\vspace{\baselineskip}
\setlength{\parindent}{0in}
\setlength{\parskip}{0in}
#1\par
}{
\vspace{\baselineskip}\par
}

\graphicspath{{./}{graphics/}}

\newcommand{\qvbar}{\mbox{$|\hspace*{-3pt}|\hspace*{-3pt}|$}}

\newcommand{\qrangle}{\mbox{$\rangle\hspace*{-4.3pt}\rangle\hspace*{-4.3pt}\rangle$}}
\newcommand{\qlangle}{\mbox{$\langle\hspace*{-4.3pt}\langle\hspace*{-4.3pt}\langle$}}

\newcommand{\sysfnt}{\mathsf}

\newcommand{\ket}[1]{\qvbar{#1}\qrangle}
\newcommand{\bra}[1]{\qlangle{#1}\qvbar}
\newcommand{\braket}[2]{\qlangle{#1}\qvbar{#2}\qrangle}
\newcommand{\ketbra}[2]{\ket{#1}\bra{#2}}
\newcommand{\kets}[2]{\qvbar{#1}\qrangle_{{}_{\!\!{\sysfnt{#2}}}}}
\newcommand{\bras}[2]{{}^{\scriptscriptstyle\sysfnt{ #2}}\!\qlangle{#1}\qvbar}
\newcommand{\brakets}[3]{{}^{\scriptscriptstyle\sysfnt{ #3}}\!\qlangle{#1}\qvbar{#2}\qrangle_{{}_{\!\!{\sysfnt{#3}}}}}
\newcommand{\ketbras}[3]{\kets{#1}{#3}\!\!\bras{#2}{#3}}
\newcommand{\slb}[2]{{#1}^{({\sysfnt{#2}})}}

\newcommand{\qavec}[2]{\left(\!\begin{array}{c}#1\\ #2\end{array}\!\right)}
\newcommand{\qaop}[4]{\left(\begin{array}{cc}#1&#2\\ #3&#4\end{array}\right)}

\newcommand{\nputbox}[3]{\put(#1){\makebox(0,0)[#2]{#3}}}
\newcommand{\nputgr}[4]{\put(#1){\makebox(0,0)[#2]{\includegraphics[#3]{#4}}}}
\newcommand{\ppbox}[3]{\begin{picture}(0,0)(0,0)\put(#1){\makebox(0,0)[#2]{#3}}\end{picture}}

\newlength{\elimdepthdim}
\newlength{\elimheightdim}
\newlength{\elimwidthdim}
\newcommand{\elimdepth}[1]{
\setlength{\elimdepthdim}{\depthof{#1}}
\setlength{\elimheightdim}{\heightof{#1}}
\setlength{\elimwidthdim}{\widthof{#1}}
\raisebox{0in}[\elimheightdim][0in]{#1}
}

\newlength{\strutdepthdim}
\newlength{\strutheightdim}
\newlength{\strutwidthdim}
\newcommand{\strutlike}[1]{
\setlength{\strutdepthdim}{\depthof{#1}}
\setlength{\strutheightdim}{\heightof{#1}}
\rule[-\strutdepthdim]{0in}{\strutheightdim+\strutdepthdim}
}

\newcommand{\iboxlike}[1]{%
\setlength{\strutdepthdim}{\depthof{#1}}%
\setlength{\strutheightdim}{\heightof{#1}}%
\setlength{\strutwidthdim}{\widthof{#1}}%
\rule[-\strutdepthdim]{0in}{\strutheightdim+\strutdepthdim}%
\rule{\strutwidthdim}{0in}%
}

\newcommand{\mod}{\;\mathrm{mod}\;}

\newcommand{\cO}{{\cal O}}

\newcommand{\tensor}{\otimes}
\newcommand{\trace}{\mbox{tr}}

\def\id{{\mathchoice {\rm 1\mskip-4mu l} {\rm 1\mskip-4mu l} {\rm
1\mskip-4.5mu l} {\rm 1\mskip-5mu l}}}

\newcommand{\mb}[1]{\mathbf{#1}}

\newcommand{\bitzero}{\mathfrak{0}}
\newcommand{\bitone}{\mathfrak{1}}
\newcommand{\stfnt}{\mathfrak}
\newcommand{\idop}{\id}

\newcommand{\problb}{\{}
\newcommand{\probrb}{\}}
\newcommand{\probplus}{,}
\newcommand{\probmul}{{:}}

\newcounter{herefignum}
\newenvironment{herefig}{\begin{center}\refstepcounter{herefignum}}{\end{center}}
\newcommand{\herefigcap}[1]{\\\begin{minipage}{\textwidth}{FIG.~\theherefignum: #1}\end{minipage}}

\begin{document}

\title{Introduction to Quantum Information Processing}
\author{E. Knill, R. Laflamme, H. Barnum, D. Dalvit, J. Dziarmaga,\\
J. Gubernatis, L. Gurvits, G. Ortiz, L. Viola and W. H. Zurek}
\date{\today}

\maketitle

\begin{latexonly}
\makecnts
\end{latexonly}

\ignore{
As a result of the capabilities of quantum information, the science of
quantum information processing is now a prospering, interdisciplinary
field focused on better understanding the possibilities and
limitations of the underlying theory, on developing new applications
of quantum information and on physically realizing controllable
quantum devices.  The purpose of this primer is to provide an
elementary introduction to quantum information processing, and then to
briefly explain how we hope to exploit the advantages of quantum
information.  These two sections can be read independently.  For
reference, we have included a glossary of the main terms of quantum
information.  
}

\begin{quote}
\textsc{\textbf{Quantum information processing, science of}} - The
theoretical, experimental and technological areas covering the
use of quantum mechanics for communication and
computation.

\hspace*{\fill}Kluwer Encyclopedia of Mathematics, Supplement III
\end{quote}

Research of the last few decades has established that quantum
information, or information based on quantum mechanics, has
capabilities that exceed those of traditional ``classical''
information.  For example, in communication, quantum information
enables quantum cryptography~\cite{wiesner:qc1983a,bennett:qc1982a},
which is a method for communicating in secret. Secrecy is guaranteed
because eavesdropping attempts necessarily disturb the exchanged
quantum information without revealing the content of the
communication.  In computation, quantum information enables efficient
simulation of quantum physics~\cite{feynman:qc1982a}, a task for which
general purpose, efficient, classical algorithms are not known to
exist.  Quantum information also leads to efficient algorithms for
factoring of large numbers~\cite{shor:qc1994a,shor:qc1995a}, which is
believed to be difficult for classical computers. An efficient
factoring algorithm would break the security of commonly used public
key cryptographic codes used for authenticating and securing internet
communications.  A fourth application of quantum information improves
the efficiency with which unstructured search problems can be
solved~\cite{grover:qc1995a}.  Quantum unstructured search may make it
possible to solve significantly larger instances of optimization
problems such as the scheduling and traveling salesman problems.

As a result of the capabilities of quantum information, the science of
quantum information processing is now a prospering, interdisciplinary
field focused on better understanding the possibilities and
limitations of the underlying theory, on developing new applications
of quantum information and on physically realizing controllable
quantum devices.  The purpose of this primer is to provide an
elementary introduction to quantum information processing
(Sect.~\ref{sect:qi}), and then to briefly explain how we hope to
exploit the advantages of quantum information (Sect.~\ref{sect:aqi}).
These two sections can be read independently.  For reference, we have
included a glossary (Sect.~\ref{sect:glossary}) of the main terms of
quantum information.

When we use the word ``information'', we generally think of the things
we can talk about, broadcast, write down, or otherwise record. Such
records can exist in many forms, such as sound waves, electrical signals
in a telephone wire, characters on paper, pit patterns on an optical
disk, or magnetization on a computer hard disk.  A crucial property of
information is that it is ``fungible'': It can be represented in many
different physical forms and easily converted from one form to
another without changing its meaning.  In this sense information
exists independently of the devices used to represent it, but requires
at least one physical representation to be useful.

We call the familiar information stored in today's computers
``classical'' or ``deterministic'' to distinguish it from quantum
information. It is no accident that classical information is the basis
of all human knowledge. Any information passing through our senses is
best modeled by classical discrete or continuous
information. Therefore, when considering any other kind of
information, we need to provide a method for extracting classically
meaningful information.  We begin by recalling the basic ideas of
classical information in a way that illustrates the general procedure
for building an information processing theory.

\section{Classical Information}
\label{sec:class}

The basic unit of classical deterministic information is the ``bit''.
A bit is an abstract entity or ``system'' that can be in one of the
two states symbolized by $\bitzero$ and $\bitone$. At this point, the
symbols for the two states have no numeric meaning. That is why
we have used a font different from that for the numbers $0$
and $1$.  By making a clear distinction between the bit
and its states we emphasize that a bit should be
physically realized as a system or device whose states correspond to
the ideal bit's states.  For example, if you are reading this primer
on paper, the system used to realize a bit is a reserved location on
the surface, and the state depends on the pattern of ink ($\bitzero$
or $\bitone$) in that location. In a computer, the device realizing a
bit can be a combination of transistors and other integrated circuit
elements with the state of the bit determined by the distribution of
charge.

In order to make use of information it must be possible to manipulate
(or ``process'') the states of information units.  The elementary
operations that can be used for this purpose are called ``gates''. Two
one-bit gates are the $\mb{not}$ and the $\mb{reset}$ gates. Applying
the $\mb{not}$ gate to a bit has the effect of ``flipping'' the state
of the bit.  For example, if the initial state of the bit is
$\bitzero$, then the state after applying $\mb{not}$ is
$\mb{not}(\bitzero)=\bitone$. We can present the effect of the gate in
the following form:
\begin{equation}
\begin{array}{ccc}
  \textrm{Initial State}&&\textrm{Final State}\\
  \bitzero &\rightarrow & \mb{not}(\bitzero)  =  \bitone,\\
  \bitone &\rightarrow & \mb{not}(\bitone)  =  \bitzero.
\end{array}
\end{equation}
The $\mb{reset}$ gate sets the state to $\bitzero$ regardless
of the input:
\begin{equation}
\begin{array}{ccc}
  \textrm{Initial State}&&\textrm{Final State}\\
  \bitzero &\rightarrow & \mb{reset}(\bitzero)  =  \bitzero,\\
  \bitone &\rightarrow & \mb{reset}(\bitone)  =  \bitzero.
\end{array}
\end{equation}
By applying a combination of $\mb{not}$ and $\mb{reset}$ gates one can
transform the state of a bit in every possible way.

Information units can be combined to represent more information.  Bits
are typically combined into sequences. The states of such a sequence
are symbolized by strings of state symbols for the constituent bits.
For example a two-bit sequence can be in one of the following
four states: $\bitzero\bitzero,\bitzero\bitone,\bitone\bitzero$ and
$\bitone\bitone$. The different bits are distinguished by their
position in the sequence.

The one-bit gates can be applied to any bit in a sequence.
For example, the $\mb{not}$ gate applied to the second bit
of a three-bit sequence in the state $\bitzero\bitone\bitone$ changes
the state to $\bitzero\bitzero\bitone$.

One-bit gates act independently on each bit. To compute with multiple
bits, we need gates whose action can correlate the states of two or
more bits.  One such gate is the $\mb{nand}$ (``not and'') gate, which
acts on two bits in a bit sequence. Its effect is to set the state of
the first bit to $\bitzero$ if both the first and the second bit are
$\bitone$, otherwise it sets it to $\bitone$. Here is what happens
when $\mb{nand}$ is applied to two consecutive bits:
\begin{equation}
  \begin{array}{ccc}
    \textrm{Initial State}&&\textrm{Final State}\\
\bitzero\bitzero&\rightarrow&\mb{nand}(\bitzero\bitzero)=\bitone\bitzero,\\
\bitzero\bitone&\rightarrow&\mb{nand}(\bitzero\bitone)=\bitone\bitone,\\
\bitone\bitzero&\rightarrow&\mb{nand}(\bitone\bitzero)=\bitone\bitzero,\\
\bitone\bitone&\rightarrow&\mb{nand}(\bitone\bitone)=\bitzero\bitone.
  \end{array}
\end{equation}
The $\mb{nand}$ gate can be applied to any two bits in a sequence.
For example, it can be applied to the fourth and second bits (in this
order) of four bits, in which case the initial state
$\bitone\bitone\bitzero\bitone$ is transformed to
$\bitone\bitone\bitzero\bitzero$, setting the fourth bit to
$\bitzero$.

Other operations on bit sequences include adding a new bit to the
beginning ($\mb{prepend}$) or end ($\mb{append}$) of a bit sequence.
The new bit is always initialized to $\bitzero$.  It is also possible
to discard the first or last bit, regardless of its state.  Versions
of these operations that are conditional on the state of another bit
may also be used. An example is the conditional append operation: ``if
the $k$'th bit is in the state $\bitzero$ then append a bit.''

The gates just introduced suffice for implementing arbitrary
state transformations of a given bit sequence.  Instructions
for applying gates in a particular order are called a ``circuit''.
An important part of investigations in information
processing is to determine the minimum resources required to perform
information processing tasks. For a given circuit, the two primary
resources are the number of gates and the total number
of bits used.  The ``circuit complexity'' of a desired transformation
is the minimum number of gates needed to implement it.

The model of computation defined by the ability to apply gates in a
fixed sequence is called the ``circuit model''.  Classical computation
extends the circuit model by providing a means for repeating blocks of
instructions indefinitely or until a desired condition is achieved.
In principle, it is possible to conceive of a general purpose computer
as a device that repeatedly applies the same circuit to the beginnings
of several bit sequences.  In this introduction, we take for granted a
traditional programmable computer based on classical information.
Thus a ``quantum algorithm'' is a program written for such a computer
with additional instructions for applying gates to quantum
information.  The computational power of this model is equivalent to that of
other general purpose models of quantum computation, such as
quantum Turing machines~\cite{yao:qc1993a}.

For an introduction to algorithms and their analysis,
see~\cite{cormen:qc1990a}.  A useful textbook on computational complexity
with an introduction to classical computation and computational
machine models is~\cite{papadimitriou:qc1994a}.

\section{Quantum Information}
\label{sect:qi}

The foundations of an information processing theory
can be constructed by the procedure we followed in the previous section:
\begin{itemize}
\item[1.] Define the basic unit of information.
\item[2.] Give the means for processing one unit.
\item[3.] Describe how multiple units can be combined.
\item[4.] Give the means for processing multiple units.
\item[5.] Show how to convert the content of any of the extant units
to classical information.
\end{itemize}
Note that the last step was not required for classical information processing.

In this section, we follow the general procedure for defining an
information processing theory to introduce quantum information
processing. A simple example that exhibits the advantages of
quantum information is given in Sect.~\ref{sect:parity}. A version
of the quantum factoring algorithm is described in Sect.~\ref{sec:factor}.

\subsection{The Quantum Bit}

The fundamental resource and basic unit of quantum information is the
quantum bit (qubit), which behaves like a classical bit enhanced by
the superposition principle (see below). From a physical point of
view, a qubit is represented by an ideal two-state quantum
system. Examples of such systems include photons (vertical and
horizontal polarization), electrons and other spin-${1\over 2}$
systems (spin up and down), and systems defined by two energy levels
of atoms or ions. From the beginning the two-state system played a
central role in studies of quantum mechanics. It is the most simple
quantum system, and in principle all other quantum systems can be
modeled in the state space of collections of qubits.

From the information processing point of view, a qubit's state space
contains the two ``logical'', or ``computational'', states
$\ket{\bitzero}$ and $\ket{\bitone}$. The so-called ``ket'' notation
for these states was introduced by P.~Dirac, and its variations are
widely used in quantum physics. One can think of the pair of symbols
``$\qvbar$'' and ``$\qrangle$'' as representing the qubit
system. Their content specifies a state for the system.  In this
context $\bitzero$ and $\bitone$ are system-independent state labels.
When, say, $\bitzero$ is placed within the ket, the resulting
expression $\ket{\bitzero}$ represents the corresponding state of a
specific qubit.

The initial state of a qubit is always one of the logical states.
Using operations to be introduced later, we can obtain states which
are ``superpositions'' of the logical states. Superpositions can be
expressed as sums $\alpha\ket{\bitzero}+\beta\ket{\bitone}$ over the
logical states with complex coefficients. The complex numbers $\alpha$
and $\beta$ are called the ``amplitudes'' of the superposition.  The
existence of such superpositions of distinguishable states of quantum
systems is one of the basic tenets of quantum theory called the
``superposition principle''. Another way of writing a general
superposition is as a vector
\begin{equation}
\alpha\ket{\bitzero}+\beta\ket{\bitone}
\leftrightarrow\qavec{\alpha}{\beta},
\end{equation}
where the two-sided arrow ``$\leftrightarrow$'' is used to denote
the correspondence between expressions that mean the same thing.

The qubit states that are superpositions of the logical states are
called ``pure'' states: A superposition
$\alpha\ket{\bitzero}+\beta\ket{\bitone}$ is a pure state if the
corresponding vector has length $1$, that is
$|\alpha|^2+|\beta|^2=1$. Such a superposition or vector is said to be
``normalized''.  (For a complex number given by $\gamma=x+ iy$, one
can evaluate $|\gamma|^2=x^2+y^2$.  Here, $x$ and $y$ are the real and
imaginary part of $\gamma$, and the symbol $i$ is a square root of
$-1$, that is, $i^2=-1$. The conjugate of $\gamma$ is
$\overline\gamma=x-iy$. Thus $|\gamma|^2=\overline\gamma
\gamma$.)  Here are a few examples of states given in both the ket and
the vector notation:
\begin{eqnarray}
\ket{\psi_1}&=&\Big(\ket{\bitzero}+\ket{\bitone}\Big)/\sqrt{2} \leftrightarrow
  \qavec{1/\sqrt{2}}{1/\sqrt{2}},\\
\ket{\psi_2}&=&{3\over 5}\ket{\bitzero}-{4\over 5}\ket{\bitone} \leftrightarrow
  \qavec{3/5}{-4/5},\\
\ket{\psi_3}&=&{i3\over 5}\ket{\bitzero}-{i4\over 5}\ket{\bitone} \leftrightarrow
  \qavec{i3/5}{-i4/5}.
\end{eqnarray}
The state $\ket{\psi_3}$ is obtained from $\ket{\psi_2}$ by
multiplication with $i$. It turns out that two states cannot be
distinguished if one of them is obtained by multiplying the other by a
``phase'' $e^{i\theta}$.  Note how we have
generalized the ket notation by introducing expressions such as
$\ket{\psi}$ for arbitrary states.

The superposition principle for quantum information means that we can
have states that are sums of logical states with complex coefficients.
There is another, more familiar type of information whose states are
combinations of logical states.  The basic unit of this type of
information is the probabilistic bit (pbit). Intuitively, a pbit can
be thought of as representing the as-yet-undetermined outcome of a
coin flip.  Since we need the idea of probability to understand how
quantum information converts to classical information, we briefly
introduce pbits.

A pbit's state space is a probability distribution over the states of
a bit. One very explicit way to symbolize such a state is by using the
expression $\problb
p\probmul\bitzero\probplus(1-p)\probmul\bitone\probrb$, which means
that the pbit has probability $p$ of being $\bitzero$ and $1-p$ of
being $\bitone$. Thus a state of a pbit is a ``probabilistic''
combination of the two logical states, where the coefficients are
nonnegative real numbers summing to $1$.  A typical example is the
unbiased coin in the process of being flipped. If ``tail'' and
``head'' represent $\bitzero$ and $\bitone$, respectively, the
coin's state is $\problb {1\over 2}\probmul\bitzero\probplus {1\over
2}\probmul\bitone\probrb$.  After the outcome of the flip is known,
the state ``collapses'' to one of the logical states $\bitzero$ and
$\bitone$. In this way, a pbit is converted to a classical bit. If the
pbit is probabilistically correlated with other pbits, the collapse
associated with learning the pbit's logical state changes the overall
probability distribution by a process called ``conditioning on the
outcome''.

A consequence of the conditioning process is that we never actually
``see'' a probability distribution. We only see classical
deterministic bit states. According to the frequency interpretation of
probabilities, the original probability distribution can only be
inferred after one looks at many independent pbits in the same state
$\problb p\probmul\bitzero\probplus(1-p)\probmul\bitone\probrb$: In
the limit of infinitely many pbits, $p$ is given by the fraction of
pbits seen to be in the state $\bitzero$.  As we will explain, we can
never ``see'' a general qubit state either. For qubits there is a
process analogous to conditioning. This process is called
``measurement'' and converts qubit states to classical information.

Information processing with pbits has many advantages over
deterministic information processing with bits. One advantage is that
algorithms are often much easier to design and analyze if they are
probabilistic. Examples include many optimization and physics
simulation algorithms.  In some cases, the best available
probabilistic algorithm is more efficient than any known
deterministic algorithm. An example is an algorithm for determining
whether a number is prime or not. It is not known whether every
probabilistic algorithm can be ``derandomized'' efficiently.  There
are important communication problems that can be solved
probabilistically but not deterministically. For a survey,
see~\cite{gupta:qc1994a}.

What is the difference between bits, pbits and qubits?
One way to visualize the difference and see the enrichment
provided by pbits and qubits is shown in Fig.~\ref{fig:bit_comparison}.
%
\begin{herefig}
\begin{picture}(5,2.8)(0,-.2)
\nputbox{.8,2.5}{t}{\large Bit}
\nputbox{.8,2.2}{t}{$\mathbf{0}$}
\nputgr{.8,.8}{b}{height=1.2in}{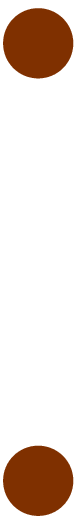}
\nputbox{.8,.65}{b}{$\mathbf{1}$}
\nputbox{2.5,2.5}{t}{\large Pbit}
\nputbox{2.5,2.2}{t}{$\mathbf{0}$}
\nputgr{2.5,.8}{b}{height=1.2in}{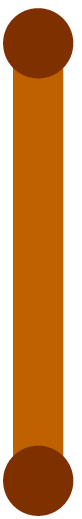}
\nputbox{2.5,.65}{b}{$\mathbf{1}$}
\nputbox{4.2,2.5}{t}{\large Qubit}
\nputbox{4.2,2.2}{t}{$\mathbf{0}$}
\nputgr{4.2,.8}{b}{height=1.2in,width=1.2in}{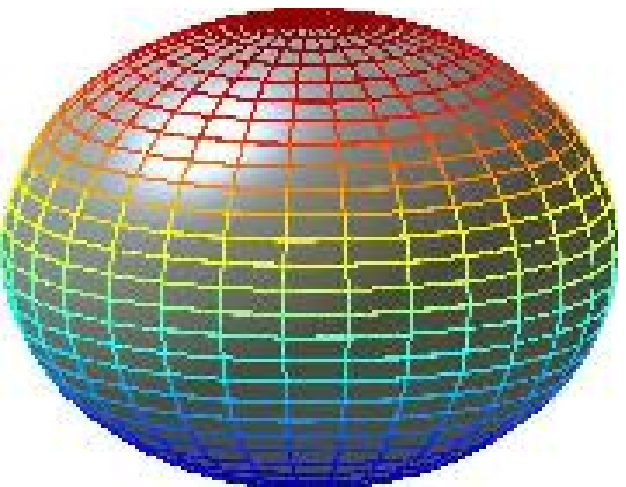}
\nputbox{4.2,.65}{b}{$\mathbf{1}$}
\nputbox{0,.4}{lb}{States:}
\nputbox{.8,.4}{b}{\elimdepth{$\bitzero$ or $\bitone$}}
\nputbox{2.5,.4}{b}{\elimdepth{$\problb p\probmul\bitzero\probplus(1-p)\probmul\bitone\probrb$}}
\nputbox{4.2,.4}{b}{\elimdepth{$\alpha\ket{\bitzero}+\beta\ket{\bitone}$}}
\nputbox{4.2,.25}{lt}{\elimdepth{$|\alpha|^2+|\beta|^2=1$}}
\end{picture}
\herefigcap{Visual comparison of the state spaces of 
different information units.  The states of a bit correspond to two
points. The states of a pbit can be thought of as ``convex''
combinations of the states of a bit and therefore can be visualized as
lying on the line connecting the two bit states.  A qubit's pure states
correspond to the surface of the unit sphere in three dimensions, where
the logical states correspond to the poles. This representation of
qubit states is called the ``Bloch sphere''. The explicit
correspondence is discussed at the end of Sect.~\ref{sec:mixt}. See
also the definition and use of the Bloch sphere
in~\cite{knill:qc2001f}. The correspondence between the pure states
and the sphere is physically motivated and comes from a way of viewing
a spin-${1\over 2}$ system as a small quantum magnet. Intuitively, a
state is determined by the direction of the north pole of the
magnet. }
\label{fig:bit_comparison}
\end{herefig}

\subsection{Processing One Qubit}

The quantum version of the $\mb{not}$ gate for bits exchanges
the two logical states. That is, using ket notation,
\begin{equation}
\mb{not}\Big(\alpha\ket{\bitzero}+\beta\ket{\bitone}\Big) = 
  \alpha\ket{\bitone}+\beta\ket{\bitzero} =
  \beta\ket{\bitzero}+\alpha\ket{\bitone}.
\end{equation}
In vector notation this equation becomes
\begin{equation}
\mb{not}\qavec{\alpha}{\beta} = \qavec{\beta}{\alpha}.
\label{eq:not}
\end{equation}
Another way of expressing the effect of $\mb{not}$ 
is by multiplying the vector by a matrix representing $\mb{not}$:
\begin{equation}
\mb{not}\qavec{\alpha}{\beta} = 
 \qaop{0}{1}{1}{0}\qavec{\alpha}{\beta} = \qavec{\beta}{\alpha},
\end{equation}
so we that can identify the action of $\mb{not}$ with the matrix
$\sigma_x=\qaop{0}{1}{1}{0}$. 
An even simpler gate is the one that does nothing. We call this
the $\mb{noop}$ gate, and its matrix form is the identity matrix
as shown in the following equation:
\begin{equation}
  \mb{noop}\qavec{\alpha}{\beta} = 
  \qaop{1}{0}{0}{1}\qavec{\alpha}{\beta} = \qavec{\alpha}{\beta}.
\end{equation}

The $\mb{noop}$ and $\mb{not}$ gates are ``reversible''.  In other
words, we can undo their actions by applying other
gates.  For example, the action of the $\mb{not}$ gate can be undone
by another $\mb{not}$ gate.  The action of every reversible
quantum gate can be represented by matrix multiplication, where the
matrix has the additional property of preserving the length of
vectors.  Such matrices are called ``unitary'' and are
characterized by the equation $A^\dagger A = \id$, where $A^\dagger$
is the conjugate transpose of $A$ and $\id$ is the identity
matrix. (The conjugate transpose of a matrix is computed by flipping
the matrix across the main diagonal and conjugating the complex
numbers.) For gates
represented by a matrix, the unitarity condition is necessary and
sufficient for ensuring that pure states get mapped to pure states.

Because qubit states can be represented as points on a sphere,
reversible one-qubit gates can be thought of as rotations of the Bloch
sphere.  This is why such quantum gates are often called
``rotations''. As explained in detail in~\cite{knill:qc2001f},
rotations around the $x$, $y$ and $z$ axis are in a sense generated by
the three Pauli matrices
\begin{equation}
\sigma_x = \qaop{0}{1}{1}{0}\;,\;\;
\sigma_y = \qaop{0}{-i}{i}{0}\;,\;\;
\sigma_z = \qaop{1}{0}{0}{-1}\;,
\label{eq:paulidef}
\end{equation}
each of which represents a one-qubit gate.  For example, a rotation
around the $x$-axis by an angle $\phi$ is given by
$e^{-i\sigma_x\phi/2}=\cos(\phi/2)\id - i\sin(\phi/2)\sigma_x$.  To
obtain this identity, one can use the power series for $e^A$,
$e^{A}=\sum_{k=0}^\infty{1\over k!} A^k$, and exploit the fact that
$\sigma_x^2=\id$ to simplify the expression.  Here are some gates that
can be defined with the help of rotations:
\begin{equation}
\begin{array}{lrcl}
\mbox{$90^\circ$ $x$-rotation}: &\mb{rotx}_{90^\circ}&=&
  {1\over\sqrt{2}}\qaop{1}{-i}{-i}{1}\\\\
\mbox{$90^\circ$ $y$-rotation}: &\mb{roty}_{90^\circ}&=&
  {1\over\sqrt{2}}\qaop{1}{-1}{1}{1}\\\\
\mbox{$\phi$ $z$-rotation}: &\mb{rotz}_{\phi}&=&
  \qaop{e^{-i\phi/2}}{0}{0}{e^{i\phi/2}}\\\\
\mbox{Hadamard gate}: &\mb{H} &=&
  {1\over\sqrt{2}}\qaop{1}{1}{1}{-1}
\end{array}
\label{eq:hadamard}
\end{equation}
The rotation gates often show up in controlling spins or ions
with radio-frequency pulses or lasers. The Hadamard gate is used primarily
by quantum programmers. It can be expressed
as a product of a $90^\circ$ $y$-rotation and $\sigma_z$.

To check directly that the rotation gates are reversible one can
determine their inverses. In this case and as expected, the inverse of
a rotation is the rotation around the same axis in the opposite
direction.  For example, the inverses of the $\mb{roty}_{90^\circ}$
and $\mb{rotz}_{\phi}$ gates are given by
\begin{equation}
\begin{array}{rcl}
\mb{roty}_{-90^\circ}&=&
  {1\over\sqrt{2}}\qaop{1}{1}{-1}{1}\\\\
\mb{rotz}_{-\phi}&=&
  \qaop{e^{i\phi/2}}{0}{0}{e^{-i\phi/2}}
\end{array}
\end{equation}
Another useful property of the rotation gates is that the angles add
when rotations are applied around the same axis. For example,
$\mb{rotz}_{\phi}\mb{rotz}_{\theta} = \mb{rotz}_{\phi+\theta}$.

The ket notation can be extended so that we can write gates in a
compact form that readily generalizes to multiple qubits.  To do so we
have to introduce expressions such as
$\bra{\psi}=\alpha\bra{\bitzero}+\beta\bra{\bitone}$. This is called
the ``bra'' notation. The terminology comes from the term ``bracket'':
The `bra'' is the left and the ``ket'' is the right part of a matched
pair of brackets.  From the vector point of view, $\bra{\psi}$
corresponds to the row vector $(\alpha,\beta)$.  Note that a column
vector multiplied by a row vector yields a matrix. In the bra-ket
notation, this corresponds to multiplying a ket $\ket{\psi}$ by a bra
$\bra{\phi}$, written as $\ketbra{\psi}{\phi}$. Since this represents
an operator on states, we expect to be able to compute the effect of
$\ketbra{\psi}{\phi}$ on a state $\ket{\varphi}$ by forming the
product. To be able to evaluate such products with one-qubit kets and
bras, we need the following two rules.
\begin{itemize}
\item[] \textbf{Distributivity.}
You can rewrite sums and products using distributivity.
For example,
\begin{equation}
\Big({3\over 5}\bra{\bitzero}+{4\over 5}\bra{\bitone}\Big)
  i\ket{\bitone} =
{i3\over 5}\bra{\bitzero}\ket{\bitone} + {i4\over 5}\bra{\bitone}\ket{\bitone}.
\label{eq:distex}
\end{equation}
Observe that we can combine the amplitudes of terms, but we cannot
rearrange the order of the bras and kets in a product.
\item[] \textbf{Inner product evaluation.}
The product of a logical ``bra'' and a logical ``ket'' is evaluated according
to the identities
\begin{eqnarray}
\bra{\bitzero}\ket{\bitzero} &=& 1,\nonumber\\
\bra{\bitzero}\ket{\bitone} &=& 0,\nonumber\\
\bra{\bitone}\ket{\bitzero} &=& 0,\nonumber\\
\bra{\bitone}\ket{\bitone} &=& 1.
\end{eqnarray}
It follows that for logical states, if a bra multiplies a ket, the
result cancels unless the states match, in which case the answer is
$1$. Applying inner product evaluation to the example (Eq.~\ref{eq:distex}) results in
\begin{equation}
{i3\over 5}\bra{\bitzero}\ket{\bitone} + {i4\over 5}\bra{\bitone}\ket{\bitone}
 = {i3\over 5} 0 + {i4\over 5} 1 = {i4\over 5}.
\end{equation}
\end{itemize}
To simplify the notation, we can omit one of the two vertical bars
in products such as $\bra{a}\ket{b}$ and write $\braket{a}{b}$.

To understand inner product evaluation,
think of the expressions
as products of row and column vectors. For example,
\begin{equation}
\braket{\bitzero}{\bitone} \leftrightarrow\;\;\;\;
  \begin{array}{@{}c@{}}(\;1\;\;\;0\;)\\\strutlike{(} \end{array}\qavec{0}{1} = 0,
\end{equation}
That is, as vectors the two states $\ket{\bitzero}$ and
$\ket{\bitone}$ are orthogonal. In general, if $\ket{\phi}$ and
$\ket{\psi}$ are states, then $\braket{\phi}{\psi}$ is the ``inner
product'' or ``overlap'' of the two states. In the expression for the
overlap, $\bra{\phi}$ is computed from
$\ket{\phi}=\alpha\ket{\bitzero}+\beta\ket{\bitone}$ by conjugating
the coefficients and converting the logical kets to bras: $\bra{\phi}
= \overline\alpha\bra{\bitzero}+\overline\beta\bra{\bitone}$. In the vector
representation, this is the conjugate transpose of the column vector
for $\ket{\phi}$, so the inner product agrees with the usual one. Two
states are orthogonal if their overlap is zero.  We write
$\ket{\phi}^\dagger = \bra{\phi}$ and $\bra{\phi}^\dagger =
\ket{\phi}$.

Every linear operator on states can be expressed with the bra-ket
notation.
For example, the bra-ket expression for the $\mb{noop}$ gate
is $\mb{noop} = \ketbra{\bitzero}{\bitzero}+\ketbra{\bitone}{\bitone}$.
To apply $\mb{noop}$ to a qubit, you multiply its state on the left
by the bra-ket expression:
\begin{eqnarray}
\mb{noop}\Big(\alpha\ket{\bitzero}+\beta\ket{\bitone}\Big) &=&
\Big(\ketbra{\bitzero}{\bitzero}+\ketbra{\bitone}{\bitone}\Big)\Big(\alpha\ket{\bitzero}+\beta\ket{\bitone}\Big) \nonumber\\
&=&
\ketbra{\bitzero}{\bitzero}\Big(\alpha\ket{\bitzero}+\beta\ket{\bitone}\Big)
 +
\ketbra{\bitone}{\bitone}\Big(\alpha\ket{\bitzero}+\beta\ket{\bitone}\Big) 
\nonumber\\
&=&
\alpha\ket{\bitzero}\braket{\bitzero}{\bitzero} +
\beta\ket{\bitzero}\braket{\bitzero}{\bitone} +
\alpha\ket{\bitone}\braket{\bitone}{\bitzero} +
\beta\ket{\bitone}\braket{\bitone}{\bitone} \nonumber\\
&=&
\alpha\ket{\bitzero} 1 +
\beta\ket{\bitzero} 0 +
\alpha\ket{\bitone} 0 +
\beta\ket{\bitone} 1 \nonumber\\
&=&
\alpha\ket{\bitzero}+\beta\ket{\bitone}
\end{eqnarray}
One way to think about an operator such as $\ketbra{a}{b}$ is to notice
that when it is used to operate on a ket expression, the $\bra{b}$ picks
out the matching kets in the state, which are then changed to
$\ket{a}$.  For example, we can write the $\mb{not}$
operation as $\mb{not} =
\ketbra{\bitzero}{\bitone}+\ketbra{\bitone}{\bitzero}$.

The coefficients of the $\ketbra{a}{b}$ in a bra-ket representation
of a gate correspond to matrix entries in the matrix representation.
The relationship is defined by
\begin{equation}
\alpha_{00}\ketbra{\bitzero}{\bitzero}
+
\alpha_{01}\ketbra{\bitzero}{\bitone}
+
\alpha_{10}\ketbra{\bitone}{\bitzero}
+
\alpha_{11}\ketbra{\bitone}{\bitone}
\leftrightarrow
\qaop{\alpha_{00}}{\alpha_{01}}{\alpha_{10}}{\alpha_{11}}.
\end{equation}

\subsection{Two Quantum Bits}

Some states of two quantum bits can be symbolized by the juxtaposition
(or multiplication) of states of each quantum bit. In particular, the
four logical states
$\ket{\bitzero}\ket{\bitzero},\ket{\bitzero}\ket{\bitone},\ket{\bitone}\ket{\bitzero},$
and $\ket{\bitone}\ket{\bitone}$ are acceptable pure states for two
quantum bits.  In these expressions, we have distinguished the qubits
by position (first or second). It is easier to manipulate state
expressions if we explicitly name the qubits, say $\sysfnt{A}$ and
$\sysfnt{B}$.  We can then distinguish the kets by writing, for
example, $\kets{\psi}{A}$ for a state of qubit $\sysfnt{A}$. Now the
state $\ket{\bitzero}\ket{\bitone}$ can be written with explicit qubit
names (or ``labels'') as
\begin{equation}
\kets{\bitzero}{A}\kets{\bitone}{B} =
\kets{\bitone}{B}\kets{\bitzero}{A} = \kets{\bitzero\bitone}{AB} = \kets{\bitone\bitzero}{BA}.
\end{equation}
Having explicit labels allows us to unambiguously reorder the states
in a product of states belonging to different qubits.  We say that kets for
different qubits ``commute''.

So far we have seen four states of two qubits, which are the logical states
that correspond to the states of two bits. As in the case of one
qubit, the superposition principle can be used to get all the other
pure states. Each state of two qubits is therefore of the form
\begin{equation}
\alpha\kets{\bitzero\bitzero}{AB}+\beta\kets{\bitzero\bitone}{AB}
+\gamma\kets{\bitone\bitzero}{AB}
+\delta\kets{\bitone\bitone}{AB},
\end{equation}
where $\alpha,\beta,\gamma,$ and $\delta$ are complex numbers. Again,
there is a column vector form for the state:
\begin{equation}
\left(\begin{array}{c}\alpha\\\beta\\\gamma\\\delta\end{array}\right),
\end{equation}
and this vector has to be of unit length, that is
$|\alpha|^2+|\beta|^2+|\gamma|^2+|\delta|^2=1$.  When using the vector
form for qubit states, one has to be careful about the convention used
for ordering the coefficients.

Other examples of two-qubit states in ket notation are the following:
\begin{eqnarray}
\kets{\psi_1}{AB} &=& 
  {1\over\sqrt{2}}\Big(\kets{\bitzero}{A}+\kets{\bitone}{A}\Big)
     \kets{\bitone}{B}, \nonumber \\
\kets{\psi_2}{AB} &=&
{1\over\sqrt{2}}\Big(\kets{\bitzero}{A}-\kets{\bitone}{A}\Big)
{1\over\sqrt{2}}\Big(\kets{\bitzero}{B}+i\kets{\bitone}{B}\Big) \nonumber \\
 &=&
{1\over 2}\Big(\kets{\bitzero\bitzero}{AB} +i\kets{\bitzero\bitone}{AB} - \kets{\bitone\bitzero}{AB} -i\kets{\bitone\bitone}{AB}\Big) \nonumber \\
\kets{\psi_3}{AB} &=& {1\over \sqrt{2}}\Big(\kets{\bitzero\bitzero}{AB} + \kets{\bitone\bitone}{AB}\Big), \nonumber \\
\kets{\psi_4}{AB} &=& {1\over \sqrt{2}}\Big(\kets{\bitzero\bitone}{AB} - \kets{\bitone\bitzero}{AB}\Big).
\end{eqnarray}
The first two of these states have the special property that they
can be written as a product $\kets{\phi_1}{A}\kets{\phi_2}{B}$ 
of a state of qubit $A$ and a state of qubit $B$. The second
expression for $\ket{\psi_2}$ shows that the product decomposition
is not always easy to see. Such states are called ``product'' states.
The last two states, $\kets{\psi_3}{AB}$ and $\kets{\psi_4}{AB}$ are
two of the famous Bell states. They have no such representation as a
product of independent states of each qubit.  They are said to be
``entangled'' because they contain a uniquely quantum correlation
between the two qubits. Pbits can also have correlations that
cannot be decomposed into product states, but the entangled states
have additional properties that make them very useful. For example, if
$\sysfnt{A}$lice and $\sysfnt{B}$ob each have one of the qubits that
together are in the state $\kets{\psi_3}{AB}$, they can use them to
create a secret bit for encrypting their digital communications.

\subsection{Processing Two Qubits}

The simplest way of modifying the state of two qubits is
to apply one of the one-qubit gates. If the gates are expressed
in the bra-ket notation, all we need to do is add qubit labels so
that we know which qubit each bra or ket  belongs to. For example, the
$\mb{not}$ gate for qubit $\sysfnt{B}$ is written as
\begin{equation}
\slb{\mb{not}}{B} = \ketbras{\bitzero}{\bitone}{B}+\ketbras{\bitone}{\bitzero}{B}.
\end{equation}
The labels for bra expressions occur as left superscripts.
To apply expressions like this to states, we need one more rule:
\begin{itemize}
\item[] \textbf{Commutation.}
Kets and bras with different labels can be interchanged in products
(they ``commute''). This is demonstrated by the following example:
\begin{eqnarray}
\left(\ketbras{\bitzero}{\bitone}{B}\right)\kets{\bitzero\bitone}{AB} &=&
  \ketbras{\bitzero}{\bitone}{B}\kets{\bitzero}{A}\kets{\bitone}{B} \nonumber \\
&=&  \kets{\bitzero}{A}\;\ketbras{\bitzero}{\bitone}{B}\;\kets{\bitone}{B} \nonumber\\
&=&  \kets{\bitzero}{A}\;\kets{\bitzero}{B}\brakets{\bitone}{\bitone}{B}\nonumber\\
 &=& \kets{\bitzero}{A}\kets{\bitzero}{B} = \kets{\bitzero\bitzero}{AB}.
\end{eqnarray}
\end{itemize}
Note that we cannot merge the two vertical bars in expressions such as
$\bras{\bitone}{B}\kets{\bitzero}{A}$ because the two terms
belong to different qubits.  The bars can only be merged when the
expression is an inner product, which requires that the two terms
belong to the same qubit.

With the rules for bra-ket expressions in hand, we can apply the
$\mb{not}$ gate to one of our Bell states to see how it acts:
\begin{eqnarray}
\slb{\mb{not}}{B}{1\over\sqrt{2}}
  \Big(\kets{\bitzero\bitzero}{AB}+\kets{\bitone\bitone}{AB}\Big) 
  &=&
     \Big(\ketbras{\bitzero}{\bitone}{B}+\ketbras{\bitone}{\bitzero}{B}\Big)
     {1\over\sqrt{2}}\Big(\kets{\bitzero\bitzero}{AB}+\kets{\bitone\bitone}{AB}\Big)
\nonumber \\
  &=&
    {1\over\sqrt{2}}\Bigg(
      \ketbras{\bitzero}{\bitone}{B}
      \Big(\kets{\bitzero\bitzero}{AB}+\kets{\bitone\bitone}{AB}\Big) 
         +
      \ketbras{\bitone}{\bitzero}{B}
      \Big(\kets{\bitzero\bitzero}{AB}+\kets{\bitone\bitone}{AB}\Big) \Bigg)
\nonumber \\
  &=&
    {1\over\sqrt{2}}\Big(
      \ketbras{\bitzero}{\bitone}{B}\kets{\bitzero\bitzero}{AB}
        +
      \ketbras{\bitzero}{\bitone}{B}\kets{\bitone\bitone}{AB}
        +
      \ketbras{\bitone}{\bitzero}{B}\kets{\bitzero\bitzero}{AB}
        +
      \ketbras{\bitone}{\bitzero}{B}\kets{\bitone\bitone}{AB}\Big)
\nonumber \\
  &=&
    {1\over\sqrt{2}}\Big(
      \kets{\bitzero}{A}\ketbras{\bitzero}{\bitone}{B}\kets{\bitzero}{B}
        +
      \kets{\bitone}{A}\ketbras{\bitzero}{\bitone}{B}\kets{\bitone}{B}
        +
      \kets{\bitzero}{A}\ketbras{\bitone}{\bitzero}{B}\kets{\bitzero}{B}
        +
      \kets{\bitone}{A}\ketbras{\bitone}{\bitzero}{B}\kets{\bitone}{B}\Big)
\nonumber \\
  &=&
    {1\over\sqrt{2}}\Big(
      \kets{\bitzero}{A}\kets{\bitzero}{B}\; 0
        +
      \kets{\bitone}{A}\kets{\bitzero}{B}\; 1
        +
      \kets{\bitzero}{A}\kets{\bitone}{B}\; 1
        +
      \kets{\bitone}{A}\kets{\bitone}{B}\; 0 \Big)
\nonumber \\
  &=&
    {1\over\sqrt{2}}\Big(
      \kets{\bitone}{A}\kets{\bitzero}{B}
        +
      \kets{\bitzero}{A}\kets{\bitone}{B} \Big)
  = {1\over\sqrt{2}}\Big(\kets{\bitzero\bitone}{AB}+\kets{\bitone\bitzero}{AB}\Big).
\end{eqnarray}
The effect of the gate was to flip the state symbols for qubit $\sysfnt{B}$,
which results in another Bell state.

The gate $\slb{\mb{not}}{B}$ can also be written as a $4\times 4$
matrix acting on the vector representation of a two-qubit
state. However, the relationship between this matrix and the one-qubit
matrix is not as obvious as for the bra-ket expression. The matrix
is
\begin{equation}
\slb{\mb{not}}{B} =
  \left(\begin{array}{cccc}
    0&1&0&0\\1&0&0&0\\0&0&0&1\\0&0&1&0
  \end{array}\right),
\end{equation}
which swaps the top two and the bottom two entries of a state vector.

One way to see the relationship between the one and the two-qubit
representation of the gate $\slb{\mb{not}}{B}$ is to notice that
because the $\mb{noop}$ gate acts as the identity, and because we can act on
different qubits independently,
$\slb{\mb{noop}}{A}\slb{\mb{not}}{B}\simeq\slb{\mb{not}}{B}$.  The matrix
for $\slb{\mb{not}}{B}$ can be expressed as a ``Kronecker product''
(``$\tensor$'') of the matrices for $\mb{noop}$ and $\mb{not}$:
\begin{eqnarray}
\slb{\mb{noop}}{A}\slb{\mb{not}}{B} &=&
    \qaop{1}{0}{0}{1}\tensor\qaop{0}{1}{1}{0}.
\nonumber \\
  &=&
    \left(\begin{array}{cc}
       1\qaop{0}{1}{1}{0} & 0\qaop{0}{1}{1}{0} \\\\
       0\qaop{0}{1}{1}{0} & 1\qaop{0}{1}{1}{0} 
    \end{array}\right)
\nonumber \\
  &=&
    \left(\begin{array}{cccc}
      0&1&0&0\\1&0&0&0\\0&0&0&1\\0&0&1&0
    \end{array}\right).
\end{eqnarray}
The Kronecker product of two matrices expands the first matrix by
multiplying each entry by the second matrix. A disadvantage of the
matrix representation of quantum gates is that it depends on the
number and order of the qubits. However, it is often
easier to visualize what the operation does by writing down the
corresponding matrix.

One cannot do much with one-bit classical gates. Similarly, the
utility of one-qubit gates is limited. In particular, it is not
possible to obtain a Bell state starting from
$\kets{\bitzero\bitzero}{AB}$ or any other product state.  We
therefore need to introduce at least one two-qubit gate not
expressible as the product of two one-qubit gates. The best known such
gate is the ``controlled-not'' ($\mb{cnot}$) gate. Its action can be
described by the statement, ``if the first bit is $\bitone$, flip the
second bit, otherwise do nothing''.  The bra-ket and matrix
representations for this action are
\begin{eqnarray}
\slb{\mb{cnot}}{AB} &=&
\ketbras{\bitzero}{\bitzero}{A}
 + \ketbras{\bitone}{\bitone}{A}\Big(\ketbras{\bitzero}{\bitone}{B}+
                                      \ketbras{\bitone}{\bitzero}{B}\Big)
\nonumber \\
 &=& 
\left(\begin{array}{cccc}
  1&0&0&0\\
  0&1&0&0\\
  0&0&0&1\\
  0&0&1&0
      \end{array}
\right).
\end{eqnarray}
The $\mb{cnot}$ gate is reversible because its action
is undone if a second $\mb{cnot}$ is applied.
This outcome is easy to see by computing the square of the matrix for
$\mb{cnot}$, which yields the identity matrix. As an exercise in
manipulating bras and kets, let us calculate the product of two
$\mb{cnot}$ gates by using the bra-ket representation.
\begin{equation}
\slb{\mb{cnot}}{AB}\slb{\mb{cnot}}{AB} =
\Bigg(\ketbras{\bitzero}{\bitzero}{A}
 + \ketbras{\bitone}{\bitone}{A}\Big(\ketbras{\bitzero}{\bitone}{B}+
                                      \ketbras{\bitone}{\bitzero}{B}\Big)\Bigg)
\Bigg(\ketbras{\bitzero}{\bitzero}{A}
 + \ketbras{\bitone}{\bitone}{A}\Big(\ketbras{\bitzero}{\bitone}{B}+
                                      \ketbras{\bitone}{\bitzero}{B}\Big)\Bigg).
\end{equation}
The first step is to expand this expression by multiplying
out. Expressions such as
$\ketbras{\bitzero}{\bitzero}{A}\,\ketbras{\bitone}{\bitone}{A}$
cancel because of the inner product evaluation rule,
$\brakets{\bitzero}{\bitone}{A}=0$.  One can also reorder bras and kets
with different labels and rewrite
$\ketbras{\bitzero}{\bitzero}{A}\,\ketbras{\bitzero}{\bitzero}{A} =
\ketbras{\bitzero}{\bitzero}{A}$ to get
\begin{eqnarray}
\slb{\mb{cnot}}{AB}\slb{\mb{cnot}}{AB} &=&
 \ketbras{\bitzero}{\bitzero}{A} + \ketbras{\bitone}{\bitone}{A}
    \Big(\ketbras{\bitzero}{\bitone}{B}+\ketbras{\bitone}{\bitzero}{B}\Big)
    \Big(\ketbras{\bitzero}{\bitone}{B}+\ketbras{\bitone}{\bitzero}{B}\Big)
\nonumber \\
 &=&
 \ketbras{\bitzero}{\bitzero}{A} + \ketbras{\bitone}{\bitone}{A}
    \Big(\ketbras{\bitzero}{\bitzero}{B}+\ketbras{\bitone}{\bitone}{B}\Big)
\nonumber \\
 &=&
   \ketbras{\bitzero}{\bitzero}{A} + \ketbras{\bitone}{\bitone}{A}\slb{\mb{noop}}{B}
\nonumber \\
 &\simeq&
   \ketbras{\bitzero}{\bitzero}{A} + \ketbras{\bitone}{\bitone}{A}
\nonumber \\
 &=&
   \slb{\mb{noop}}{A} 
\nonumber \\
 &\simeq& 1,
\end{eqnarray}
where we used the fact that when the bra-ket expression for
$\mb{noop}$ is applied to the ket expression for a state it acts the
same as (here denoted by the symbol ``$\simeq$'') multiplication by
the number $1$.

\subsection{Using Many Quantum Bits}

To use more than two, say five, qubits, we can just start with the
state
$\kets{\bitzero}{A}\kets{\bitzero}{B}\kets{\bitzero}{C}\kets{\bitzero}{D}\kets{\bitzero}{E}$
and apply gates to any one or two of these qubits. For example,
$\slb{\mb{cnot}}{DB}$ applies the $\mb{cnot}$ operation from qubit $D$
to qubit $B$. Note that the order of $\sysfnt{D}$ and $\sysfnt{B}$ in
the label for the $\mb{cnot}$ operation matters. In the bra-ket
notation, we simply multiply the state with the bra-ket form of
$\slb{\mb{cnot}}{DB}$ from the left. One can express everything in
terms of matrices and vectors, but now the vectors have length
$2^5=32$, and the Kronecker product expression for
$\slb{\mb{cnot}}{DB}$ requires some reordering to enable inserting the
operation so as to act on the intended qubits. Nevertheless, to
analyze the properties of all reversible (that is, unitary) operations
on these qubits, it is helpful to think of the matrices, because a lot
of useful properties about unitary matrices are known. One important
result from this analysis is that every matrix that represents a
reversible operation on quantum states can be expressed as a product
of the one- and two-qubit gates introduced so far. We say that this
set of gates is ``universal''.

For general purpose computation, it is necessary to have access to
arbitrarily many qubits. Instead of assuming that there are infinitely
many from the start, it is convenient to have an operation to add a
new qubit, namely, $\mb{add}$. To add a new qubit labeled $\sysfnt{X}$ in the
state $\kets{\bitzero}{X}$, apply $\slb{\mb{add}}{X}$ to the current
state.  This operation can only be used if there is not already a
qubit labeled $\sysfnt{X}$. In the bra-ket notation, we implement the
$\slb{\mb{add}}{X}$ operation by multiplying the ket
expression for the current state by $\kets{\bitzero}{X}$.

\subsection{Qubit Measurements}

In order to classically access information about the state of qubits
we use the measurement operation $\mb{meas}$.  This is an
intrinsically probabilistic process that can be applied to any extant
qubit. For information processing, one can think of $\mb{meas}$ as a
subroutine or function that returns either $\bitzero$ or $\bitone$ as
output.  The output is called the ``measurement outcome''.  The
probabilities of the measurement outcomes are determined by the
current state.  The state of the qubit being measured is ``collapsed''
to the logical state corresponding to the outcome.  Suppose we have
just one qubit, currently in the state
$\ket{\psi}=\alpha\ket{\bitzero}+\beta\ket{\bitone}$. Measurement of
this qubit has the effect
\begin{equation}
\mb{meas}\Big(\alpha\ket{\bitzero}+\beta\ket{\bitone}\Big) =
\left\{\begin{array}{ll}\bitzero\probmul\ket{\bitzero}&\mbox{with probability $|\alpha|^2$},\\ \\
                        \bitone\probmul\ket{\bitone}& \mbox{with probability $|\beta|^2$}.              
       \end{array}
\right.
\end{equation}
The classical output is given before the new state for each possible
outcome.  This measurement behavior explains why the amplitudes
have to define unit length vectors: Up to a phase, they are associated
with square roots of probabilities. 

For two qubits the process is more involved. Because of possible
correlations between the two qubits, the measurement affects the state
of the other one too, similar to conditioning for pbits after
one ``looks'' at one of them.  As an example, consider the state
\begin{equation}
\kets{\psi}{AB} = 
{2\over 3}\kets{\bitzero\bitone}{AB} + {i2\over 3}\kets{\bitone\bitzero}{AB} +
{1\over 3}\kets{\bitzero\bitzero}{AB}.
\label{eq:exref1}
\end{equation}
To figure out what happens when we measure qubit $\sysfnt{A}$, we
first rewrite the current state in the form
$\alpha\kets{\bitzero}{A}\kets{\phi_0}{B}+
\beta\kets{\bitone}{A}\kets{\phi_1}{B}$, where $\kets{\phi_0}{B}$ and
$\kets{\phi_1}{B}$ are pure states for qubit $\sysfnt{B}$. It is
always possible to do that. For the example of Eq.~\ref{eq:exref1}:
\begin{eqnarray}
\kets{\psi}{AB} &=&
  {2\over 3}\kets{\bitzero}{A}\kets{\bitone}{B} +
  {1\over 3}\kets{\bitzero}{A}\kets{\bitzero}{B} +
  {i2\over 3}\kets{\bitone}{A}\kets{\bitzero}{B}
\nonumber \\
&=&  \kets{\bitzero}{A}
     \Big({2\over 3}\kets{\bitone}{B} + {1\over 3}\kets{\bitzero}{B}\Big)
   +
   \kets{\bitone}{A} {i2\over 3}\kets{\bitzero}{B} 
\nonumber \\
&=& {\sqrt{5}\over 3}\kets{\bitzero}{A}\Big(
       {1\over \sqrt{5}} \kets{\bitzero}{B}+{2\over \sqrt{5}}\kets{\bitone}{B}
   \Big)
   +
   {i2\over 3}\kets{\bitone}{A} \Big(\kets{\bitzero}{B}\Big),
\end{eqnarray}
so $\alpha={\sqrt{5}\over 3}$, $\beta={i2\over 3}$,
$\kets{\phi_0}{B}={1\over \sqrt{5}} \kets{\bitzero}{B}+{2\over
\sqrt{5}}\kets{\bitone}{B}$ and $\kets{\phi_1}{B} = \kets{\bitzero}{B}$.
The last step required pulling out the factor of ${\sqrt{5}\over 3}$
to make sure that $\kets{\phi_0}{B}$ is properly normalized for a pure state.
Now that we have rewritten the state, the effect of measuring qubit
$\sysfnt{A}$ can be given as follows:
\begin{equation}
\slb{\mb{meas}}{A}\Big(\alpha\kets{\bitzero}{A}\kets{\phi_0}{B}+
\beta\kets{\bitone}{A}\kets{\phi_1}{B}\Big) =
\left\{
\begin{array}{ll}
  \bitzero\probmul\kets{\bitzero}{A}\kets{\phi_0}{B} & \mbox{with probability $|\alpha|^2$,}\\\\
  \bitone\probmul\kets{\bitone}{A}\kets{\phi_1}{B}  & \mbox{with probability $|\beta|^2$.}
\end{array}
\right.
\label{eq:measdef}
\end{equation}
For the example, the measurement outcome is $\bitzero$ with
probability ${5\over 9}$, in which case the state collapses to
$\kets{\bitzero}{A}\Big({1\over \sqrt{5}} \kets{\bitzero}{B}+{2\over
\sqrt{5}}\kets{\bitone}{B}\Big)$. The outcome is $\bitone$ with
probability ${4\over 9}$, in which case the state collapses to
$\kets{\bitone}{A} \kets{\bitzero}{B}$. The probabilities add up
to $1$ as they should.

The same procedure works for figuring out the effect of measuring one
of any number of qubits. Say we want to measure qubit
$\sysfnt{B}$ among qubits $\sysfnt{A,B,C,D}$, currently in state
$\kets{\psi}{ABCD}$.  First rewrite the state in the form
$\alpha\kets{\bitzero}{B}\kets{\phi_0}{ACD}+\beta\kets{\bitone}{B}\kets{\phi_1}{ACD}$,
making sure that the $\sysfnt{ACD}$ superpositions are pure
states. Then the outcome of the measurement is $\bitzero$ with
probability $|\alpha|^2$ and $\bitone$ with probability
$|\beta|^2$. The collapsed states are $\kets{\bitzero}{B}\kets{\phi_0}{ACD}$
and $\kets{\bitone}{B}\kets{\phi_1}{ACD}$, respectively.

Probabilities of the measurement outcomes and the new states can
be calculated systematically. For example, to compute the probability
and state for outcome $\bitzero$ of $\slb{\mb{meas}}{A}$ 
given the state $\kets{\psi}{AB}$,
one can first obtain the unnormalized ket expression $\kets{\phi'_0}{B} =
\bras{\bitzero}{A}\kets{\psi}{AB}$ by using the rules for multiplying
kets by bras.  The probability is given by
$p_0=\brakets{\phi'_0}{\phi'_0}{B}$, and the collapsed, properly
normalized pure state is
\begin{equation}
\kets{\bitzero}{A}\kets{\phi'_0}{B}/\sqrt{p_0}
 = \ketbras{\bitzero}{\bitzero}{A}\kets{\psi}{AB}/\sqrt{p_0},
\end{equation}
The operator $P_{\bitzero}=\ketbras{\bitzero}{\bitzero}{A}$ is called
a ``projection operator'' or ``projector'' for short. If we perform
the same computation for the outcome $\bitone$, we find the projector
$P_{\bitone}=\ketbras{\bitone}{\bitone}{A}$. The two operators satisfy
${P_{\stfnt{a}}}^2=P_{\stfnt{a}}$, ${P_{\stfnt{a}}}^\dagger =
P_{\stfnt{a}}$ and $P_{\bitzero}+P_{\bitone}=\idop$.
In terms of the projectors, the measurement's effect can be written
as follows:
\begin{equation}
\slb{\mb{meas}}{A}\kets{\psi}{AB} =
\left\{
\begin{array}{ll}
  \bitzero\probmul P_{\bitzero}\kets{\psi}{AB}/\sqrt{p_0}
       & \mbox{with probability $p_0$},\\\\
  \bitone\probmul P_{\bitone}\kets{\psi}{AB}/\sqrt{p_1} & \mbox{with probability $p_1$},
\end{array}
\right.
\end{equation}
where $p_0=\bras{\psi}{AB}P_{\bitzero}\kets{\psi}{AB}$ and $p_1 =
\bras{\psi}{AB}P_{\bitone}\kets{\psi}{AB}$.  In quantum mechanics, any
pair of projectors satisfying the properties given above is associated
with a potential measurement whose effect can be written in the same
form. This is called a binary ``von Neumann'', or ``projective'',
measurement.

\subsection{Mixtures and Density Operators}
\label{sec:mixt}

The measurement operation ``reads out'' information from qubits to
pbits.  What if we discard the pbit that contains the measurement
outcome? The result is that the qubits are in a probabilistic
``mixture'' of two pure states. Such mixtures are a generalization of
pure states. The obvious way to think about a mixture is that we have
a probability distribution over pure quantum states.  For example,
after discarding the pbit and qubit $\sysfnt{A}$ in
Eq.~\ref{eq:measdef}, we can write the state of $\sysfnt{B}$ as
$\rho=\problb|\alpha|^2\probmul\kets{\phi_0}{B}\;\probplus\;|\beta|^2\probmul\kets{\phi_1}{B}\probrb$,
using the notation for probability distributions introduced earlier.

Mixtures frequenty form when using irreversible operations such as
measurement.  Except for measurement, the quantum gates that we have
introduced so far are reversible and therefore transform pure states
to pure states, so that no mixtures can be formed.  One of the
fundamental results of reversible classical and quantum computation is
that there is no loss in power in using only reversible gates.
Specifically, it is possible to change a computation that includes
irreversible operations to one that accomplishes the same goal, has
only reversible operations and is efficient in the sense that it uses
at most polynomial additional resources. However, the cost of using
only reversible operations is not negligible. In particular, for ease
of programming, and more importantly, when performing repetitive
error-correction tasks (see~\cite{knill:qc2001d}), the inability to
discard or reset qubits can be very inconvenient.  We therefore
introduce additional operations that enable resetting and discarding.

Although resetting has a so-called ``thermodynamic'' cost (think of
the heat generated by a computer), it is actually a simple
operation. The $\mb{reset}$ operation applied to qubit $\sysfnt{A}$
can be thought of as the result of first measuring $\sysfnt{A}$, then
flipping $\sysfnt{A}$ if the measurement outcome is $\ket{\bitone}$,
and finally discarding the measurement result.  Using the notation of
Eq.~\ref{eq:measdef}, the effect on a pure state $\kets{\psi}{AB}$ is
given by:
\begin{equation}
\slb{\mb{reset}}{A}\kets{\psi}{AB} =
  \problb|\alpha|^2 \probmul\kets{\bitzero}{A}\kets{\phi_0}{B}\;\probplus\;
    |\beta|^2 \probmul \kets{\bitzero}{A}\kets{\phi_1}{B}\probrb.
\label{eq:resetmix}
\end{equation}
To apply $\mb{reset}$ to an arbitrary probability distribution, you
apply it to each of the distribution's pure states and combine the
results to form an expanded probability distribution.  The
$\slb{\mb{discard}}{A}$ operation is $\slb{\mb{reset}}{A}$ followed by
discarding qubit $\sysfnt{A}$. Therefore, in the expression for the state after
$\slb{\mb{reset}}{A}$, all the $\kets{\bitzero}{A}$
are removed. It is an important fact that every physically realizable
quantum operation, whether reversible or not, can be expressed as
a combination of $\mb{add}$ operations, gates from the universal set and
$\mb{discard}$ operations.

The representation of mixtures using probability distributions over
pure states is redundant. That is, there are many probability
distributions that are physically indistinguishable.  A non-redundant
description of a quantum state can be obtained by using ``density
operators''. The density operator for the mixture $\rho$ given in
Eq.~\ref{eq:resetmix} is given by
\begin{equation}
\hat\rho= |\alpha|^2\ketbras{\phi_0}{\phi_0}{B}+
           |\beta|^2\ketbras{\phi_1}{\phi_1}{B}.
\end{equation}
The general rule for calculating the density operator from a probability
distribution is as follows: For each pure state $\ket{\phi}$ in the
distribution, calculate the operators $\ketbra{\phi}{\phi}$ and sum them
weighted by their probabilities.

There is a way to apply gates to the density operators defined by
states. If the gate acts by the unitary operator $U$, then the effect
of applying it to $\hat\rho$ is given by $U\hat\rho U^\dagger$, where
$U^\dagger$ is the conjugate transpose of $U$. (In the bra-ket
expression for $U$, $U^\dagger$ is obtained by replacing all complex
numbers by their conjugates, and terms such as $\ketbra{\phi}{\varphi}$ by
$\ketbra{\varphi}{\phi}$.)

The relationship between a qubit's state space and a sphere can be
explained in terms of the qubit's density operators. In matrix form,
this operator is a $2\times 2$ matrix, which can be written uniquely as
a sum $(\idop+x\sigma_x+y\sigma_y+z\sigma_z)/2$.  One can
check that if the density operator $\ketbra{\psi}{\psi}$ for
a qubit's pure state is written as such a sum,
\begin{equation}
\ketbra{\psi}{\psi} = (\idop+x\sigma_x+y\sigma_y+z\sigma_z)/2,
\label{eq:blochrep}
\end{equation}
then the vector $(x,y,z)$ thus obtained is on the surface of the unit
sphere in three dimensions. In fact, for every vector $(x,y,z)$ on the
unit sphere, there is a unique pure state satisfying
Eq.~\ref{eq:blochrep}. Since the density operators for mixtures are
arbitrary, convex (that is probabilistic) sums of pure states, the set
of $(x,y,z)$ thus obtained for mixtures fills out the unit ball.  The
rotations introduced earlier modify the vector $(x,y,z)$ in the
expected way, by rotation of the vector around the appropriate axis.
See~\cite{knill:qc2001f} for more details.

\subsection{Quantum Computation}
\label{sect:parity}

The model of computation defined by the one- and two-qubit gates and
the operations of adding ($\mb{add}$), measuring ($\mb{meas}$) and
discarding ($\mb{discard}$) qubits is called the ``quantum network
model''. A sequence of instructions for applying these operations is
called a ``quantum network''.  Quantum computation extends the network
model by providing a means for repeating blocks of instructions.  Such
means can be specified by a formal machine model of computation.
There are several such models of classical and quantum computers.  One
of the best known is the Turing machine, which has a quantum analogue,
the quantum Turing machine. This model has its uses for formal studies
of computation and complexity, but is difficult to program.
Fortunately, as mentioned in Sect.~\ref{sec:class}, there is no loss
of computational power if the means for repeating instructions is
provided by a classical computer that can apply gates and other
operations to qubits. A general quantum algorithm is a program written
for such a computer.

There are three practical methods that can be used to write quantum
networks and algorithms. The first is to use the names for the
different operations and algebraically multiply them.  The second is
to draw quantum networks, which are pictorial representations of the
sequence of steps in time, somewhat like flowcharts without loops. The
third is to use a generic programming language enhanced with
statements for accessing and modifying quantum bits.  The first two
methods work well as long as the sequence is short and we do not use
many operations that depend on measurement outcomes or require loops.
They are often used to describe subroutines of longer algorithms
presented either in words or by use of the third method.

To see how to use the different methods and also to illustrate
the power of quantum computation, we work out a short
quantum algorithm that solves the following problem:

\noindent{\textbf{The Parity Problem:}} Given is a ``black box'' quantum
operation $\slb{\mb{BB}}{ABC}$ that has the following effect
when applied to a logical basis state:
\begin{equation}
\slb{\mb{BB}}{ABC}\kets{a_{\sysfnt{A}}a_{\sysfnt{B}}}{AB}\kets{a_{\sysfnt{C}}}{C} = \kets{a_{\sysfnt{A}}a_{\sysfnt{B}}}{AB}\kets{a_{\sysfnt{C}}\oplus({b}_{\sysfnt{A}}a_{\sysfnt{A}}\oplus {b}_{\sysfnt{B}}a_{\sysfnt{B}})}{C},
\end{equation}
where $b_{\sysfnt{A}}$ and $b_{\sysfnt{B}}$ are $0$ or $1$.  The
actual values of $b_{\sysfnt{A}}$ and $b_{\sysfnt{B}}$ are unknown.
The problem is to 
determine what ${b}_{\sysfnt{A}}$ and ${b}_{\sysfnt{B}}$ are by using
the black box only once.

The terminology and the definition of the operation
$\slb{\mb{BB}}{ABC}$ require explanation. In computation, we say that
an operation is a black box or an ``oracle'' if we have no access
whatsoever to how the operation is implemented. In a black box
problem, we are promised that the black box implements an operation
from a specified set of operations. In the case of the parity problem,
we know that the operation is to add one of four possible parities
(see below).  The problem is to determine which parity is added by
using the black box in a network. Black box problems serve many
purposes.  One is to study the differences between models of
computation, just as we are about to do.  In fact, black box problems
played a crucial role in the development of quantum algorithms by
providing the first and most convincing examples of the power of
quantum computers~\cite{bernstein:qc1993b,simon:qc1994a}. Some of
these examples involve generalizations of the parity problem.  Another purpose
of black box problems is to enable us to focus on what can be learned
from the ``input/output'' behavior of an operation without having to
analyze its implementation. This is useful because in many cases of
interest, it is very difficult to exploit knowledge of the
implementation to determine a desirable property of the operation. A
classical example is the well-known satisfiability problem, in which
we are given a classical circuit with one output bit and we need to
determine whether there is an input for which the output is
$\bitone$. Instead of trying to analyze the circuit, one can look for
and use a general purpose black-box search algorithm to find the
``satisfying'' input.

In the definition of the effect of $\slb{\mb{BB}}{ABC}$, the operation
``$\oplus$'' is addition modulo $2$, so $1\oplus 1=0$, and all the
other sums are as expected. As the state symbols now have a numeric
meaning, we use the number font for states.  To see what $\mb{BB}$
does, suppose that ${b}_{\sysfnt{A}}$ and ${b}_{\sysfnt{B}}$ are both
$1$.  Then $\mb{BB}$ adds (modulo $2$) the parity of the logical state
in $\sysfnt{AB}$ to the logical state of $\sysfnt{C}$. The parity of a
logical state is $0$ if the number of $1$'s is even and $1$ if
it is odd. The action of $\mb{BB}$ for this example is given by:
\begin{eqnarray}
\slb{\mb{BB}}{ABC}\kets{{00}}{AB}\kets{{0}}{C}
 &=& \kets{{00}}{AB}\kets{{0}}{C} 
\nonumber \\
\slb{\mb{BB}}{ABC}\kets{{01}}{AB}\kets{{0}}{C}
 &=& \kets{{01}}{AB}\kets{{0\oplus1}}{C} 
\nonumber \\
 &=& \kets{{01}}{AB}\kets{{1}}{C} 
\nonumber \\
\slb{\mb{BB}}{ABC}\kets{{10}}{AB}\kets{{1}}{C}
 &=& \kets{10}{AB}\kets{{1\oplus1}}{C} 
\nonumber \\
 &=& \kets{{10}}{AB}\kets{{0}}{C} 
\nonumber \\
\slb{\mb{BB}}{ABC}\kets{{11}}{AB}\kets{{0}}{C}
 &=& \kets{{11}}{AB}\kets{{0}}{C} 
\end{eqnarray}
The action of the black box is extended to superpositions by ``linear
extension''.  This means that to apply $\mb{BB}$ to a superposition of
the logical states, simply apply it to each logical summand and add the
results.  Different values of
$b_{\sysfnt{A}}$ and $b_{\sysfnt{B}}$ correspond to different parities.  For
example, if $b_{\sysfnt{A}}=1$ and $b_{\sysfnt{B}}=0$, then the parity of
the state in $\sysfnt{A}$ is added to the state in $\sysfnt{C}$. In
this sense, what is added is the parity of a subset of the two qubits
$\sysfnt{AB}$.  Thus, one way of thinking about the problem is that we
wish to find out which subset's parity the black box is using.

We can give an algorithm that solves the parity problem using each of
the three methods for describing quantum networks.  Here is an
algebraic description of a solution, $\slb{\mb{qparity}}{ABC}$, given
as a product of quantum gates that involves one use of the black box.
We defer the explanation of why this solution works until after we
show how to represent the algorithm pictorially using quantum
networks.
\begin{eqnarray}
\slb{\mb{qparity}}{ABC} = 
\slb{\mb{meas}}{B}
\slb{\mb{H}}{B}
\slb{\mb{meas}}{A}
\slb{\mb{H}}{A}
\slb{\mb{BB}}{ABC}
\slb{\mb{H}}{C}
\slb{\mb{not}}{C}
\slb{\mb{add}}{C}
\slb{\mb{H}}{B}
\slb{\mb{add}}{B}
\slb{\mb{H}}{A}
\slb{\mb{add}}{A} .\nonumber\\
\mbox{}\label{eq:qparrl}
\end{eqnarray}
The output of the algorithm is given by the classical outputs of the
measurements of qubit $\sysfnt{A}$, which yields $b_{\sysfnt{A}}$, and
qubit $\sysfnt{B}$, which yields $b_{\sysfnt{B}}$.  As is
conventional, in writing products of linear operators, the order of
application in Eq.~\ref{eq:qparrl} is right to left, as in a product
of matrices applied to a column vector.  This order of terms in a
product is, however, counterintuitive, particularly for operations to
be performed one after the other.  It is therefore convenient to use
left to right notation, as is done in describing laser or
radio-frequency pulse sequences.  One way to make it clear that left
to right order is used involves putting dots between gates as in the
following version of Eq.~\ref{eq:qparrl}:
\begin{eqnarray}
\slb{\mb{qparity}}{ABC} = 
\slb{\mb{add}}{A} .
\slb{\mb{H}}{A} .
\slb{\mb{add}}{B} .
\slb{\mb{H}}{B} .
\slb{\mb{add}}{C} .
\slb{\mb{not}}{C} .
\slb{\mb{H}}{C} .
\slb{\mb{BB}}{ABC} .
\slb{\mb{H}}{A} .
\slb{\mb{meas}}{A} .
\slb{\mb{H}}{B} .
\slb{\mb{meas}}{B} .\nonumber\\
\mbox{}
\end{eqnarray}
In this representation, the first operation is $\slb{\mb{add}}{A}$,
the second is $\slb{\mb{H}}{A}$ (the Hadamard gate on qubit
$\sysfnt{A}$) and so on.

The algebraic specification of the algorithm as products of gates does
not make it easy to see why the algorithm works.  It is also difficult
to see which operations depend on each other. Such dependencies
are used to
determine whether the operations can be ``parallelized''.
Quantum networks make these tasks simpler. The quantum network for the
above sequence is shown in Fig.~\ref{fig:parity_network}.

\begin{herefig}
\begin{picture}(7,3.2)(-3.5,-3)
\nputgr{0,0}{t}{width=7in}{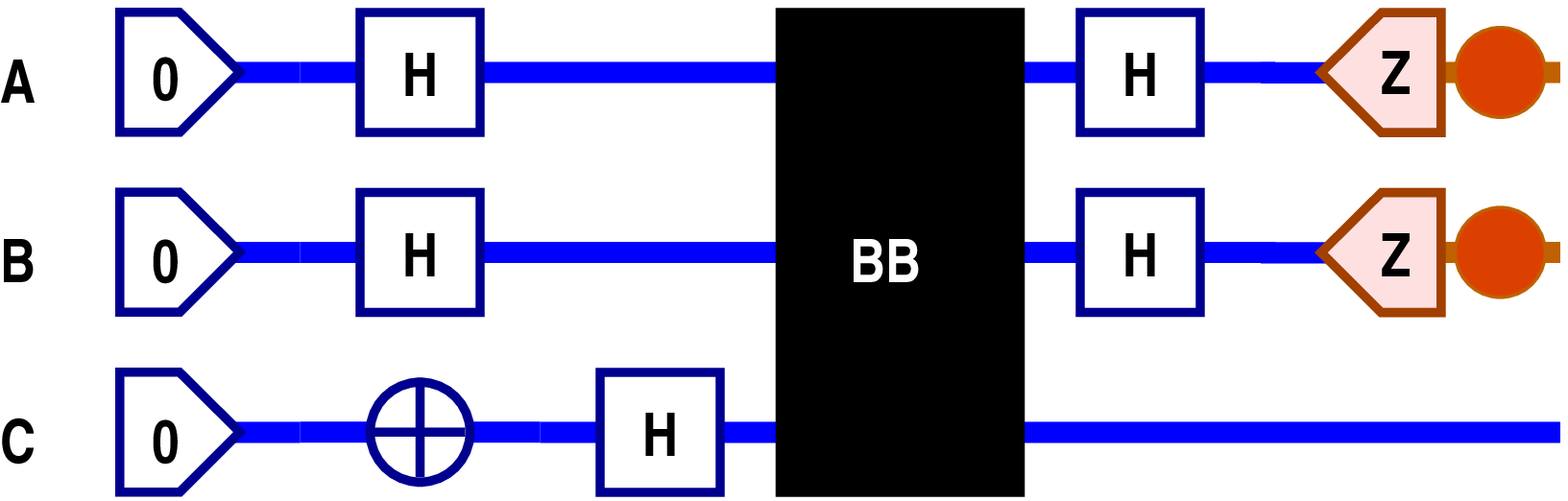}

\nputbox{3.15,-.3}{c}{\Large $b_{\sysfnt{A}}$}
\nputbox{3.15,-1.1}{c}{\Large $b_{\sysfnt{B}}$}

\nputbox{-2.05,-2.16}{t}{\Large \textbf{1}}
\nputbox{-1.28,-2.16}{t}{\Large \textbf{2}}
\nputbox{-.18,-2.16}{t}{\Large \textbf{3}}
\nputbox{1.2,-2.16}{t}{\Large \textbf{4}}
\nputbox{2,-2.16}{t}{\Large \textbf{5}}


\end{picture}
\herefigcap{Quantum network for solving the parity problem.
A quantum network has a (horizontal in this case) line for each qubit.
The line can be thought of as the time-line for the qubit and is shown
in blue. Each gate is drawn as a box, circle, or other element
intercepting the lines of the qubits it acts on. In
this case, time runs from left to right. Each
qubit's time-line starts at the point where it is added. In this
example, the qubits' time-lines end when they are measured, at which
point a classical bit (brown time line) containing the measurement
outcome is introduced. The operation $\mb{BB}$ is illustrated as a black
box.  The numbers underneath the network refer to checkpoints used to
explain how the network solves the parity problem.}
\label{fig:parity_network}
\end{herefig}

To understand how the quantum network of Fig~\ref{fig:parity_network}
solves the parity problem, we
can follow the states as the network is ``executed'' from left to
right, using the indicated checkpoints.
Using vector notation for the
states, at checkpoint \textbf{1}~the state is
\begin{equation}
\ket{\psi}_1 = \qavec{1}{0}\tensor\qavec{1}{0}\tensor\qavec{1}{0},
\end{equation}
where we used Kronecker product notation to denote the states of
$\sysfnt{A}$, $\sysfnt{B}$ and $\sysfnt{C}$, in this order.  In the
next time step, the network involves applying Hadamard gates
(Eq.~\ref{eq:hadamard}) to $\sysfnt{A}$ and $\sysfnt{B}$ and a
$\mb{not}$ gate (Eq.~\ref{eq:not}) to $\sysfnt{C}$. At checkpoint
\textbf{2}, this operation results in the state
\begin{equation}
\ket{\psi}_2 = \qavec{1/\sqrt{2}}{1/\sqrt{2}}\tensor\qavec{1/\sqrt{2}}{1/\sqrt{2}}\tensor\qavec{0}{1}.
\end{equation}
Next, a Hadamard gate is applied to $\sysfnt{C}$, so that
at checkpoint \textbf{3}~we have,
\begin{equation}
\ket{\psi}_3 = \qavec{1/\sqrt{2}}{1/\sqrt{2}}\tensor\qavec{1/\sqrt{2}}{1/\sqrt{2}}
\tensor\qavec{1/\sqrt{2}}{-1/\sqrt{2}}.
\end{equation}
The next event involves applying the black box.
To understand what happens, note that the effect of
the black box can be described as ``conditional on each
logical state of $\sysfnt{AB}$, if the parity according to $b_{\sysfnt{A}}$ and $b_{\sysfnt{B}}$
is $1$, then apply $\mb{not}$ to
$\sysfnt{C}$'' The current state of $\sysfnt{C}$
is such that if $\mb{not}$ is applied, only the sign changes:
\begin{eqnarray}
\mb{not}\qavec{1/\sqrt{2}}{-1/\sqrt{2}} &=&
  \qaop{0}{1}{1}{0}\qavec{1/\sqrt{2}}{-1/\sqrt{2}}\nonumber\\
 &=& -\qavec{1/\sqrt{2}}{-1/\sqrt{2}}.
\end{eqnarray}
Now $\sysfnt{AB}$ is in a superposition of each of the logical states,
and conditional on the logical state and the (hidden) parity, the sign
changes. As a result, although the state of $\sysfnt{C}$ does not
change, a phase is ``kicked back'' to $\sysfnt{AB}$. A generalization
of this effect is at the heart of A.~Kitaev's version
of P.~Shor's quantum factoring algorithm (Sect.~\ref{sec:factor}).
At the next checkpoint, and after some arithmetic to
check which logical states change sign, we can write the state as
\begin{equation}
\ket{\psi}_4 = \qavec{1/\sqrt{2}}{(-1)^{b_{\sysfnt{A}}}/\sqrt{2}}
              \tensor\qavec{1/\sqrt{2}}{(-1)^{b_{\sysfnt{B}}}/\sqrt{2}}
\tensor\qavec{1/\sqrt{2}}{-1/\sqrt{2}}.
\end{equation}
Notice that qubits $\sysfnt{A}$ and $\sysfnt{B}$ are in
orthogonal states for different values of $b_{\sysfnt{A}},b_{\sysfnt{B}}$. It suffices
to apply the Hadamard transform again to $\sysfnt{A}$ and
$\sysfnt{B}$ to get
\begin{equation}
\ket{\psi}_4 = \qavec{1-b_{\sysfnt{A}}}{b_{\sysfnt{A}}}
              \tensor\qavec{1-b_{\sysfnt{B}}}{b_{\sysfnt{B}}}
              \tensor\qavec{1/\sqrt{2}}{-1/\sqrt{2}}.
\end{equation}
Measurements of $\sysfnt{A}$ and $\sysfnt{B}$ now reveal
the previously unknown $b_{\sysfnt{A}}$ and $b_{\sysfnt{B}}$.

As can be seen, the visual representation of a quantum network eases
the tasks of following what happens. This is why it is used
extensively for presenting basic subroutines and algorithms in quantum
computation.  A guide to the commonly used network elements is given
in Fig.~\ref{fig:gate_list}.
\begin{herefig}
\begin{tabular}{|p{55pt}|p{79pt}|p{44pt}|p{160pt}|p{150pt}|}
\hline
Name & Gate & Symbols & Algebraic & Matrix\\
\hline
\hline
Add/prepare&
  \begin{tabular}[c]{@{}r@{}}
  \includegraphics[width=.7in]{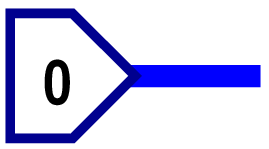}
  \end{tabular}
  & $\mb{add}$& 
  \begin{tabular}{@{}l@{}}
   If applied to existing qubit:  \\
  $\problb\ketbra{\bitzero}{\bitzero}\probplus
  \ketbra{\bitzero}{\bitone}\probrb$ \\ \hspace*{.75in} (operator mixture)
  \end{tabular}
  & $\qaop{1}{0}{0}{0}$ \probplus $\qaop{0}{1}{0}{0}$\\ \hline
Measure&
  \begin{tabular}[c]{@{}r@{}}
  \includegraphics[width=.7in]{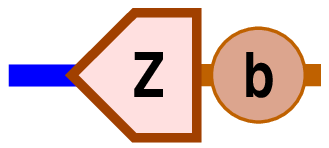}
  \end{tabular}
  & $\mb{meas}$ & 
  $\problb\bitzero\probmul\ketbra{\bitzero}{\bitzero}\probplus
  \bitone\probmul\ketbra{\bitone}{\bitone}\probrb$
  & $\qaop{1}{0}{0}{0}$ \probplus $\qaop{0}{0}{0}{1}$\\
\hline
\end{tabular}
\begin{tabular}{|p{55pt}|p{79pt}|p{44pt}|p{160pt}|p{150pt}|}
\hline
Not& 
  \begin{tabular}[c]{r@{}}
  \ppbox{0,-.06}{c}{or}\includegraphics[width=.5in]{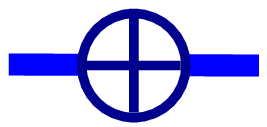}\\
  \includegraphics[width=.5in]{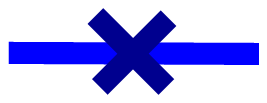}
  \end{tabular}
  & $\mb{not},\;\sigma_x$
  & $\ketbra{\bitzero}{\bitone}+\ketbra{\bitone}{\bitzero}$
  & $\qaop{0}{1}{1}{0}$\\\hline
Hadamard& 
  \begin{tabular}[c]{@{}r@{}}
  \includegraphics[width=.7in]{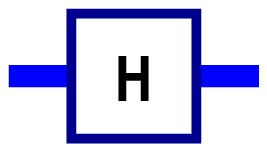}
  \end{tabular}
  & $\mb{H}$
  & $e^{-i\sigma_y\pi/4}\sigma_z$
  & ${1\over\sqrt{2}}\qaop{1}{1}{1}{-1}$\\\hline
\begin{tabular}[c]{l}Phase\\ change\end{tabular}
& 
  \begin{tabular}[c]{@{}r@{}}
  \includegraphics[width=.7in]{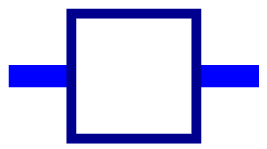}\ppbox{-.35,.2}{c}{\scalebox{1.1}{$e^{i\phi}$}}
  \end{tabular}
  & $\mb{S}(e^{i\phi})$
  & $e^{i\phi/2}e^{-i\sigma_z\phi/2}$
  & $\qaop{1}{0}{0}{e^{i\phi}}$\\\hline
$z$-Rotation& 
  \begin{tabular}[c]{@{}r@{}}
  \includegraphics[width=.7in]{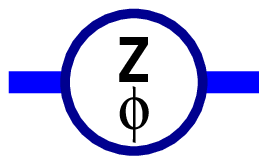}
  \end{tabular}
  & $\mb{Z}_{\phi}$
  & $e^{-i\sigma_z\phi/2}$ & $\qaop{e^{-i\phi/2}}{0}{0}{e^{i\phi/2}}$\\\hline
$y$-Rotation&
  \begin{tabular}[c]{@{}r@{}}
  \includegraphics[width=.7in]{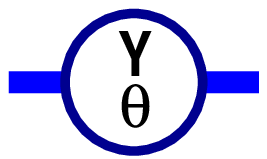}
  \end{tabular}
  &$\mb{Y}_{\theta}$
  & $e^{-i\sigma_y\theta/2}$
  & $\qaop{\cos(\theta/2)}{-\sin(\theta/2)}{\sin(\theta/2)}{\cos(\theta/2)}$\\\hline
$x$-Rotation& 
  \begin{tabular}[c]{@{}r@{}}
  \includegraphics[width=.7in]{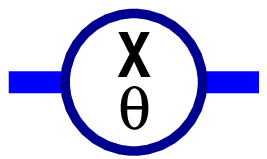}
  \end{tabular}
  &$\mb{X}_{\theta}$
  & $e^{-i\sigma_x\theta/2}$
  & $\qaop{\cos(\theta/2)}{-i\sin(\theta/2)}{-i\sin(\theta/2)}{\cos(\theta/2)}$\\\hline
\end{tabular}

\begin{tabular}{|p{55pt}|p{79pt}|p{44pt}|p{160pt}|p{150pt}|}
\hline
\begin{tabular}[c]{l}Controlled\\ not\end{tabular}&
  \begin{tabular}[c]{@{}c@{}c@{}c@{}}
  \includegraphics[width=.5in]{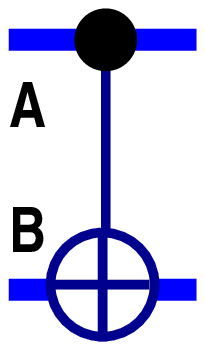}
  &\raisebox{.4in}{or}&
  \includegraphics[width=.5in]{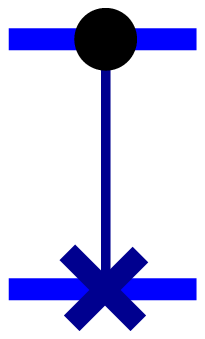}
  \end{tabular}
  & $\mb{cnot}$ 
  & $\begin{array}{c}
     \ketbras{\bitzero}{\bitzero}{A}+\ketbras{\bitone}{\bitone}{A}\slb{\sigma_x}{B}\\ \\
     e^{-i\slb{\sigma_z}{A}\pi/4}e^{-i{1\over 2}(\idop - \slb{\sigma_z}{A})\slb{\sigma_x}{B}\pi/2}
     \end{array}$
  & 
  {$\left(\begin{array}{cccc}
   1&0&0&0\\
   0&1&0&0\\
   0&0&0&1\\
   0&0&1&0
   \end{array}\right)$}\\\hline
\begin{tabular}[c]{l}$ZZ$\\ rotation\end{tabular}& 
  \begin{tabular}[c]{@{}r@{}}
  \includegraphics[width=.7in]{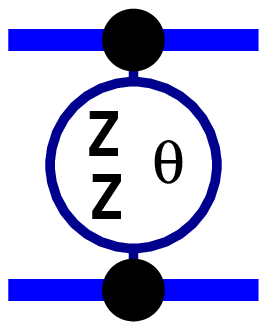}
  \end{tabular}
  & $(\mb{ZZ})_{\theta}$
  & $e^{-i\slb{\sigma_z}{A}\slb{\sigma_z}{B}\theta/2}$.
  & 
  {$\left(\!\begin{array}{cccc}
   e^{-i\theta/2}&0&0&0\\
   0&e^{i\theta/2}&0&0\\
   0&0&e^{i\theta/2}&0\\
   0&0&0&e^{-i\theta/2}
   \end{array}\!\right)$}\\\hline
\begin{tabular}[c]{l}Controlled\\ rotation\end{tabular}&
  \begin{tabular}[c]{@{}r@{}}
  \includegraphics[width=.7in]{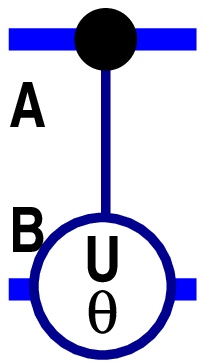}
  \end{tabular}
 & $\mb{cU}_\theta$
 & $\ketbras{\bitzero}{\bitzero}{A}+\ketbras{\bitone}{\bitone}{A}e^{-i\slb{\sigma_U}{B}\theta/2}$
 & 
  {$\left(\begin{array}{@{}cc@{}}
   \begin{array}{cc}1\;&\;0\\0\;&\;1\end{array}&\begin{array}{cc}0\;&\;0\\0\;&\;0\end{array}\\
   \begin{array}{cc}0\;&\;0\\0\;&\;0\end{array}&\mbox{\Large$e^{-i\sigma_U\theta/2}$}
   \end{array}\right)$}\\\hline
\hline
Toffoli gate&
  \begin{tabular}[c]{@{}r@{}}
  \includegraphics[width=.7in]{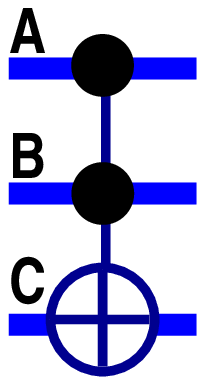}
  \end{tabular}
  & $\mb{c^2not}$
  & $\idop-\ketbras{\bitone\bitone}{\bitone\bitone}{AB} + 
           \ketbras{\bitone\bitone}{\bitone\bitone}{AB}\slb{\sigma_x}{C}$
  & \\\hline
\end{tabular}
\vspace*{\baselineskip}
\herefigcap{Quantum network elements.}
\label{fig:gate_list}
\end{herefig}

When designing or describing complicated algorithms for quantum
computers, providing everything in terms of quantum networks can
become difficult, particularly when an important part of the algorithm
consists of computations that are best done on a classical computer.
For example, a full description of Shor's algorithm for
factoring whole numbers (see~Sect.~\ref{sec:factor}) includes
a significant amount of classical preprocessing, which determines
choices made in the quantum algorithm, and classical postprocessing,
which computes a factor from the measured result by a continued
fraction algorithm. For such algorithms, one can use
a programming language similar to Pascal, BASIC or C enhanced with statements
to access quantum bits and to apply quantum operations.
For algorithm design, computer scientists often use a semi-formal
language called ``pseudocode''~\cite{cormen:qc1990a}. With
a simple extension called ``quantum pseudocode'', the algorithm
for the parity problem can be written as follows:

\begin{pseudocode}{\algname{BBparity}{$\mb{BB}$}}
\alginout{
Access to a quantum black box $\mb{BB}$ that acts
on three qubits by adding a parity function of the first
two qubits to the third.
}{
The two bits $b_{\sysfnt{A}}$ and $b_{\sysfnt{B}}$ of the parity function.
}
\begin{algtab*}
  \algforeach{$i\in\{\sysfnt{A},\sysfnt{B},\sysfnt{C}\}$}
    $\isq{\seqIndex{a}{i}}\assignfrom \ket{\bitzero}$\\
    \pcomment{Initialize three one-qubit registers 
    $\isq{\seqIndex{a}{i}}$, $i=\sysfnt{A},\sysfnt{B},\sysfnt{C}$. 
      The corner bracket annotation
      declares $\seqIndex{a}{i}$ as a quantum register.}
  \ealgend
  $\isq{\seqIndex{a}{C}}\assignfrom \sigma_x\isq{\seqIndex{a}{C}}$\\
  \algforeach{$i\in\{\sysfnt{A},\sysfnt{B},\sysfnt{C}\}$}
    $\isq{\seqIndex{a}{i}}\assignfrom \mb{H}\isq{\seqIndex{a}{i}}$\\
  \ealgend  
  $\isq{a}\assignfrom \mb{BB}\isq{a}$\\
    \pcomment{$\isq{a}$ refers to the three qubit register consisting of the
              $\isq{\seqIndex{a}{i}}$}
  \algforeach{$i\in\{\sysfnt{A},\sysfnt{B}\}$}
    $\isq{\seqIndex{a}{i}}\assignfrom \mb{H}\isq{\seqIndex{a}{i}}$\\
    $\seqIndex{b}{i} \assignfrom \mb{meas}\isq{\seqIndex{a}{i}}$\\
  \ealgend  
  \algreturn $b_{\sysfnt{A}},b_{\sysfnt{B}}$\\
\ealgend
\end{algtab*}
\end{pseudocode}

Any classical programming language can be extended with
statements to access and manipulate quantum registers.

Now that we have looked at the quantum solution of the parity problem,
let us consider the question of the least number of black-box
applications required by a classical algorithm: Each classical use of
the black box can only give us one bit of information.  In particular,
one use of the black box with input $a_{\sysfnt{A}}a_{\sysfnt{B}}$
reveals only the parity of $a_{\sysfnt{A}}a_{\sysfnt{B}}$ according to
the hidden parameters $b_{\sysfnt{A}}$ and $b_{\sysfnt{B}}$.  Each use
of the black box can therefore only help us distinguish between two
subsets of the four possible parities. At least two uses of the black
box are therefore necessary.  Two uses are also sufficient: To
determine which of the four parities is involved, use the black box first with
input $a_{\sysfnt{A}}a_{\sysfnt{B}}=10$ and then with input
$a_{\sysfnt{A}}a_{\sysfnt{B}}=01$. As a result of this argument, one
can consider the parity problem as a simple example of a case in which
there is a more efficient quantum algorithm than is possible
classically. However, it is worth noting that the comparison is not
entirely fair: A truly classical oracle answering parity questions or
implementing the black box on the states of classical bits is useless
to a quantum algorithm.  To take advantage of such an algorithm it
must be possible to use superpositions that are not implicitly
collapsed. Collapse can happen if the oracle makes a measurement or
otherwise ``remembers'' the question that it was asked.

\subsection{Resource Accounting}

When trying to solve a problem using quantum information processing,
an important issue is to determine what physical resources are
available and how much of each resource is needed for the solution.
As mentioned before, in classical information, the primary resources
are bits and operations. The number of bits used by an algorithm is
called its ``space'' requirement. The number of operations used is
called its ``time'' requirement. If parallel computation is available,
one can distinguish between the total number of operations (``work'')
and the number of parallel steps (``time'').

When quantum information processing is used, the classical resources
are still relevant for running the computer that controls the quantum
system and performs any pre- and post-processing tasks.  The main
quantum resources are analogous to the classical ones: ``quantum
space'' is the number of qubits needed, and ``quantum time'' the
number of quantum gates. Because it turns out that reset operations
have a thermodynamic cost, one can count irreversible
quantum operations separately.  This accounting of the resource
requirements of algorithms and of the minimum resources needed to
solve problems forms the foundations of quantum complexity theory.

As a simple example of resource accounting, consider the algorithm for
the parity problem. No classical computation is required to decide
which quantum gates to apply, or to determine the answer from the
measurement.  The quantum network consists of a total of 11 quantum
gates (including the $\mb{add}$'s and $\mb{meas}$'s operations) and
one oracle call (the application of the black box).  In the case of
oracle problems, one usually counts the number of oracle calls first,
as we have done in discussing the algorithm.  The network is readily
parallelized to reduce the time resource to 6 steps.

\subsection{From Factoring to Phase Estimation}
\label{sec:factor}

The publication of Shor's quantum algorithm for efficiently
factoring numbers~\cite{shor:qc1994a,shor:qc1995a} was the key event that
stimulated many theoretical and experimental investigations of quantum
computation.  One of the reasons why this algorithm is so important is
that the security of widely used public key cryptographic protocols
relies on the conjectured difficulty of factoring large numbers.  An
elementary overview of these protocols and the quantum algorithm for
breaking them is in~\cite{ekert:qc1998c}. Here, we outline the
relationship between factoring and the powerful technique of phase
estimation. This relationship helps in understanding many of the
existing quantum algorithms and was first explained
in~\cite{cleve:qc1997b}, motivated by Kitaev's
version~\cite{kitaev:qc1995a} of the factoring algorithm.

The factoring problem requires writing a whole number $N$ as a product
of primes. (Primes are whole numbers greater than $1$ that are
divisible without remainder only by $1$ and themselves.)  Shor's
algorithm solves this problem by reducing it to instances of the
order-finding problem, which will be defined below.  The reduction is
based on basic number theory and involves efficient classical
computation. At the core of Shor's algorithm is a quantum algorithm
that solves the order-finding problem efficiently.  In this case, an
algorithm is considered efficient if it uses resources bounded by a
polynomial in the number of digits of $N$. For more
information on the requisite number theory, see any textbook on number
theory~\cite{bolker:qc1970a,hardy:qc1979a}.

We begin by showing that factoring reduces
to order finding.  The first observation is that to factor a whole number it
is sufficient to solve the factor-finding problem, whose statement is:
Given a whole number $N$ find a proper factor of $N$, if one exists.
A ``factor'' of $N$ is a whole number $f$ that satisfies $N=fg$ for
some whole number $g$.  The factor $f$ is ``proper'' if $f\not=1$ and
$f\not= N$. For example, if $N=15$, then $3$ and $5$ are its proper
factors. For some numbers it is easy to find a proper factor.  For
example, you can tell that $N$ is even from the least significant
digit (in decimal or binary), in which case $2$ is a proper factor
(unless $N=2$, a prime). But many numbers are not so easy. As an
example, you can try to find a proper factor of $N=149573$ by
hand\footnote{\tiny
\rotatebox{180}{149573=373*401}}.
You can complete the factorization of a whole number by recursively
applying an algorithm for the factor-finding problem to all the proper
factors found.

Before we continue the reduction of factoring to order finding, we
briefly explain modular arithmetic, which both simplifies the
discussion and is necessary to avoid computing with numbers that have
exponential numbers of digits. We say that $a$ and $b$ are ``equal
modulo $N$'', written as $a=b\mod N$, if $a-b$ is divisible by $N$
(without remainder). For example, $3=18\mod 15 = 33\mod 15$.  Equality
modulo $N$ is well-behaved with respect to addition and
multiplication. That is, if $a=b\mod N$ and $c=d\mod N$, then
$a+c=b+d\mod N$ and $ac=bd\mod N$. For factoring $N$, we will be
looking for whole numbers $a$ that are divisible by a proper factor of
$N$.  If $a$ has this property, then so does any $b$ with $b=a\mod N$.
We therefore perform all arithmetic ``modulo $N$''.  One way to think
about this is that we only use whole numbers $a$ that satisfy $0\leq
a\leq N-1$.  We can implement an arithmetic operation modulo $N$ by
first applying the operation in the usual way and then computing the
remainder after division by $N$.  For example, to obtain $ab\mod N$,
we first compute $ab$. The unique $c$ such that $0\leq c\leq N-1$ and
$c=ab\mod N$ is the remainder after division of $ab$ by $N$. Thus $c$
is the result of multiplying $a$ by $b$ modulo $N$.  Consistent with
this procedure, we can think of the expression $a\mod N$ as referring
to the remainder of $a$ after division by $N$.

The second observation in the reduction of factoring to order finding
is that it is sufficient to find a whole number $r$ with the property
that $r^2-1$ is a multiple of $N$ but $r-1$ and $r+1$ are not.  Using
the language of modular arithmetic, the property is expressed as
$r^2=1\mod N$ but $r\not=1\mod N$ and $r\not= -1\mod N$.  Because $1\mod
N$ and $-1\mod N$ are the obvious square roots of $1\mod N$, we say
that $r$ is a ``non-trivial square root of unity'' (modulo $N$).  For
such an $r$, one can write $r^2-1=(r-1)(r+1)=mN$ for some whole number
$m$. This implies that every prime factor $p$ of $N$ divides either
$(r-1)$ or $(r+1)$ so that either $(r-1)$ or $(r+1)$ is or shares a
factor with $N$.  Suppose that $r-1$ is or shares such a factor. Because
$r-1$ is not a multiple of $N$, the greatest common divisor of $r-1$
and $N$ is a proper factor of $N$.  Since there exists an efficient
classical algorithm (the ``Euclidean algorithm'') for finding the
greatest common divisor, we can easily find the desired proper factor.

The examples of $N=15$ and $N=21$ serve to illustrate the key
features of the algorithm. For $N=15$, possible choices for $r$ are
$r=4$\ \ ($4^2-1=1*15$) and $r=11$\ \ ($11^2-1 = 120 = 8*15$).  For
the first choice, the proper factors emerge immediately:
$4-1=3,4+1=5$.  For the second, it is necessary to determine the
greatest common divisors. Let $\textrm{gcd}(x,y)$ stand for the
greatest common divisor of $x$ and $y$. The proper factors are 
$\textrm{gcd}(11-1,15)=\textrm{gcd}(10,15)=5$ and
$\textrm{gcd}(11+1,15)=\textrm{gcd}(12,15)=3$.  For $N=21$, one can
take $r=8$, as $8^2-1=63=3*21$.  In this case, $8-1=7$ is a proper
factor and $\textrm{gcd}(8+1,21)=3$ is another.

For $N$ even or a power of a prime it is not always possible to find a
non-trivial square root of unity. Because both of these cases can be
handled efficiently by known classical algorithms, we can exclude
them.  In every other case, such numbers $r$ exist. One way to find
such an $r$ is to start from any whole number $q$ with $1<q<N$. If
$\textrm{gcd}(q,N)=1$, then according to a basic result in number
theory there is a smallest whole number $k>1$ such that $q^k-1=0\mod
N$.  The number $k$ is called the ``order'' of $q$ modulo $N$. If $k$
is even, say $k=2\,l$, then $(q^l)^2=1\mod N$, so $q^l$ is a (possibly
trivial) square root of unity.  For the example of $N=15$, we can try
$q=2$. The order of $2$ modulo $15$ is $4$, which gives $r=2^2=4$, 
the first of the two choices in the previous paragraph. For $N=21$, again with
$q=2$, the order is $6$: $2^6-1=63=3*21$. Thus, $r=2^3=8$.  We can also
try $q=11$, in which case with foresight it turns out that $11^6-1$ is
divisible by $21$. A possible problem appears, namely, the
powers $q^k$ that we want to compute are extremely large.
But modular arithmetic can be used to avoid this problem.
For example, to find the order of $11$ modulo
$21$ by a direct search, we can perform the following computation:
\begin{equation}
\begin{array}[b]{rcrcrcrcrcrcrcrcr}
11^2 &=& 121 &=& 5*21+16 &=& 16\mod 21\\
11^3&=&11*11^2&=& && 11*16\mod 21 &=& 11*(-5)\mod 21 \\
    &&        &=& 
    && -55\mod 21 &=& -3*21+8\mod 21 &=& 8\mod 21\\
   11^4&=&11*11^3&=& &&11*8\mod 21 &=&4*21+4\mod 21 &=& 4\mod 21\\
   11^5&=&11*11^4 &=& &&11*4\mod 21 &=& 2\mod 21\\
   11^6 &=& 11*11^5&=& &&11*2 \mod 21&=& 1\mod 21
\end{array}
\end{equation}
In general such a direct search for the order of $q$ modulo $N$ is
very inefficient, but as we will see, there is an efficient quantum
algorithm that can determine the order.

A factor-finding algorithm based on the above observations
is the following:
\begin{pseudocode}{\algname{FactorFind}{$N$}}
\alginout{
A positive, non-prime whole number $N$.
}{
A proper factor $f$ of $N$, that is $f$ is a whole number
such that $1<f<N$ and $N=fg$ for some
whole number $g$.
}
\begin{itemize}
  \item[1.] If $N$ is even, return $f=2$.
  \item[2.] If $N=p^k$ for $p$ prime, return $p$.
  \item[3.] Randomly pick $1<q<N-1$.
  \begin{itemize}
     \item[3.a.] If $f=\textrm{gcd}(q,N)>1$ return $f$.
  \end{itemize}
  \item[4.] Determine the order $k$ of $q$ modulo $N$ using
  the quantum order-finding algorithm.
     \begin{itemize}
     \item[4.a.] If $k$ is not even, repeat at step 3.
     \end{itemize}
  \item[5.] Write $k=2l$ and determine $r=q^l\mod N$ with
  $1<r<N$.
  \begin{itemize}
     \item[5.a.] If $1<f=\textrm{gcd}(r-1,N)<N$, return $f$.
     \item[5.b.] If $1<f=\textrm{gcd}(r+1,N)<N$, return $f$.
     \item[5.c.] If we failed to find a proper factor, repeat at step 3.
  \end{itemize}
\end{itemize}
\end{pseudocode}
The efficiency of this algorithm depends on the probability that a
randomly chosen $q$ at step 3 results in finding a factor. By using
an analysis of the group of numbers $q$ that satisfy
$\textrm{gcd}(q,N)=1$, it can be shown that this probability is
sufficiently large.

The main problem that remains to be solved is
that of finding the order of $q\mod N$. A direct search for
the order of $q\mod N$ involves 
computing the sequence
\begin{equation}
1\rightarrow q \rightarrow q^2\mod N \rightarrow\ldots\rightarrow q^{k-1}\mod N
 \rightarrow 1=q^{k}\mod N.
\end{equation}
This sequence can be conveniently visualized as a cycle whose length
is the order of $q\mod N$ (Fig.~\ref{fig:cycord}).

\begin{herefig}
\begin{picture}(3,3)(-1.5,-1.5)
\nputbox{0,1.5}{t}{The cycle of $q\mod N$} 
\nputbox{0,1}{c}{$1$}
\nputbox{.866,.5}{c}{$q$}
\nputbox{.866,-.5}{c}{$q^2\mod N$}
\nputbox{0,-1}{c}{$\ldots$}
\nputbox{-.866,.5}{c}{$q^{k-1}\mod N$}
\nputbox{-.866,-.5}{c}{$q^{k-2}\mod N$}

\nputgr{0,0}{c}{}{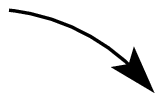}
\nputgr{0,0}{c}{angle=60}{c60arr}
\nputgr{0,0}{c}{angle=-60}{c60arr}
\nputgr{0,0}{c}{angle=120}{c60arr}
\nputgr{0,0}{c}{angle=-120}{c60arr}
\nputgr{0,0}{c}{angle=180}{c60arr}

\end{picture}\hspace*{.5in}
\begin{picture}(3,3)(-1.5,-1.5)
\nputbox{0,1.5}{t}{The cycle of $8\mod 15$} 
\nputbox{0,1}{c}{$1$}
\nputbox{1,0}{c}{$8$}
\nputbox{0,-1}{c}{$4$}
\nputbox{-1,0}{c}{$2$}

\nputgr{0,0}{c}{}{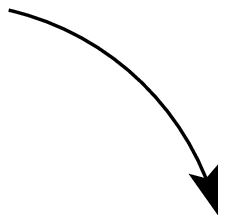}
\nputgr{0,0}{c}{angle=90}{c90arr}
\nputgr{0,0}{c}{angle=-90}{c90arr}
\nputgr{0,0}{c}{angle=180}{c90arr}
\end{picture}
\label{fig:cycord}
\herefigcap{Multiplicative cycles of $q\mod N$.
Each number on a cycle is obtained from the previous one by
multiplication by $q\mod N$.}
\end{herefig}

To introduce the quantum algorithm, we first associate the logical
quantum states $\ket{0},\ket{1},\ldots\ket{N-1}$ with the numbers
$0,1,\ldots,N-1$.  The map $f$ which takes each number on the cycle to
the next number along the cycle is given by $f(x)=qx\mod N$.  For $q$
satisfying $\textrm{gcd}(q,N)=1$, the map $f$ permutes not only the
numbers on the cycle, but all the numbers modulo $N$.  As a result,
the linear operator $\hat f$ defined by $\hat f\ket{x}=
\ket{f(x)}=\ket{qx\mod N}$ is unitary.  The quantum algorithm
deduces the length of the cycle for $q$ by making measurements to
determine properties of the action of $\hat f$ on superpositions of
the states $\ket{q^s\mod N}$.  To illustrate the basic ideas, we work
out the example of $N=15$ and $q=8$. The action of $\hat f$ on the
states $\ket{1},\ket{8},\ket{4},\ket{2}$ in the cycle of $8\mod 15$ is
completely determined by the eigenstates and eigenvalues of $\hat f$.  For
cyclicly acting permutations, a basis of eigenstates is given by the
``Fourier'' basis for the space spanned by the states in a cycle. For
the cycle of interest, the Fourier basis consists of the states
\begin{equation}
  \begin{array}{rcr@{}c@{}r@{}c@{}r@{}c@{}r}
  \ket{\psi_0} &=& {1\over 2}\Big(\ket{1}&+&\ket{8}&+&\ket{4}&+&\ket{2}\Big)\\
  \ket{\psi_1} &=& {1\over 2}\Big(\ket{1}&+&i\ket{8}&-&\ket{4}&-&i\ket{2}\Big)\\
  \ket{\psi_2} &=& {1\over 2}\Big(\ket{1}&-&\ket{8}&+&\ket{4}&-&\ket{2}\Big)\\
  \ket{\psi_3} &=& {1\over 2}\Big(\ket{1}&-&i\ket{8}&-&\ket{4}&+&i\ket{2}\Big)
  \end{array}
\label{eq:fourier4}
\end{equation}
The phases of the $l$'th state of the cycle occurring in the sum for
$\ket{\psi_m}$ can be written as $i^{lm}$.  It follows that $\hat
f\ket{\psi_m} = i^m\ket{\psi_m}$, that is, the eigenvalue of $\hat f$
for $\ket{\psi_m}$ is $i^m$.  Note that in the complex numbers, the
powers of $i$ are all the fourth roots of unity. In general, the
Fourier basis for the cycle $\ldots\rightarrow\ket{q^l\mod
N}\rightarrow\ldots$ consists of the states $\ket{\psi_m}=\sum_l
\omega^{lm}\ket{q^l\mod N}$, where $\omega=e^{i2\pi/k}$ is a primitive
$k$'th root of unity.  (The complex number $x$ is a primitive $k$'th
root of unity if $k$ is the smallest whole number $k>0$ such that
$x^k=1$. For example, both $-1$ and $i$ are fourth roots of unity, but
only $i$ is primitive.)

It is, of course, possible to express the logical state $\ket{1}$ using
the Fourier basis:
\begin{equation}
\ket{1} = 
{1\over 2}\Big(\ket{\psi_0}+\ket{\psi_1}+\ket{\psi_2}+\ket{\psi_3}\Big).
\label{eq:1sup}
\end{equation}
The key step of the quantum algorithm for order finding consists of a
measurement to estimate a random eigenvalue of $\hat f$ whose
associated eigenstate occurs in the expression for $\ket{1}$ in terms
of the Fourier basis. If the eigenvalue found is a primitive $k$'th
root of unity, we infer that the cycle length is divisible by $k$ and
check (using a classical algorithm) whether this is the order of
$q$. In the example, the random eigenvalues are $1$ (the only
primitive first root of unity), $i$ and $-i$ (primitive fourth roots
of unity) and $-1$ (the primitive second root of unity). The
order is found if the random eigenvalue is a primitive fourth root of
unity, which happens with probability $1/2$ in this case.

The quantum algorithm for obtaining an eigenvalue is called the
``phase estimation'' algorithm. It exploits a more general version of
the phase kick back we encountered in the solution of the parity
problem. The phase kick back transfers the eigenvalue of an eigenstate
of $\hat f$ to a Fourier basis on a number of additional
qubits called ``helper'' or ``ancilla'' qubits. Which Fourier state
results is then determined by a subroutine called the ``measured
quantum Fourier transform''.  We introduce these elements in the next
paragraphs.  Their combination for solving the general order-finding
problem is shown in Fig.~\ref{fig:qorder}.

Fig.~\ref{fig:phkick} shows how to kick back the eigenvalue of an
eigenstate of $\hat f$ using a network implementing the
controlled-$\hat f$ operation.
\begin{herefig}
\begin{picture}(4,2.5)(-2,-2.5)
\nputgr{0,0}{t}{}{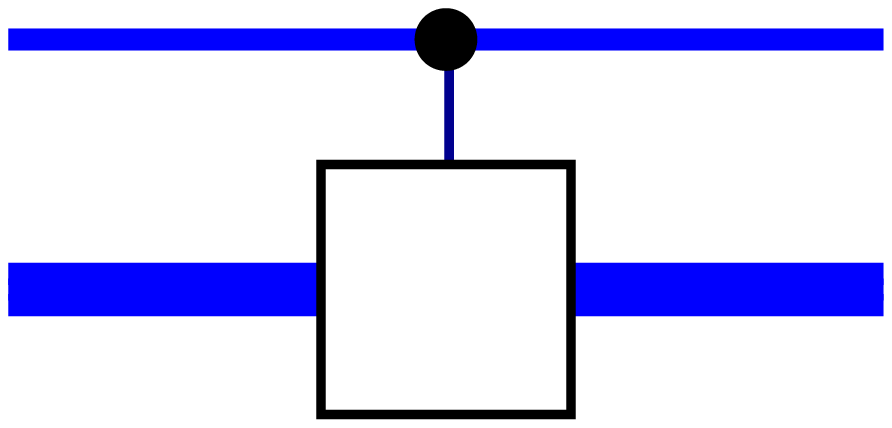}
\nputbox{0,-1.5}{c}{\scalebox{2}{$\hat f$}}
\nputbox{-1.8,-.5}{r}{\scalebox{1.4}{${1\over\sqrt{2}}\left(\ket{\bitzero}+\ket{\bitone}\right)$}}
\nputbox{-1.8,-1.5}{r}{\scalebox{1.4}{$\ket{\psi_m}$}}

\nputbox{1.8,-.5}{l}{\scalebox{1.4}{${1\over\sqrt{2}}\left(\ket{\bitzero}+ i^m\ket{\bitone}\right)$}}
\nputbox{1.8,-1.5}{l}{\scalebox{1.4}{$\ket{\psi_m}$}}
\end{picture}
\label{fig:phkick}
\herefigcap{Phase estimation with one qubit.
The input is a product state on one ancilla qubit and on a second
quantum system as shown. The state $\ket{\psi_m}$ on the second system
is an eigenstate of $\hat f$. For the example under discussion (see
Eq.~\ref{eq:fourier4}), the eigenvalue is $i^m$. A controlled-$\hat f$
operation is applied to the input, that is, $\hat f$ is applied to the
second system conditional on $\ket{\bitone}$ for the ancilla qubit. In
the bra-ket notation, the total operation can be written as
$\ketbra{\bitzero}{\bitzero}+\ketbra{\bitone}{\bitone}\hat f$ (system
labels have been omitted).  Since $\hat f$ changes only the phase of
its input, the second system is unchanged, but the phase modifies the
ancilla qubit's superposition as shown.  }
\end{herefig}
The network in Fig.~\ref{fig:phkick} can be used with
input $\ket{1}$ on the second system. From Eq.~\ref{eq:1sup}
and the superposition principle, it follows that the output correlates the
different phase kickback states with the four eigenvectors $\ket{\psi_m}$.
That is, the network implements the following transformation:
\begin{equation}
{1\over 2\sqrt{2}}\left(\ket{\bitzero}+\ket{\bitone}\right)
  \left(
  \begin{array}{@{}l@{}}
    \ket{\psi_0}\\
    \hspace*{.2in}+\ket{\psi_1}\\
    \hspace*{.4in}\iboxlike{$+$}+\ket{\psi_2}\\
    \hspace*{.6in}\iboxlike{$+$}\iboxlike{$+$}+\ket{\psi_3}
  \end{array}\right)
    \;\longrightarrow\;
  {1\over 2\sqrt{2}}
  \left(
  \begin{array}{@{}l@{}}
    \left(\ket{\bitzero}+i^0\ket{\bitone}\right)\ket{\psi_0}\\
    \hspace*{.2in}+\left(\ket{\bitzero}+i^1\ket{\bitone}\right)\ket{\psi_1}\\
    \hspace*{.4in}\iboxlike{$+$}+\left(\ket{\bitzero}+i^2\ket{\bitone}\right)\ket{\psi_2}\\
    \hspace*{.6in}\iboxlike{$+$}\iboxlike{$+$}+\left(\ket{\bitzero}+i^3\ket{\bitone}\right)\ket{\psi_3}
  \end{array}  \right)
\end{equation}
The hope is that a measurement of the first qubit can distinguish
between the four possible phases that can be kicked back.
However, because the four states are not mutually orthogonal,
they are not unambiguously distinguishable by a measurement.
To solve this problem, we use a second qubit and a controlled-$\hat f^2$
as shown in Fig.~\ref{fig:phkick2}.
\begin{herefig}
\begin{picture}(4.5,3)(-1.8,-3)
\nputgr{0,0}{t}{}{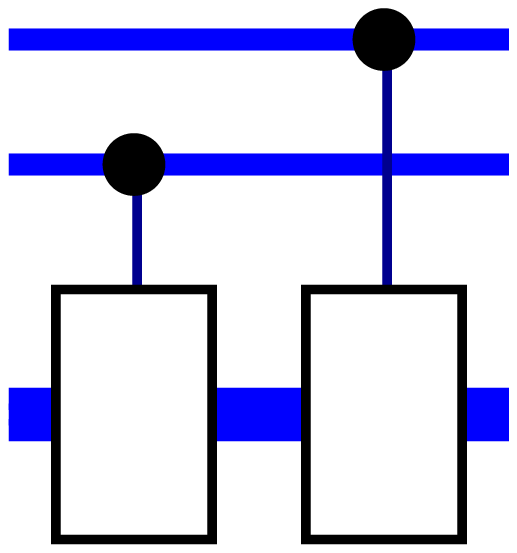}
\nputbox{-.5,-2}{c}{\scalebox{2}{$\hat f$}}
\nputbox{.5,-2}{c}{\scalebox{2}{$\hat f^2$}}
\nputbox{-1.05,-.75}{r}{$\left\{\rule{0in}{.4in}\right.$}
\nputbox{-1.25,-.75}{r}{\scalebox{1.5}{${1\over 2}$}\scalebox{1.2}{
  $\left(\begin{array}{@{}l@{}}
      \ket{0}\\\hspace*{.1in}{+}\ket{1}\\\hspace*{.2in}\iboxlike{$+$}{+}\ket{2}\\\hspace*{.3in}\iboxlike{$+$}\iboxlike{$+$}{+}\ket{3}\end{array}\right)$}}
\nputbox{-1.25,-2}{r}{\scalebox{1.4}{$\ket{\psi_m}$}}

\nputbox{1.05,-.75}{l}{$\left.\rule{0in}{.4in}\right\}$}
\nputbox{1.25,-.75}{l}{\scalebox{1.2}{$\ket{\textrm{u}_m}=$}\scalebox{1.5}{${1\over 2}$}\scalebox{1.2}{
  $\left(\begin{array}{@{}l@{}}
      i^{0m}\ket{0}\\\hspace*{.1in}{+}i^{1m}\ket{1}\\\hspace*{.2in}\iboxlike{$+$}{+}i^{2m}\ket{2}\\\hspace*{.3in}\iboxlike{$+$}\iboxlike{$+$}{+}i^{3m}\ket{3}\end{array}\right)$}}
\nputbox{1.25,-2}{l}{\scalebox{1.4}{$\ket{\psi_m}$}}
\end{picture}
\herefigcap{ Phase estimation with two qubits.
Using two qubits ensures distinguishability of the eigenvalues of
$\hat f$ for the states $\ket{\psi_m}$.  The states of the
input qubits are used to represent the numbers from $0$ to $3$ in
binary. The most significant bit (the ``two'''s digit in the binary
representation) is carried by the top qubit.  That is, we make the
following identification: $\ket{0}=\ket{\bitzero\bitzero}$,
$\ket{1}=\ket{\bitzero\bitone}$, $\ket{2}=\ket{\bitone\bitzero}$ and
$\ket{3}=\ket{\bitone\bitone}$.  It follows that the network has the
effect of applying $\hat f^m$ conditional on the input qubits'
logical state being $\ket{m}$. }
\label{fig:phkick2}
\end{herefig}
The four possible states $\ket{u_m}$ that appear on the ancilla
qubits in the network of Fig.~\ref{fig:phkick2} are the Fourier basis
for the cycle $0\rightarrow 1\rightarrow 2\rightarrow 3\rightarrow 0$
and are therefore orthonormal.  If we apply the network of
Fig.~\ref{fig:phkick2} with $\ket{1}$ instead of $\ket{\psi_m}$ at the
lower input, the output correlates the four $\ket{\psi_m}$ in the
superposition with the $\ket{u_m}$, which makes the information about
the eigenvalues of $\hat f$ available in the Fourier basis of the two
ancilla qubits. This approach has the advantage that the states are known,
whereas in the Fourier basis for the cycle of $q\mod N$, the states depend on
the numbers in the cycle, which are not known in advance (except in
very simple cases, such as the example we are working with).

To learn one of the eigenvalues of $\hat f$, the last step is to make
a ``measurement in the Fourier basis''.  For one qubit representing
the binary numbers $0$ and $1$, the Fourier basis is ${1\over
\sqrt{2}}\left(\ket{0}+\ket{1}\right)$ and ${1\over
\sqrt{2}}\left(\ket{0}-\ket{1}\right)$, which is constructed as
discussed after Eq.~\ref{eq:fourier4}, but using the square root of
unity $\omega=-1$ instead of the fourth root $i$. To make a measurement that
determines which of the two basis vectors is present, it suffices to
apply the Hadamard transform $\mb{H}$ and make a standard measurement,
just as we did twice in the network of Fig.~\ref{fig:parity_network}.
A more complicated network works with two qubits representing the
binary numbers from $0$ to $3$. Such a network is shown in
Fig.~\ref{fig:measfour4}.

\begin{herefig}
\begin{picture}(6,3)(-3,-3)
\nputgr{0,0}{t}{}{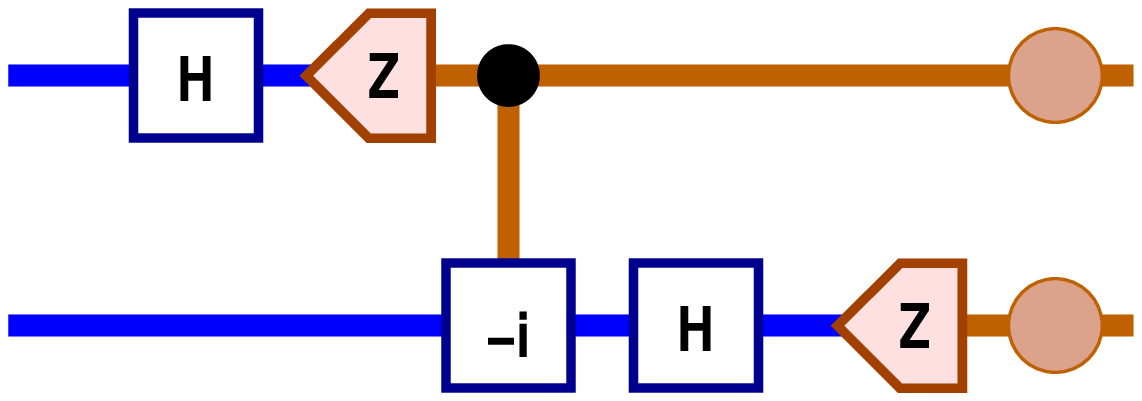}
\nputbox{-2.3,-1}{r}{$\left\{\rule{0in}{.7in}\right.$}
\nputbox{-2.5,-1}{r}{\scalebox{1.2}{$\ket{u_{2*a_1+a_0}}$}}
\nputbox{1.9375,-.5}{c}{\scalebox{1.3}{$a_0$}}
\nputbox{1.9375,-1.5}{c}{\scalebox{1.3}{$a_1$}}
\nputbox{-2,-1.7}{tr}{\Large \textbf{1}}
\nputbox{-1,-1.7}{tr}{\Large \textbf{2}}
\nputbox{0.25,-1.7}{tr}{\Large \textbf{3}}
\nputbox{1,-1.7}{tr}{\Large \textbf{4}}
\end{picture}
\herefigcap{
The measured quantum Fourier transform~\cite{griffiths:qc1995a} on two qubits
representing the numbers $0,1,2,3$.
If the input is one of the Fourier states $\ket{u_{a}}$,
where the binary digits of $a$ are determined
by $a=2*a_1+a_0$, then the measurement outcomes are $a_0$
and $a_1$, as shown. The numbers under the network
are checkpoints used for analyzing the network.}
\label{fig:measfour4}
\end{herefig}
To see how the network extracts the bits in the index
of $\ket{u_a}$, we can follow the states as the network is executed.
The input state at checkpoint \textbf{1} in Fig.~\ref{fig:measfour4}
is given by
\begin{equation}
\ket{\phi_1}=\ket{u_a} = 
{1\over 2}\left(
\begin{array}{@{}l@{}}
 i^{0*a}\ket{0}\\
 {+}i^{1*a}\ket{1}\\
 \iboxlike{$+$}{+}i^{2*a}\ket{2}\\
 \iboxlike{$+$}\iboxlike{$+$}{+}i^{3*a}\ket{3}
\end{array}\right)
  = {1\over 2}\left(
\begin{array}{@{}l@{}}
     i^{(0*2^1+0*2^0)(a_1*2^1+a_0*2^0)}\ket{\bitzero\bitzero}\\
      {+}i^{(0*2^1+1*2^0)(a_1*2^1+a_0*2^0)}\ket{\bitzero\bitone}\\
      \iboxlike{$+$}{+}i^{(1*2^1+0*2^0)(a_1*2^1+a_0*2^0)}\ket{\bitone\bitzero}\\
      \iboxlike{$+$}\iboxlike{$+$}{+}i^{(1*2^1+1*2^0)(a_1*2^1+a_0*2^0)}\ket{\bitone\bitone}
\end{array}\right).
\label{eq:cp1f}
\end{equation}
In the last sum, the relevant numbers have been fully expanded in
terms of their binary digits to give a flavor of the general
principles underlying the measured Fourier transform. The next step of
the network applies a Hadamard gate to the qubit carrying the most
significant digit.  To understand how it succeeds in extracting $a_0$,
the least significant bit of $a$, let $b$ with binary digits $b_0$ and
$b_1$ represent one of the logical states of the two qubits.  As
before, the most significant bit $b_1$ is represented by the top/first
qubit that the first Hadamard gate is applied to. The phase of
$\ket{b}$ in Eq.~\ref{eq:cp1f} is given by
$i^{(b_1*2^1+b_0*2^0)(a_1*2^1+a_0*2^0)}$.  Next, we determine how this phase
depends on $b_1$:
\begin{eqnarray}
i^{(b_1*2^1+b_0*2^0)(a_1*2^1+a_0*2^0)} &=& 
  i^{b_1*2^1*(a_1*2^1+a_0*2^0)}\;i^{b_0*2^0*(a_1*2^1+a_0*2^0)}
  \nonumber \\
   &=& i^{b_1*a_1*2^2}i^{b_1*a_0*2^1}\;i^{b_0*2^0*(a_1*2^1+a_0*2^0)}
  \nonumber \\
   &=& (i^4)^{b_1*a_1}(i^2)^{b_1*a_0}\;i^{b_0*2^0*(a_1*2^1+a_0*2^0)}
  \nonumber \\
     &=& (-1)^{b_1*a_0}\;i^{b_0*2^0*(a_1*2^1+a_0*2^0)}.
\end{eqnarray}
It follows that if $a_0=0$, the phase does not depend on $b_1$,
and if $a_0=1$, it changes sign with $b_1$. This sign change can be detected
by performing the Hadamard transform and measuring, as
can be seen explicitly by computing the state after the Hadamard
transform at checkpoint \textbf{2}:
\begin{eqnarray}
\ket{\phi_2} &=&
{1\over\sqrt{2}}\left(
 i^{0*2^0*(a_1*2^1+a_0*2^0)}\ket{a_0}\ket{\bitzero}
  + i^{1*2^0*(a_1*2^1+a_0*2^0)}\ket{a_0}\ket{\bitone}
\right) \nonumber \\
&=&
\ket{a_0}
{1\over\sqrt{2}}\left(
 i^{0*2^0*(a_1*2^1+a_0*2^0)}\ket{\bitzero}
  + i^{1*2^0*(a_1*2^1+a_0*2^0)}\ket{\bitone}
\right).
\end{eqnarray}
The phases still show a dependence on $a_0$ via
the terms $i^{b_0*2^0*a_0*2^0} = i^{b_0a_0}$.
The purpose of the phase shift gate conditioned on
the measurement outcome is to remove that dependence. The result is
the following state on the remaining qubit at checkpoint \textbf{3}:
\begin{eqnarray}
\ket{\phi_3} &=& {1\over\sqrt{2}}\left(
 i^{0*2^0*a_1*2^1}\ket{\bitzero}
  + i^{1*2^0*a_1*2^1}\ket{\bitone}
\right) \nonumber\\
 &=& {1\over\sqrt{2}}\left(\strutlike{$i^{0*2^0*a_1*2^1}$}
 (-1)^{0*a_1}\ket{\bitzero}
  + (-1)^{1*a_1}\ket{\bitone}
\right) \nonumber\\
 &=& {1\over\sqrt{2}}\left(\strutlike{$i^{0*2^0*a_1*2^1}$}
 \ket{\bitzero}
  + (-1)^{a_1}\ket{\bitone}\right).
\end{eqnarray}
The final Hadamard transform followed by a measurement therefore
results in the bit $a_1$, as desired.

The elements that we used to determine the order of $8$ modulo $15$
can be combined and generalized to determine the order of any $q$
modulo $N$ with $\textrm{gcd}(q,N)=1$.  The general network is shown
in Fig.~\ref{fig:qorder}. Two features of the generalization are not
apparent from the example. First, in order for the quantum network to
be efficient, an efficient implementation of the controlled ${\hat
f}^{2^l}$ operation is required.  To obtain such an implementation,
first note that to calculate $f^{2^l}(x)=q^{2^l}x\mod N$ it suffices
to square $q$ repeatedly modulo $N$ using $\left(q^{2^m}\right)^2\mod
N=q^{2^{m+1}}\mod N$ until we obtain $q^{2^l}\mod N$.  The result is
then multiplied by $x \mod N$.  This computation is efficient.  For
any given $q$, it can be converted to an efficient network consisting
of Toffoli and controlled-not gates acting on the binary
representation of $x$.  The conversion can be accomplished with
standard techniques from the theory of reversible classical
computation.  The result is an efficient network for ${\hat
f}^{2^l}$. Basic network theory can then be used to implement the
controlled version of this operation~\cite{barenco:qc1995a}.

The understand the second feature, note that we were lucky that the
order of $8$ modulo $15$ was a power of $2$, which nicely matched the
measured Fourier transform we constructed on two qubits.  The measured
Fourier transform on $m$ ancilla qubits can detect exactly
only eigenvalues that are powers of the $2^{m}$'th root of unity
$e^{i\pi/2^{m-1}}$.  The phase kicked back by the controlled
operations corresponds to a $k$'th root of unity. Given a Fourier
state on the cycle of $q\mod N$, the resulting state on the ancilla
qubits has phases that go as powers of a $k$'th root of
unity. Fortunately, the ancilla's Fourier basis is such that the
measured Fourier transform picks up primarily those basis states whose
generating phase is close to the kick back phase.  Thus we are likely
to detect a nearby $\omega=e^{i\,l\pi/2^{m-1}}$. It is still necessary
to infer (a divisor of) $k$ from knowledge of such an $\omega$.  Since
we know that the order $k$ is bounded by $N$, the number of possible
phases kicked back that are near the measured $\omega$ is limited.  To
ensure that there is only one possible such phase, it is necessary to
choose $m$ such that $2^m > N^2$. See also the caption of
Fig.~\ref{fig:qorder}.

\newcommand{\qordercap}{Network for quantum order finding and phase
estimation.  The number $m$ of qubits used for the phase kick back has
to be chosen such that $m>2*\log_2(k_u)$, where $k_u$ is a known upper
bound on the order $k$ of $q\mod N$. Because $N>k$, one can set
$m=2\lceil\log_2(N)\rceil$, where $\lceil x\rceil$ is the
least whole number $s\geq x$.  There is an eigenvalue
$\lambda_l=e^{i\, 2l\pi/k}$ of one of the Fourier eigenvectors
associated with the cycle of $q\mod N$ such that the number $a$ whose
binary digits are the measurement outcomes satisfies $e^{i\pi
a/2^{m-1}}\approx e^{i\,2\pi l/k}$.  More precisely, with probability
above $.405$, there exists $l$ such that $|a/2^{m}-l/k|\leq
1/2^{m+1}$~\cite{cleve:qc1997b}.  Since any two distinct rational
numbers with denominator at most $k_u$ differ by at least $1/k_u^2 >
2/2^{m+1}$, the theory of rational approximations guarantees that we
can uniquely determine the number $l/k$.  There is an efficient
classical algorithm based on continued fractions that computes $r$ and
$s$ with $r/s=l/k$ and $s=k/\textrm{gcd}(l,k)$.  The probability that
$\textrm{gcd}(l,k)=1$ is at least $1/(\log_2(k)+1)$, \ignore{

Asymptotically it is even $\Omega(1/\log(log(k)))$ (see Hardy\&Wright,
Sect. 19.4). But to get $1/\log_2(k)$, note that the desired
probability is given by $\Pi_{i=1}^l (1-{1/p_i})$, where the $p_i$ are
the distinct prime divisors of $k$.  This is at least
$\Pi_{i=2}^l(1-1/i)$, with $l!\leq k$.  The product collapses to
$1/l\geq 1/(\log_2(k)+1)$, because $2^{l-1}\leq l!\leq k$.

} in which case we
learn that $s=k$, and this is the order of $q\mod N$.  Note that the
complexity of the network depends on the complexity of implementing
the controlled ${\hat f}^{2^l}$ operations.  Because these operations
can be implemented efficiently, the network and hence the
determination of the order of $q\mod N$ are efficient in the sense that
on average, polynomial resources in $\log_2(N)$ suffice.}

\begin{latexonly}
\begin{herefig}
\refstepcounter{herefignum}
\label{fig:qorder}
\rotatebox{90}{\scalebox{.8}{
\begin{picture}(11,6)(-.5,-7)
\nputgr{0,0}{tl}{}{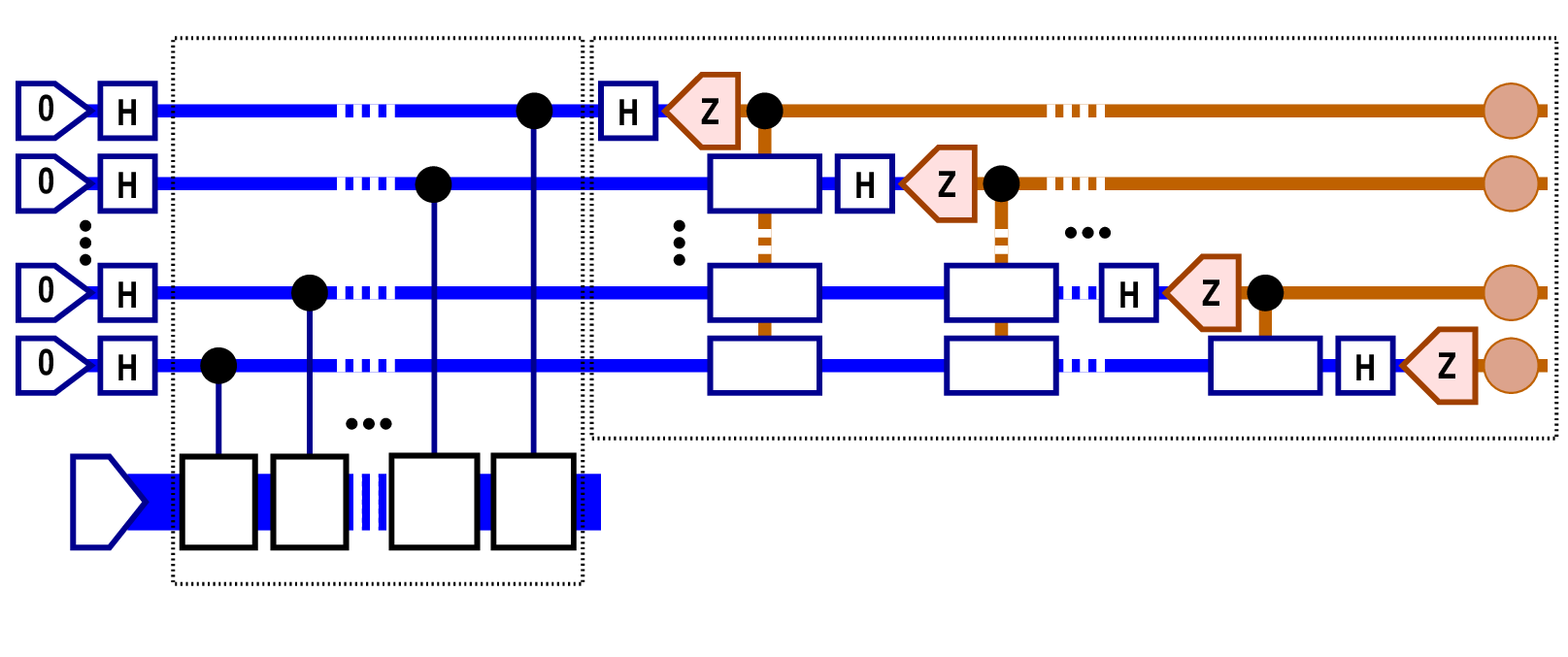}
\nputbox{0,-.5}{bl}{\scalebox{1.4}{State prep.}}
\nputbox{1.5,-.5}{bl}{\scalebox{1.4}{\strutlike{p}Phase kick back.}}
\nputbox{5.5,-.5}{bl}{\scalebox{1.4}{\strutlike{p}Measured Fourier transform.}}
\nputbox{.85,-3.4375}{r}{\scalebox{1.4}{$\ket{1}$}}
\nputbox{1.5,-3.4375}{c}{\scalebox{1.4}{$\hat f$}}
\nputbox{2.125,-3.4375}{c}{\scalebox{1.4}{${\hat f}^2$}}
\nputbox{2.975,-3.4375}{c}{\scalebox{1.4}{${\hat f}^{2^{m\!-\!2}}$}}
\nputbox{3.6625,-3.4375}{c}{\scalebox{1.4}{${\hat f}^{2^{m\!-\!1}}$}}
\nputbox{5.275,-1.25}{c}{\scalebox{1.1}{$e^{\!{-}i\pi{/}2}$}}
\nputbox{5.275,-2}{c}{\scalebox{1.1}{$e^{\!{-}i\pi{/}2^{m{-}2}}$}}
\nputbox{5.275,-2.5}{c}{\scalebox{1.1}{$e^{\!{-}i\pi{/}2^{m{-}1}}$}}
\nputbox{6.9,-2}{c}{\scalebox{1.1}{$e^{\!{-}i\pi{/}2^{m{-}3}}$}}
\nputbox{6.9,-2.5}{c}{\scalebox{1.1}{$e^{\!{-}i\pi{/}2^{m{-}2}}$}}
\nputbox{8.7125,-2.5}{c}{\scalebox{1.1}{$e^{\!{-}i\pi{/}2}$}}
\nputbox{10.375,-.75}{c}{\scalebox{1.2}{$a_0$}}
\nputbox{10.375,-1.25}{c}{\scalebox{1.2}{$a_1$}}
\nputbox{10.39,-2}{c}{\scalebox{1.2}{$a_{m\!-\!2}$}}
\nputbox{10.375,-2.5}{c}{\scalebox{1.2}{$a_{m\!-\!1}$}}
\nputbox{5.5,-4.5}{t}{\large
  \begin{minipage}{8.75in}\setlength{\baselineskip}{18pt}
FIG.~\theherefignum: \qordercap
\end{minipage}
}%
\end{picture}%
}%
}%
\end{herefig}
\end{latexonly}

\latexignore{
\begin{herefig}
\begin{picture}(11,6)(-.5,-7)
\nputgr{0,0}{tl}{}{qorder}
\nputbox{0,-.5}{bl}{\scalebox{1.4}{State prep.}}
\nputbox{1.5,-.5}{bl}{\scalebox{1.4}{\strutlike{p}Phase kick back.}}
\nputbox{5.5,-.5}{bl}{\scalebox{1.4}{\strutlike{p}Measured Fourier transform.}}
\nputbox{.85,-3.4375}{r}{\scalebox{1.4}{$\ket{1}$}}
\nputbox{1.5,-3.4375}{c}{\scalebox{1.4}{$\hat f$}}
\nputbox{2.125,-3.4375}{c}{\scalebox{1.4}{${\hat f}^2$}}
\nputbox{3,-3.4375}{c}{\scalebox{1.4}{${\hat f}^{2^{m{-}1}}$}}
\nputbox{3.6875,-3.4375}{c}{\scalebox{1.4}{${\hat f}^{2^m}$}}
\nputbox{5.275,-1.25}{c}{\scalebox{1.1}{$e^{{-}i\pi{/}2}$}}
\nputbox{5.275,-2}{c}{\scalebox{1.1}{$e^{{-}i\pi{/}2^{m-1}}$}}
\nputbox{5.275,-2.5}{c}{\scalebox{1.1}{$e^{{-}i\pi{/}2^{m}}$}}
\nputbox{6.9,-2}{c}{\scalebox{1.1}{$e^{{-}i\pi{/}2^{m-2}}$}}
\nputbox{6.9,-2.5}{c}{\scalebox{1.1}{$e^{{-}i\pi{/}2^{m-1}}$}}
\nputbox{8.7125,-2.5}{c}{\scalebox{1.1}{$e^{{-}i\pi{/}2}$}}
\nputbox{10.375,-.75}{c}{\scalebox{1.2}{$a_0$}}
\nputbox{10.375,-1.25}{c}{\scalebox{1.2}{$a_1$}}
\nputbox{10.39,-2}{c}{\scalebox{1.2}{$a_{m\!-\!1}$}}
\nputbox{10.375,-2.5}{c}{\scalebox{1.2}{$a_m$}}
\end{picture}%
\herefigcap{\qordercap}
\label{fig:qorder}
\end{herefig}
}

\pagebreak

\section{Advantages of Quantum Information}
\label{sect:aqi}

The notion of quantum information as explained in this primer was
established in the 1990s. It emerged from research focused on
understanding how physics affects our capabilities to communicate and
to process information. The recognition that usable types of
information need to be physically realizable was repeatedly emphasized
by R.~Landauer who proclaimed that ``information is
physical''~\cite{landauer:qc1991a}.  Beginning in the 1960s,
R.~Landauer studied the thermodynamic cost of irreversible operations
in computation~\cite{landauer:qc1961a}.  C.~Bennett showed that by
using reversible computation, this cost can be
avoided~\cite{bennett:qc1973a}.  Limitations of measurement in quantum
mechanics were investigated early by researchers such as J. von
Neumann~\cite{vonneumann:qc1932a,vonneumann:qc1932b}, and later by
A.~Holevo~\cite{holevo:qc1973a} and
C.~Helstrom~\cite{helstrom:qc1976a}.  A.~Holevo introduced the idea of
quantum communication channels and found bounds on their capacity for
transmitting classical information~\cite{holevo:qc1973b}.  Initially,
most work focused on determining the
physical limitations placed on classical information
processing.  The fact that pairs of two-level systems can have
correlations not possible for classical systems was proven by
J.~Bell~\cite{bell:qc1964a} in 1964.  Subsequently, indications that
quantum mechanics offers advantages to information processing came
from S.~Wiesner's studies of cryptographic
applications~\cite{wiesner:qc1983a} in the late 1960s. S.~Wiesner's
work was not recognized until the 1980s, when C.~Bennett, G.~Brassard,
S. Breidbart and S.~Wiesner~\cite{bennett:qc1982a} introduced the idea
of quantum cryptography, which can be used to communicate in
secret.

Initially, the term ``quantum computation'' was mostly used to refer
to classical computers realized using quantum mechanical systems. In
the 1980s, P.~Benioff~\cite{benioff:qc1980a},
R.~Feynman~\cite{feynman:qc1982a} and Y. I. Manin~\cite{manin:qc1980a}
introduced the idea of a quantum computer based on quantum
information.  They noted that the apparent exponential complexity of
simulating quantum mechanics on a classical computer might be overcome
if we could use a computer that is itself based on quantum mechanics.
A formal model of quantum Turing machines was soon defined by
D.~Deutsch~\cite{deutsch:qc1985a}, who later also introduced quantum
networks~\cite{deutsch:qc1989a}. D.~Deutsch and
R.~Jozsa~\cite{deutsch:qc1992a} were the first to introduce a black
box problem that can be solved deterministically on a quantum
computer in fewer steps than on a classical computer.

In spite of suggestions that it could lead to large efficiency
improvements in simulating physics, quantum information processing was
still largely an academic subject. Based on work by E.~Bernstein and
U.~Vazirani~\cite{bernstein:qc1993b} that formalized quantum
complexity theory, D.~Simon~\cite{simon:qc1994a} showed that, for
black-box problems, quantum computers can be exponentially more
efficient than classical deterministic or probabilistic computers,
giving the first indication of a strong advantage for quantum
information processing.  It was Shor's algorithm for factoring large
whole numbers~\cite{shor:qc1994a,shor:qc1995a} that finally convinced
a larger community that quantum information was more than just a tool
for realizing classical computers.  This change in attitudes was in no
small part due to the fact that the security of commonly used
cryptographic protocols is based on the hardness of factoring.

At that point, it was still generally believed that the
fragility of quantum states made it unlikely for reasonably large
quantum computers to be realized in practice.  But the discovery by
Shor~\cite{shor:qc1995b} and A.~Steane~\cite{steane:qc1995a} that
quantum error-correction was possible soon changed that view,
see~\cite{knill:qc2001d} for an introductory overview.

As a result of the recognition of the utility and realizability of
quantum information, the science of quantum information processing is
a rapidly growing field. As quantum information becomes increasingly
accessible by technology, its usefulness will be more
apparent. The next few sections briefly discuss what we currently know about
applications of quantum information processing.  A useful reference
text on quantum computation and information with historical
notes is the book by M.~Nielsen and I.~Chuang~\cite{nielsen:qc2001a}.

\subsection{Quantum Algorithms}

Shor's factoring algorithm, which precipitated much of the current
work in quantum information processing, is based on a quantum
realization of the fast Fourier transform.  The most powerful version
of this technique is now represented by the phase-estimation algorithm
of A.~Kitaev~\cite{kitaev:qc1995a} as formalized by R.~Cleve \emph{et
al.}~\cite{cleve:qc1997b}. See Sect.~\ref{sec:factor} for an
explanation of the factoring algorithm and phase estimation.  The best
known application of quantum factoring is to cryptanalysis, where it
can be used to efficiently break the currently used public-key
cryptographic codes.  Whether there are any constructive applications
of quantum factoring and its generalizations remains to be determined.
For users of public key cryptography, a crucial question is: ``How long
can public key codes based on factoring continue to be used safely?'' To
attempt to answer this question, one can note that to break a code
with a typical key size of $1000$ bits requires more than $3000$
qubits and $10^8$ quantum gates, which is well out of reach of current
technology.  However, it is conceivable that a recording of encrypted
information transmitted in 2000 can be broken in the next ``few''
decades.

Shor's quantum factoring algorithm was not the first with a
significant advantage over classical algorithms.  The first quantum
algorithms to be proposed with this property were algorithms for
simulating quantum mechanical systems. These algorithms simulate the
evolution of a reasonably large number of interacting quantum
particles, for example, the electrons and nuclei in a molecule. The
algorithms' outputs are what would be measurable physical quantities
of the system being simulated.  The known methods for obtaining these
quantities on classical computers scale exponentially with the number
of particles, except in special cases.

The idea of using quantum computers for simulating quantum physics
spurred the work that eventually lead to the quantum factoring
algorithm.  However, that idea did not have the broad scientific
impact that the quantum factoring algorithm had. One reason is that because
of its cryptographic applications, factoring is a heavily studied
problem in theoretical computer science and cryptography.  Because so
many people have tried to design efficient algorithms for factoring
and failed, the general belief that factoring is hard for classical
computers has a lot of credibility. In contrast, the problem of
quantum physics simulation has no simple formulation as an algorithmic
problem suitable for study in theoretical computer
science. Furthermore, many researchers still believe that the
physically relevant questions can be answered with efficient classical
algorithms, requiring only more cleverness on the part of the
algorithms designers.  Another reason for the lack of impact is that
many of the fundamental physical quantities of interest are not known
to be efficiently accessible even on quantum computers. For example,
one of the first questions about a physical system with a given
Hamiltonian (energy observable), is: What is the ground state energy?
It is known that the ability to efficiently answer this question for
physically reasonable Hamiltonians leads to efficient algorithms for
hard problems such as the traveling salesman or the scheduling
problems. In spite of occasional claims to the contrary, an efficient
quantum solution to these problems is widely considered unlikely.

Most quantum algorithms for physics simulations are based on a direct
emulation of the evolution of a quantum mechanical system.  The focus
of the original proposals by Feynman and others was on how to implement
the emulation using a suitable formulation of general-purpose quantum
computers. After such computers were formalized by
Deutsch, the implementation of the emulation was generalized and
refined by S.~Lloyd~\cite{lloyd:qc1996a},
Wiesner~\cite{wiesner:qc1996a} and C.~Zalka~\cite{zalka:qc1996b}.
The ability to emulate the evolution of quantum systems is actually
widely used by classical ``Monte-Carlo'' algorithms for simulating physics,
where the states amplitudes are, in effect, represented by
expectations of random variables that are computed during the
simulation. As in the case of the quantum algorithms for physics emulation,
the Monte-Carlo algorithms efficiently evolve the representation of
the quantum system. The inefficiency of the classical algorithm arises
only in determining a physical quantity of interest.  In the case of
Monte-Carlo algorithms, the ``measurement'' of a physical quantity
suffers from the so-called ``sign problem'', often resulting in
exponentially large, random errors that can be reduced only by
repeating the computation extremely many times.  In contrast,
the quantum algorithms for emulation can determine many (but not all)
of the interesting physical quantities with polynomially bounded
statistical errors. How to efficiently implement measurements of these
quantities has been the topic of more recent work in this area, much
of which is based on variants of the phase estimation
algorithm~\cite{terhal:qc1998a,knill:qc1998c,abrams:qc1999a,ortiz:qc2001a,miquel:qc2001a}.

Although several researchers have suggested that there are interesting
quantum physics simulations that can be implemented with well below
100 qubits, one of the interesting problems in this area of research
is to come up with a specific simulation algorithm that uses small numbers
of qubits and quantum gates, and that computes an interesting physical
quantity not easily obtainable using available classical computers.

Another notable algorithm for quantum computers, unstructured quantum
search, was described by L.~Grover~\cite{grover:qc1995a}.  Given is a
black box that computes a binary function $f$ on inputs $x$ with
$0\leq x<N$. The function $f$ has the property that there is a unique
input $a$ for which $f(a)=1$.  The standard quantum version of this
black box implements the transformation $\hat
f\ket{x}\ket{b}=\ket{x}\ket{b\oplus f(x)}$, where $b$ is a bit and
$b\oplus f(x)$ is computed modulo $2$.  Unstructured quantum search
finds $a$ quadratically faster, that is, in time of order $N^{1/2}$,
than the best classical black-box search, which requires time of order
$N$.  The context for this algorithm is the famous $P\not=NP$
conjecture, which is captured by the following algorithmic problem:
Given is a classical circuit $C$ that computes an output. Is there an
input to the circuit for which the circuit's output is $\bitone$?
Such an input is called a ``satisfying'' input or ``assignment''. For
any given input, it is easy to check the output, but an efficient
algorithm that finds a satisfying input is conjectured to be
impossible. This is the $P\not=NP$ conjecture. Generalizations of
Grover's search algorithm (``amplitude
amplification''~\cite{brassard:qc1998a}) can be used to find
satisfying inputs faster than the naive classical search, which tries
each possible input in some, possibly random, order.  It is worth
noting, howoever, that if sufficient classical parallelism is
available, quantum search loses many of its advantages.

The three algorithms just described capture essentially all the known
algorithmic advantages of quantum computers. Almost all algorithms
that have been described are applications of phase estimation or of
amplitude amplification.  These algorithms well justify developing
special purpose quantum information processing technology. Will
general purpose quantum computers be useful?  More specifically, what
other algorithmic advantages do quantum computers have?

\subsection{Quantum Communication}

Quantum communication is an area in which quantum information has proven
(rather than conjectured) advantages. The best known application is
quantum cryptography, whereby two parties, Alice and Bob, can generate a
secret key using a quantum communication channel (for example, photons
transmitted in optical fiber) and an authenticated classical channel
(for example, a telephone line). Any attempt at learning the key by
eavesdropping is detected. A quantum protocol for generating a secret
key is called a ``quantum key exchange'' protocol. There are no equally
secure means for generating a secret key by using only classical deterministic
channels.  Few quantum operations are needed to implement quantum key
exchange, and as a result there are working prototype
systems~\cite{hughes:qc2000a,townsend:qc1998a,ribordy:qc2001a}.  To
overcome the distance limitations (tens of kilometers) of current
technology requires the use of quantum error-correction and hence more
demanding quantum technology.

Quantum key exchange is one of an increasing number of multi-party
problems that can be solved more efficiently with quantum information.
The area of research concerned with how several parties at different
locations can solve problems while minimizing communication resources
is called ``communication complexity''.  For quantum communication
complexity (R. Cleve and H. Burhman~\cite{cleve:qc1997c}), the
communication resources include either shared entangled qubits or a
means for transmitting quantum bits.  A seminal paper by Burhman,
Cleve and W.~Van~Dam~\cite{buhrman:qc2000a} shows how the
non-classical correlations present in maximally entangled states lead
to protocols based on such states that are more efficient than any
classical deterministic or probabilistic protocol achieving the same
goal.  R. Raz~\cite{raz:qc1999a} showed that there is an exponential
improvement in communication resources for a
problem in which Alice and Bob have to answer a question about the
relationship between a vector known to Alice and a matrix known to
Bob.  Although this problem is artificial, it suggests that there are
potentially useful advantages to be gained from quantum information in
this setting.

\subsection{Quantum Control}

According to G.~Moore's law of semiconductor technology, the size of
transistors is decreasing exponentially, by a factor of about $.8$
every year. If this trend continues, then over the next few decades
devices will inevitably be built whose behavior will be primarily
quantum mechanical. For the purpose of classical computation, the goal
is to remove the quantum behavior and stabilize classical
information. But quantum information offers an alternative: It is
possible to directly use the quantum effects to advantage.  Whether or
not this advantage is useful (and we believe it is), the ideas of quantum
information can be used to systematically understand and control
quantum mechanical systems.

The decreasing size of semiconductor components is a strong motivation
to strive for better understanding the behavior of condensed
matter quantum mechanical systems. But there is no reason to wait for
Moore's law: There are a rapidly increasing number of experimental
systems in which quantum mechanical effects are being used and
investigated.  Examples include many optical devices (lasers,
microwave cavities, entangled photon pairs), nuclear magnetic
resonance with molecules or in solid state, trapped ion or atom
systems, Rydberg atoms, superconducting devices (Josephson junctions,
SQUIDs) and spintronics (electron spins in semiconductor devices).
Many of these systems are being considered as candidates for realizing
quantum information processing.  Yet, regardless of the future of quantum
information processing, there is ample motivation for studying these
systems.

\subsection{Outlook}

The science of quantum information processing is promising to have a
significant impact on how we process information, solve algorithmic
problems, engineer nano-scale devices and model fundamental physics.
It is already changing the way we understand and control matter at the
atomic scale, making the quantum world more familiar, accessible and
understandable. Whether or not we do most of our everyday computations 
by using the classical model, it is likely that the physical devices
that support these computations will exploit quantum mechanics and
integrate the ideas and tools that have been developed for quantum
information processing.

\vspace*{\baselineskip}

\noindent{\bf Acknowledgements}: We thank Nikki Cooper and Ileana Buican
for their extensive encouragement and editorial help.

\vspace*{\baselineskip}

\makeaddress

\bibliographystyle{unsrt}
\bibliography{journalDefs,qc}

\section{Glossary}
\label{sect:glossary}
\begin{description}
\setlength{\itemsep}{0pt}\setlength{\parskip}{0pt}\setlength{\parsep}{0pt}

\item[\textbf{Algorithm}.]
A set of instructions to be executed by a computing device.
What instructions are available depends on the computing device.
Typically, instructions include commands for manipulating
the contents of memory and means for repeating blocks of
instructions indefinitely or until a desired condition
is met.

\item[\textbf{Amplitude}.]
A quantum system with a chosen orthonormal basis of ``logical'' states
$\ket{i}$ can be in any superposition $\sum_i\alpha_i\ket{i}$ of these
states, where $\sum_i|\alpha_i|^2=1$. In such a superposition, the
complex numbers $\alpha_i$ are called the amplitudes.  Note that the
amplitudes depend on the chosen basis.

\item[\textbf{Ancillas}.]
Helper systems used to assist in a computation involving other
information systems. 

\item[\textbf{Bell basis}.]
For two qubits $\sysfnt{A}$ and $\sysfnt{B}$,
the Bell basis consists of
the four states ${1\over\sqrt{2}}\left(\kets{\bitzero\bitzero}{AB}\pm
\kets{\bitone\bitone}{AB}\right)$ and
${1\over\sqrt{2}}\left(\kets{\bitzero\bitone}{AB}\pm
\kets{\bitone\bitzero}{AB}\right)$.

\item[\textbf{Bell states}.]
The members of the Bell basis.

\item[\textbf{Bit}.]
The basic unit of deterministic information. It is a system that can
be in one of two possible states, $\bitzero$ and $\bitone$.

\item[\textbf{Bit sequence}.] A way of combining
bits into a larger system whose constituent bits are in a specific
order.

\item[\textbf{Bit string}.]
A sequence of $\bitzero$'s and $\bitone$'s that represents a state of a
bit sequence. Bit strings are the words of a binary alphabet.

\item[\textbf{Black box}.]
A computational operation whose implementation is unknown.  Typically,
a black box implements one of a restricted set
of operations, and the goal is to determine which of these operations
it implements by using it with different inputs. Each use of the black
box is called a ``query''.  The smallest number of queries required to
determine the operation is called the ``query complexity'' of the
restricted set. Determining the query complexity of sets of operations
is an important problem area of computational complexity.

\item[\textbf{Bloch sphere}.]
The set of pure states of a qubit represented as points
on the surface of the unit sphere in three dimensions.

\item[\textbf{Bra}.]
A state expression of the form $\bra{\psi}$, which
is considered to be the conjugate transpose of
the ket expression $\ket{\psi}$.

\item[\textbf{Bra-ket notation}.]
A way of denoting states and operators of quantum systems with
kets (for example, $\ket{\psi}$) and bras (for example, $\bra{\phi}$).

\item[\textbf{Circuit}.]
A combination of gates to be applied to information units in a
prescribed order.  To draw circuits, one often uses a convention for
connecting and depicting gates. See also ``network''.

\item[\textbf{Circuit complexity}.]
The circuit complexity of an operation on a fixed number of
information units is the smallest number of gates required to
implement the operation.

\item[\textbf{Classical information}.]
The type of information based on bits and bit strings and more
generally on words formed from finite alphabets. This is the
information used for communication between people.  Classical
information can refer to deterministic or probabilistic information,
depending on the context.

\item[\textbf{Computation}.]
The execution of the instructions provided by an algorithm.

\item[\textbf{Computational states}.] See the entry
for ``logical states''.

\item[\textbf{Computer}.]
A device that processes information.

\item[\textbf{Density matrix or operator}.]
A representation of pure and mixed states without redundancy.
For a pure state $\ket{\psi}$, the corresponding
density operator is $\ketbra{\psi}{\psi}$. A general density operator
is a probabilistic combination $\sum_i\lambda_i\ketbra{\psi_i}{\psi_i}$,
with $\sum_i\lambda_i=1$.

\item[\textbf{Deterministic information}.]
The type of information that is based on bits and bit strings. Deterministic
information is classical, but it explicitly excludes probabilistic
information.

\item[\textbf{Distinguishable states}.]
In quantum mechanics, two states are considered distinguishable
if they are orthogonal. In this case, a measurement exists
that is guaranteed to determine which of the two states a system
is in.

\item[\textbf{Efficient computation}.]
A computation is efficient if it requires at most polynomially many
resources as a function of input size.  For example, if the
computation returns the value $f(x)$ on input $x$, where $x$ is a
bit string, then it is efficient if there exists a power $k$ such
that the number of computational steps used to obtain $f(x)$ is
bounded by $|x|^k$, where $|x|$ is the length (number of bits) of $x$.

\item[\textbf{Entanglement}.]
A non-classical correlation between two quantum systems most strongly
exhibited by the maximally entangled states such as the Bell states
for two qubits, and considered to be absent in mixtures of product
states (which are called ``separable'' states).  Often states that are
not separable are considered to be entangled.  However, nearly
separable states do not exhibit all the features of maximally
entangled states. As a result, studies of different types of
entanglement are an important component of quantum information theory.

\item[\textbf{Gate}.]
An operation applied to information for the purpose
of information processing.

\item[\textbf{Global phase}.]
Two quantum states are indistinguishable if they differ only by a
global phase. That is, $\ket{\psi}$ and $e^{i\phi}\ket{\psi}$ are in
essence the same state. The global phase difference is the factor
$e^{i\phi}$.  The equivalence of the two states
is apparent from the fact that their density matrices are the
same.

\item[\textbf{Hilbert space}.]
An $n$-dimensional Hilbert space consists of all complex
$n$-dimensional vectors. A defining operation in a Hilbert space is
the inner product.  If the vectors are thought of as column vectors,
then the inner product $\langle x,y\rangle$ of $x$ and $y$ is obtained
by forming the conjugate transpose $x^\dagger$ of $x$ and calculating
$\langle x,y\rangle=x^\dagger y$. The inner product induces the usual
squared norm $|x|^2 = \langle x,x\rangle$.

\item[\textbf{Information}.]
Something that can be recorded, communicated, and computed
with. Information is fungible; that is, its meaning can be identified
regardless of the particulars of the physical realization. Thus,
information in one realization (such as ink on a sheet of paper) can
be easily transferred to another (for example, spoken words). Types of
information include deterministic, probabilistic and quantum
information. Each type is characterized by ``information units'',
which are abstract systems whose states represent the simplest
information of each type. The information units define the ``natural''
representation of the information.  For deterministic information the
information unit is the bit, whose states are symbolized by $\bitzero$
and $\bitone$. Information units can be put together to form larger
systems and can be processed with basic operations acting on a small
number of them at a time.

\item[\textbf{Inner product}.]
The defining operation of a Hilbert space. In a finite dimensional
Hilbert space with a chosen orthonormal basis $\{e_i: 1\leq
i\leq n\}$, the inner product of two vectors $x=\sum_i x_i e_i$ and
$y=\sum_i y_i e_i$ is given by $\sum_i \overline x_i y_i$.  In the standard
column representation of the two vectors, this is the number obtained
by computing the product of the conjugate transpose of $x$ with
$y$. For real vectors, this agrees with the usual ``dot'' product.
The inner product of $x$ and $y$ is often written in the form $\langle
x, y\rangle$. Pure quantum states are unit vectors in a Hilbert
space. If $\ket{\phi}$ and $\ket{\psi}$ are two quantum states
expressed in the ket-bra notation, there inner product is given by
$\left(\ket{\phi}\right)^\dagger\ket{\psi}=\braket{\phi}{\psi}$.

\item[\textbf{Ket}.]
A state expression of the form $\ket{\psi}$ representing a quantum
state. Usually $\ket{\psi}$ is thought of as a superposition of
members of a logical state basis $\ket{i}$.  One way to think about
the notation is to consider the two symbols ``$\qvbar$'' and
``$\qrangle$'' as delimiters denoting a quantum system and $\psi$ as a
symbol representing a state in a standard Hilbert space.  The
combination $\ket{\psi}$ is the state of the quantum system associated
with $\psi$ in the standard Hilbert space via a fixed isomorphism.  In
other words, one can think of $\psi\leftrightarrow \ket{\psi}$ as an
identification of the quantum system's state space with the standard
Hilbert space.

\item[\textbf{Linear extension of an operator}.]
The unique linear operator that implements a map defined on a
basis. Typically, we define an operator $U$ on a quantum system only
on the logical states $U:\ket{i}\mapsto\ket{\psi_i}$.  The linear
extension is defined by
$U(\sum_i\alpha_i\ket{i})=\sum_i\alpha_i\ket{\psi_i}$.

\item[\textbf{Logical states}.]
For quantum systems used in information processing, the logical states
are a fixed orthonormal basis of pure states. By convention, the
logical basis for qubits consists of $\ket{\bitzero}$ and
$\ket{\bitone}$.  For larger dimensional quantum systems, the logical
basis is often indexed by the whole numbers,
$\ket{0},\ket{1},\ket{2},\ldots$.  The logical basis is often also
called the ``computational'' basis, or sometimes, the ``classical''
basis.

\item[\textbf{Measurement}.]
The process used to extract classical information from a quantum
system. A general projective measurement is defined by a set of
projectors $P_i$ satisfying $\sum_iP_i=\id$ and
$P_iP_j=\delta_{ij}P_i$. Given the quantum state $\ket{\psi}$, the
outcome of a measurement with the set $\{P_i\}_i$ is one of the classical
indeces $i$ associated with a projector $P_i$.  The index $i$ is the
measurement outcome. The probability of outcome $i$ is
$p_i=|P_i\ket{\psi_i}|^2$, and given outcome $i$, the quantum state
``collapses'' to $P_i\ket{\psi_i}/\sqrt{p_i}$.

\item[\textbf{Mixture}.]
A probabilistic combination of pure states of a quantum system.
Mixtures can be represented without redundancy with density
operators. Thus a mixture
is of the form $\sum_i\lambda_i\ketbra{\psi_i}{\psi_i}$,
with $\lambda_i\geq 0$, $\sum_i\lambda_i=1$ being the probabilities
of the states $\ket{\psi_i}$. This expression for mixtures defines the set
of density operators, which can also be characterized 
as the set of operators $\rho$ satisfying $\trace(\rho)=1$ and
for all $\ket{\psi}$, $\bra{\psi}\rho\ket{\psi}\geq 0$ (``positive
semidefinite operator'').

\item[\textbf{Network}.]
In the context of information processing, a network is a sequence of
gates applied to specified information units.  We visualize networks 
by drawing horizontal lines to denote the time line of an
information unit. The gates are represented by graphical elements
that intercept the lines at specific points. A realization of the
network requires applying the gates to the information units in the
specified order (left to right).

\item[\textbf{Operator}.]
A function that transforms the states of a system. Operators may
be restricted depending on the system's properties. For example,
in talking about operators acting on quantum systems, one always
assumes that they are linear.

\item[\textbf{Oracle}.]
An information processing operation that can be applied.  A use of the
oracle is called a ``query''.  In the oracle model of computation, a
standard model is extended to include the ability to query an oracle.  Each
oracle query is assumed to take one time unit. Queries can reduce the
resources required for solving problems.  Usually, the oracle
implements a function or solves a problem not efficiently
implementable by the model without the oracle.  Oracle models are used
to compare the power of two models of computation when the oracle can
be defined for both models. For example, in 1994, D. Simon showed that
quantum computers with a specific oracle $\cO$ could efficiently solve
a problem that had no efficient solution on classical computers with
access to the classical version of $\cO$. At the time, this result was
considered to be the strongest evidence for an exponential gap in
power between classical and quantum computers.

\item[\textbf{Overlap}.]
The inner product between two quantum states.

\item[\textbf{Pauli operators}.]
The Hermitian matrices $\sigma_x,\sigma_y,\sigma_z$ acting on qubits,
which are two-level quantum systems. They are defined in
Eq.~\ref{eq:paulidef}. It is often convenient to consider the
identity operator to be included in the set of Pauli operators.

\item[\textbf{Polynomial resources}.]
To say that an algorithm computing the function $f(x)$, where $x$ is a
bit string, uses polynomial resources (in orther words, ``is
efficient'') means that the number of steps required to compute $f(x)$
is bounded by $|x|^k$ for some fixed $k$. Here $|x|$ denotes the
length of the bit string $x$.

\item[\textbf{Probabilistic bit.}]
The basic unit of probabilistic information.  It is a system whose
state space consists of all probability distributions over the two
states of a bit. The states can be thought of as describing the
outcome of a biased coin flip before the coin is flipped.

\item[\textbf{Probabilistic information}.]
The type of information obtained by extending the state
spaces of deterministic information to include arbitrary
probability distributions over the deterministic states.
This is the main type of classical information to which
quantum information is compared.

\item[\textbf{Probability amplitude}.]
The squared norm of an amplitude with respect to a chosen orthonormal
basis $\{\ket{i}\}$. Thus, the probability amplitude is the
probability with which the state $\ket{i}$ is measured in a complete
measurement that uses this basis.

\item[\textbf{Product state}.]
For two quantum systems $\sysfnt{A}$ and $\sysfnt{B}$,
product states are of the form $\kets{\psi}{A}\kets{\phi}{B}$.
Most states are not of this form.

\item[\textbf{Program}.]
An algorithm expressed in a language that can be understood
by a particular type of computer.

\item[\textbf{Projection operator}.]
A linear operator $P$ on a Hilbert space that satisfies $P^2=P^\dagger
P=P$. The projection onto a subspace $V$ with orthogonal complement $W$
is defined as follows: If $x\in V$ and $y\in W$, then $P(x+y)=x$.

\item[\textbf{Pseudo-code}.]
An semi-formal computer language that is intended to be executed by a
standard ``random access machine'', which is a machine model with a
central processing unit and access to a numerically indexed unbounded
memory. This machine model is representative of the typical
one-processor computer.  Pseudo-code is similar to programming
languages such as BASIC, Pascal, or C, but does not have
specialized instructions
for human interfaces, file management, or other ``external'' devices.
Its main use is to describe algorithms and enable machine-independent
analysis of the algorithms' resource usage.

\item[\textbf{Pure state}.]
A state of a quantum system that corresponds to a unit vector in the
Hilbert space used to represent the system's state space.  In the ket
notation, pure states are written in the form
$\ket{\psi}=\sum_i\alpha_i\ket{i}$, where the $\ket{i}$ form a logical
basis and $\sum_i|\alpha_i|^2=1$.

\item[\textbf{Quantum information}.]
The type of information obtained when the state space of
deterministic information is extended by normalized superpositions of
deterministic states.  Formally, each deterministic state is
identified with one of an orthonormal basis vector in a Hilbert space
and normalized superpositions are unit-length vectors that are
expressible as complex linear sums of the chosen basis vectors.  It is
convenient to extend this state space further by permitting
probability distributions over the quantum states (see the entry for
``mixtures''). This extension is still called quantum information.

\item[\textbf{Qubit}.]
The basic unit of quantum information. It is the quantum extension of
the deterministic bit, which implies that its state space consists of
the unit-length vectors in a two dimensional Hilbert space.

\item[\textbf{Read-out}.]
A method for obtaining human-readable information from the state
of a computer. For quantum computers, read-out refers to a measurement
process used to obtain classical information about a quantum
system.

\item[\textbf{Reversible gate}.]
A gate whose action can be undone by a sequence of gates.

\item[\textbf{Separable state}.]
A mixture of product states.

\item[\textbf{States}.]
The set of states for a system characterizes the system's behavior and
possible configurations.

\item[\textbf{Subspace}.]
For a Hilbert space, a subspace is a linearly closed subset of the
vector space. The term can be used more generally for a system
$\sysfnt{Q}$ of any information type: A subspace of $\sysfnt{Q}$ or,
more specifically, of the state space of $\sysfnt{Q}$ is a subset of
the state space that preserves the properties of the information type
represented by $\sysfnt{Q}$.

\item[\textbf{Superposition principle}.]
One of the defining postulates of quantum mechanics according
to which if states $\ket{1},\ket{2},\ldots$ are distinguishable
then $\sum_i\alpha_i\ket{i}$ with $\sum_i|\alpha_i|^2=1$ is
a valid quantum state. Such a linear combination is called
a normalized superposition of the states $\ket{i}$.

\item[\textbf{System}.]
An entity that can be in any of a specified number of states. An
example is a desktop computer whose states are determined by the
contents of its various memories and disks. Another example is a
qubit, which can be thought of as a particle whose state space is
identified with complex, two-dimensional, length-one vectors.
Here, a system is always associated with a type of information
that determines the properties of the state space.
For example, for quantum information the state space
is a Hilbert space. For deterministic information,
it is a finite set called an alphabet.

\item[\textbf{Unitary operator}.]
A linear operator $U$ on a Hilbert space that preserves
the inner product. That is, $\langle Ux,Uy\rangle=\langle x,y\rangle$.
If $U$ is given in matrix form, then this expression is equivalent
to $U^\dagger U = \idop$.

\item[\textbf{Universal set of gates}.]
A set of gates that satisfies the requirement that every allowed
operation on information units can be implemented by a network of
these gates.  For quantum information, it means a set of gates that
can be used to implement every unitary operator. More generally, a set
of gates is considered universal if for every operator $U$, there are
implementable operators $V$ arbitrarily close to $U$.

\end{description} 

\end{document}